\documentclass[11pt]{article}
\usepackage{silence}
\pdfoutput=1
\usepackage{jheppub}
\usepackage{import}
\usepackage{hyperref}
\numberwithin{equation}{section}
\usepackage{hyperref}

\newcommand{\be}{\begin{equation}}
\newcommand{\ee}{\end{equation}}
\newcommand\bea{\begin{eqnarray}}
\newcommand\eea{\end{eqnarray}}
\newcommand{\nn}{\nonumber}

\begin{document}

\title{Minimal Basis for Exact Time Dependent Kernels in Cosmological Perturbation Theory and Application to $\Lambda$CDM and $w_0w_a$CDM}

\author{Michael Hartmeier, }
\author{Mathias Garny}

\affiliation{\small Physik Department T31, Technische Universit\"at M\"unchen \\
James-Franck-Stra\ss e 1, D-85748 Garching, Germany}

\emailAdd{michael.hartmeier@tum.de, mathias.garny@tum.de}

\preprint{\today, TUM-HEP-1466/23}

\abstract{We derive a minimal basis of kernels furnishing the perturbative expansion of the density contrast and velocity divergence in powers of the initial density field that is applicable to cosmological models with arbitrary expansion history, thereby relaxing the commonly adopted Einstein-de-Sitter (EdS) approximation. 
For this class of cosmological models, the non-linear kernels are at every order given by a sum of terms, each of which factorizes into a time-dependent growth factor and a wavenumber-dependent basis function. We show how to reduce the set of basis functions to a minimal amount, and give explicit expressions up to order $n=5$. We find that for this minimal basis choice, each basis function \emph{individually} displays the expected scaling behaviour due to momentum conservation, being non-trivial at $n\geq 4$. This is a highly desirable property for numerical evaluation of loop corrections. In addition, it allows us to match the density field to an effective field theory (EFT) description for cosmologies with an arbitrary expansion history, which we explicitly derive at order four. 
We evaluate the differences to the EdS approximation for $\Lambda$CDM and $w_0w_a$CDM, paying special attention to the irreducible cosmology dependence that cannot be absorbed into EFT terms for the one-loop bispectrum.
Finally, we provide algebraic recursion relations for a special generalization of the EdS approximation that retains its simplicity and is relevant for mixed hot and cold dark matter models.

}

\maketitle

\section{Introduction}
\label{sec:intro}

The interpretation of ongoing and future galaxy surveys requires precise theoretical predictions of cosmological correlation functions on weakly non-linear scales, that can be
efficiently evaluated for a large set of model parameters.
An important tool in this context is the perturbative description of large-scale structure formation~\cite{goroff1986,Bernardeau_2002}. It is based on an expansion of the matter density contrast $\delta=\rho_m/\bar\rho_m-1$ in powers of the
underlying linear density field $\delta_0$. In Fourier space the $n$th order contribution is characterized by kernels $F_n(\veck_1,\veck_2,\dots,\veck_n,\eta)$, that
describe how the product of $n$ linear fields evaluated at wavevectors $\veck_i$ generate a non-linear density contrast $\delta(\veck,\eta)$ at wavevector $\veck =\sum_i\veck_i$ and time parameterized by $\eta=\ln D_1(a)$,
where $D_1$ is the linear growth factor, $a$ the scale-factor, and $\bar\rho_m(a)$ the average matter density. The kernels $F_n$ are instrumental for computing non-linear corrections to the power- and bispectrum, both
within the framework of standard perturbation theory (SPT) as well as extensions such as effective field theory (EFT)~\cite{Baumann:2010tm,Carrasco:2012cv}.

The kernels are commonly evaluated within the so-called Einstein-de-Sitter (EdS) approximation, for which the ratio $\Omega_m(a)/f(a)^2$ is approximated by unity, where $f=d\ln D_1/d\ln a$ is the growth rate
and $\Omega_m(a)$ the time-dependent matter density parameter. While the ratio deviates from unity by about $15\%$ for a realistic $\Lambda$CDM cosmology, and can vary even more in extended models, the
EdS approximation is in practice convenient since it leads to a separation of time- and scale-dependence in the kernels at any order in perturbation theory, 
$F_n(\veck_1,\veck_2,\dots,\veck_n,\eta) \mapsto [D_1(a)]^nF_n^{\text{EdS}}(\veck_1,\veck_2,\dots,\veck_n)$. This allows one to compute the $F_n^{\text{EdS}}$ via algebraic recursion relations~\cite{goroff1986,Bernardeau_2002}.

Given the increasing precision of galaxy survey data, it is timely to investigate the quantitative impact of relaxing the EdS approximation, and implement efficient algorithms for evaluating
proper cosmology-dependent kernels $F_n$. Here one can discriminate two fundamentally different cases: \emph{(i)} for models leading to a time- and scale-dependent growth rate $D_1(a,k)$, such as
massive neutrino cosmologies, the evolution equations for the $F_n$ acquire a non-trivial scale-dependence as well. This complicates their evaluation. For the case of massive neutrinos, the kernels up to $F_5$ have been evaluated
numerically in~\cite{taule,Garny:2022fsh}\footnote{The same is true when including vorticity, velocity dispersion and higher cumulants in the perturbative description~\cite{Garny:2022tlk,Garny:2022kbk}, leading to a non-trivial scale-dependence.}, and for
modified gravity models up to $F_3$ in~\cite{Taruya:2016jdt}. We do not deal with this class of models in this work. \emph{(ii)} for models with (arbitrary) time-dependent but scale-independent growth $D_1(a)$, the kernels can be represented in the form~\cite{Skewness_and_kurtosis,Takahashi_2008}
\be\label{eq:Fnintro}
  F_n(\veck_1,\dots,\veck_n;\eta) = \sum_i d_{n,i}^{F}(\eta) h_{n,i}^{F}(\veck_1,\dots,\veck_n)\,,
\ee
with a set of generalized growth functions $d_{n,i}^{F}(\eta)$ and (algebraic) basis functions $h_{n,i}^{F}(\veck_1,\dots,\veck_n)$.
This includes cosmological models where clustering is dominated by standard cold matter on the relevant scales, while allowing for (almost) homogeneous additional species contributing
to the Hubble expansion rate $H(a)$ with an, in principle, arbitrary time-dependence. This encompasses for example $\Lambda$CDM with massless neutrinos, dynamical dark energy models (when neglecting clustering of the dark energy component(s)), or mixed hot and cold dark matter models, with
a small fraction of hot dark matter that is (approximately) homogeneously distributed on the scales probed by galaxy surveys.
The growth factors $d_{n,i}^{F}(\eta)$ in general have to be determined numerically, by solving ordinary differential equations similar as for the
linear growth function $D_1(a)$. The basis functions $h_{n,i}^{F}$ are independent of the cosmology (within the class of models where the decomposition from above can be used).

The representation of Eq.\,\eqref{eq:Fnintro} up to $n=3$ has been derived in~\cite{Skewness_and_kurtosis} and applied and discussed for various models for example in~\cite{Takahashi_2008,Fasiello_2016,Lewandowski:2016yce,Lewandowski:2017kes,Donath:2020abv,DAmico:2021rdb,Piga:2022mge,Rampf:2022tpg,Joyce:2022oal,Amendola:2023awr}.
It is in principle straightforward to obtain a decomposition of the form of Eq.\,\eqref{eq:Fnintro} also for higher $n$, following for example the algorithm from~\cite{Takahashi_2008}, see e.g.~\cite{zvonimir}.
Alternatively, one may also resort to computing the entire kernel numerically at higher order, as done for example in~\cite{taule,baldauf2021twoloop,Garny:2022fsh} for the one-loop bispectrum (requiring $n=4$) and two-loop power spectrum (requiring $n=5$) within $\Lambda$CDM.

Obtaining the representation of the form of Eq.\,\eqref{eq:Fnintro} ``naively'', following e.g. the algorithm from~\cite{Takahashi_2008} or straightforward variations of it,
actually yields a \emph{redundant set of growth factors and basis functions} (that we denote by capital $D_{n,i}^X(\eta)$ and $H_{n,i}^X(\veck_1,\dots,\veck_n)$, respectively, see below for details). One important consequence of this redundancy is that each individual basis function 
$H_{n,i}^X$ does not necessarily obey the the scaling $F_n(\veck_1,\dots,\veck_{n-2},\vecq,-\vecq)\propto \veck^2/\vecq^2$ for $\veck^2\ll\vecq^2$, where $\veck=\sum_i\veck_i$, that is guaranteed by mass and momentum conservation.
That is, the ``naive'' basis functions $H_{n,i}^X$ generically approach a non-vanishing value for $\veck^2/\vecq^2\ll 1$. The correct scaling is only recovered when summing over all $i$~\cite{zvonimir}.
This feature occurs for kernels with $n\geq 4$.

This technical shortcoming is inconvenient and unsatisfactory, both from a conceptual point of view as well as regarding the numerical computation of loop corrections:
\begin{itemize}
\item The cancellation of terms of order $(k/q)^0$ among the summands contributing to the time-dependent kernels $F_n$ becomes problematic for numerical evaluation especially at two-loop order, where it
requires to evaluate the kernels with a much higher accuracy than what is actually desired for the final result. This is also inconvenient in practice since the growth functions are computed from numerically solving a set of differential
equations, with an in general finite numerical accuracy.\footnote{A workaround for the two-loop power spectrum has been presented in~\cite{zvonimir}.}
\item Within the EFT setup, the UV limit of the kernels for $|\vecq|\to\infty$ can in general be matched to EFT corrections. The terms scaling as $(k/q)^0$ basically prohibit this matching to be performed in practice, since
their cancellation relies on implicit relations among the summands that are not manifest. This is relevant for evaluating the cosmology-dependence of EFT parameters, and for determining their ``running'' when varying the
smoothing scale $\Lambda$. It is also problematic when using EFT schemes for which the relative size of various EFT parameters is assumed to be given by the UV limit of the corresponding kernels, an assumption that has been demonstrated to work
well in the EdS context for the two-loop power spectrum~\cite{doublehardlim}, and to a certain extent for the one- and two-loop bispectrum~\cite{Steele_2021,baldauf2021twoloop}.
\item On a more practical side, the number of summands rapidly grows with $n$, and a reduction of terms is also desirable from this point of view.
\end{itemize}

In this work we derive non-trivial, universal relations among the sets of (naively constructed) growth functions at order $n=4$ and $n=5$, and use these to obtain a minimal, reduced basis
with a significantly smaller number of terms contributing to the sum over $i$ in Eq.\,\eqref{eq:Fnintro}. We demonstrate that, after using these relations, the resulting basis functions
$h_{n,i}^{F}(\veck_1,\dots,\veck_n)$ \emph{individually} satisfy all constraints dictated by mass/momentum conservation as well as Galilean invariance, such that for example for
each $h_{n,i}^{F}(\veck_1,\dots,\veck_{n-2},\vecq,-\vecq)\propto \veck^2/\vecq^2$ in the UV limit $\veck^2\ll\vecq^2$. This remedies the shortcomings of the ``naive'' approach discussed above.

We then show explicitly that each $4$th order basis function $h_{4,i}^{F}(\veck_1,\veck_{2},\vecq,-\vecq)$ can in the UV limit be matched to a linear combination of the set of shape functions
following from the four independent EFT operators contributing at this order in perturbation theory. Apart from confirming that the EFT description works also at $n=4$ for the class of cosmological models
with arbitrary time-dependent growth, this result allows us to derive ``renormalization group'' equations for how the EFT parameters change with the smoothing scale, and to assess their cosmology-dependence.

In addition, we give explicit EFT expressions for the one-loop power- and bispectrum for cosmology-dependent kernels, and show that the UV sensitivity is captured by the EFT corrections in general.
We formulate the one-loop contribution in terms of ``renormalized'' kernels and EFT parameters, with the former having a smaller UV sensitivity than their ``bare'' counterparts, and the latter
capturing mostly the physical effect of UV modes. Finally, we use this formulation to assess the irreducible cosmology-dependence of the one-loop power- and bispectrum, that cannot be absorbed
into a shift of the EFT parameters. Our results are complementary to those of~\cite{Choustikov:2023uyk}, that aims primarily at a more efficient numerical evaluation of the growth functions,
while not discussing a basis reduction as done here.

This work is organized as follows: In Sec.\,\ref{sec:setup}, we set up our notation and review the ``naive'' algorithm leading, in general, to a redundant basis
for cosmology-dependent kernels $F_n$ and $G_n$. In Sec.\,\ref{sec:red} we derive relations among the set of ``naive'' growth functions and basis functions and show how
to use them in order to arrive at a minimal, ``reduced'' basis. Furthermore we demonstrate the scaling in accordance with mass/momentum conservation for each basis
function within the reduced basis. In Sec.\,\ref{sec:eft} we discuss EFT corrections for the case of cosmology-dependent kernels, with a focus on the
one-loop power- and bispectrum, and in Sec.\,\ref{application} we present results for the impact of taking the precise time-dependence into account,
as compared to the EdS approximation, within \lcdm and \ocdm models. We conclude in Sec.\,\ref{sec:conclusion}. The appendices contain a summary of
all technical ingredients and explicit expressions for the kernels up to order $n=5$. A {\sc Mathematica} notebook generating all results contained in this
work can be provided on request.

\section{General time dependent perturbation theory}\label{sec:setup}

In this section, we first briefly review the perturbative approach to large-scale structure formation, and then discuss the generalization to cosmological models with arbitrary expansion history, as given by the (conformal) Hubble parameter $\m{H}(a)=aH(a)$ and the scale-factor dependent matter density parameter $\Omega_m(a)$.

\subsection{Standard perturbation theory}

Within the standard formulation of perturbation theory~\cite{Bernardeau_2002}, cosmic evolution is described by fluid equations for the density contrast $\delta(\veck,\tau)$ and the velocity divergence $\theta(\veck,\tau)=\nabla \cdot \vek{u}(\veck,\tau)$ in Fourier space,
\begin{align}
    \partial_\tau \delta(\veck,\tau)+\theta(\veck,\tau)&=-\int_{\veck_1}\int_{\veck_2} (2\pi)^3\delta_D(\veck-\veck_{12})\alpha(\veck_1,\veck_2)\theta(\veck_1,\tau)\delta(\veck_2,\tau),\\
    \partial_\tau\theta(\veck,\tau)+\m{H}\theta(\veck,\tau)+\frac{3}{2}\m{H}^2\Omega_m\delta(\veck,\tau)&= - \int_{\veck_1}\int_{\veck_2}(2\pi)^3\delta_D(\veck-\veck_{12})\beta(\veck_1,\veck_2)\theta(\veck_1,\tau)\theta(\veck_2,\tau) \label{euler_nl}.
\end{align}
Here $\m{H}\equiv (\partial_\tau a)/a=aH(a)$ is the conformal Hubble rate, $\tau=\int dt/a(t)$ the conformal time, $a$ the scale factor, $\Omega_m=8\pi G\rho_m(a)/(3H(a)^2)$ is the matter density parameter with Newton constant $G$, matter density $\rho_m(a)$ and Hubble rate $H(a)$, and $\delta_D$ is Dirac's delta and we use the notations $\int_{\veck}=\int \frac{\dd^3\veck}{(2\pi)^3}$ and $\veck_{12}=\veck_1+\veck_2$.  The vertices are defined as 
\begin{align}
    &\alpha(\veck_1,\veck_2)=\frac{(\veck_1+\veck_2)\cdot\veck_1}{k_1^2}, 
    &\beta(\veck_1,\veck_2)=\frac{(\veck_1+\veck_2)^2\veck_1\cdot\veck_2}{2k_1^2k_2^2}.
\end{align}
It is convenient to substitute conformal time $\tau$ by $\eta=\ln D_1$, where $D_1$ is the linear growth factor  following from the growing solution of the \mesza equation~\cite{meszaros1980},
\begin{equation}
    \frac{\partial^2 D_1(a)}{\partial \ln^2 a} + \left(\frac{\dd \ln \m{H}(a)}{\dd \ln a}+1\right)\frac{\partial D_1(a)}{\partial \ln a} - \frac{3}{2}\Omega_m(a) D_1(a)=0.
    \label{meszaroslcdm}
\end{equation}
The growth rate $f$ is defined as $f=\dd\ln D_1/\dd\ln a$. Then, with 
\be
  \psi_a=\left(\delta(\veck,\eta), -\theta(\veck,\eta)/(\m{H}(\eta)f(\eta))\right),
\ee
the fluid equations can \emph{equivalently} be written as 
\begin{equation}
    \partial_\eta \psi_a(\veck,\eta) + {\Omega_a}^b(\veck,\eta)\psi_b(\veck,\eta)= \int_{\veck_1}\int_{\veck_2}(2\pi)^3\delta_D(\veck-\veck_{12}){\gamma_a}^{bc}(\veck_1,\veck_2)\psi_b(\veck_1,\eta)\psi_c(\veck_2,\eta).
    \label{masterequation}
\end{equation}
The only non-vanishing elements of $\gammabc$ are ${\gamma_{1}}^{21}=\alpha(\veck_1,\veck_2) $ and ${\gamma_{2}}^{22}=\beta(\veck_1,\veck_2)$ and the matrix is defined as
\begin{align}
    {\Omega_a}^b(\eta)= 
    \begin{pmatrix}
        0 & -1\\- x(\eta) &  x(\eta)-1
    \end{pmatrix} \,.
\end{align}
Using this parameterization, the expansion history enters only via the function
\be\label{eq:xdef}
  x(\eta)\equiv\frac{3}{2}\frac{\Omega_m(\eta)}{f^2(\eta)}\,,
\ee
that is in turn completely determined by the time-evolution of $\m{H}$ and $\Omega_m$, i.e. the expansion rate and the total matter density~\cite{Scoccimarro:1997st}.
Within a matter dominated Einstein-de-Sitter  universe, the matter density parameter and linear growth factor would both be equal to unity and time-independent, such that
\be\label{eq:edslimit}
  x(\eta)\stackrel{\mbox{\scriptsize EdS}}{=}\frac{3}{2},
\ee
is constant in time. Since the solutions simplify considerably in this case,
this {\it EdS approximation} for $x(\eta)$ is often adopted also when considering other cosmological models. For $\Lambda$CDM this is motivated by the observation that, although both $\Omega_m$ and $f$ decrease considerably at low redshift, their ratio as in $x(\eta)$ is somewhat closer to the EdS value. Nevertheless, even for $\Lambda$CDM $x$ deviates from $3/2$ by about $10-15\%$ at $z=0$.
In the following we relax the EdS assumption, and obtain solutions that are valid for an {\it arbitrary time-dependent function} $x(\eta)$.

Note that the case with time-dependent $x(\eta)$ is, apart from $\Lambda$CDM, also relevant for extensions that modify the expansion history, such as dynamical dark energy models. The deviation of $x$ from the EdS value can in this case be even larger.
We remark however that not all models can be captured by a time-dependent $x(\eta)$. In particular, massive neutrinos lead to a time- and scale-dependent growth, which goes beyond the class of models captured in this work (see e.g.~\cite{taule,Garny:2022fsh,Chen:2022cgw} for numerical treatments of this case), unless assuming all modes are far above the neutrino free-streaming scale~\cite{Joyce:2022oal}, see below. The same is true when taking the clustering of dynamical dark energy into account~\cite{Sefusatti:2011cm,Piga:2022mge}, which is however suppressed when deviations from a cosmological constant are sufficiently small.

Perturbatively expanding the density contrast and velocity divergence leads to the formal solution,
\begin{equation}
  \left(\delta(\veck,\eta) \atop -\theta(\veck,\eta)/(\m{H}f)\right)
    =\sum_{n=1}^{\infty}\int_{\veck_1}\dots\int_{\veck_n}(2\pi)^3\delta_{D}(\veck-\veck_{1\dots n}) \left( F_n(\veck_1,\dots,\veck_n;\eta) \atop G_n(\veck_1,\dots,\veck_n;\eta)\right) \prod_{i=1}^n\delta_0(\veck_i),
    \label{psieq}
\end{equation}
expanded in powers of the initial density field $\delta_0$, and with $\veck_{1\dots n}\equiv\sum_{i=1}^n \veck_i$. 
The $F_n$ and $G_n$ denote the {\it kernels} for the $n$th order contribution to the non-linear density and rescaled velocity divergence fields, and in general depend on all $n$ wavevectors of the initial density field {\it as well as on time}.
We also use the notation 
\be
  F_{n,1}\equiv F_n\qquad \text{and} \qquad F_{n,2}\equiv G_n\,,
\ee
for convenience.
Within the EdS approximation, the time- and wavevector dependence factorizes, and one obtains
\bea
  F_n(\veck_1,\dots,\veck_n;\eta) & \stackrel{\mbox{\scriptsize EdS}}{=} & e^{n\eta}F_n^{\mbox{\scriptsize EdS}}(\veck_1,\dots,\veck_n),\nn\\
  G_n(\veck_1,\dots,\veck_n;\eta) & \stackrel{\mbox{\scriptsize EdS}}{=} & e^{n\eta}G_n^{\mbox{\scriptsize EdS}}(\veck_1,\dots,\veck_n).
\eea
Here $F_n^{\mbox{\scriptsize EdS}}$ and $G_n^{\mbox{\scriptsize EdS}}$ are the conventional non-linear kernels in EdS approximation, that can be constructed from well-known algebraic recursion relations~\cite{goroff1986}.

\subsection{Relaxing the EdS approximation}\label{sec:relax}

In this section, we go beyond the EdS approximation and present kernel recursion relations for a general time dependence as described by the function 
\be
  x(\eta)\equiv\frac{3}{2}\frac{\Omega_m(\eta)}{f^2(\eta)}\,,
\ee
defined in Eq.\,\eqref{eq:xdef}. This approach works for all cosmologies characterized by time-dependent but scale-independent growth of perturbations at the level of the linear approximation. 
More details can be found in App.\,\ref{propderiv}. 

When relaxing the EdS approximation, the kernels depend non-trivially on time. In this case the recursion relation takes the form of a coupled set of ordinary differential equations obtained from inserting Eq.\,\eqref{psieq} into the equation of motion Eq.\,\eqref{masterequation},
\begin{equation}
    (\partial_\eta + {\Omega_{a}}^b(\eta)) F_{n,b}(\veck_1,\dots,\veck_n;\eta) = S_a^{(n)}(\veck_1,\dots,\veck_n;\eta),
    \label{sourcetermdiffe}
\end{equation}
with source term that depends on all lower order kernels with $m,n-m<n$,
\begin{equation}
    S_a^{(n)}(\veck_1,\dots,\veck_n;\eta)=\sum_{m=1}^{n-1}\left[ \gammabc(\vecq_{1\dots m},\vecq_{m+1\dots n}) F_{m,b}(\vecq_1,\dots,\vecq_m;\eta)F_{n-m,c}(\vecq_{m+1},\dots,\vecq_n;\eta)\right]_{sym} \label{sourcetermdeff}.
\end{equation}
Here $[\dots]_{sym}$ stands for summing over all possibilities to choose two sets $\vecq_1,\dots,\vecq_m$ and $\vecq_{m+1},\dots,\vecq_n$ out of the wavenumbers $\veck_1,\dots,\veck_n$, and dividing by the number of possibilities given by $n!/(m!(n-m)!)$.

The formal solution of Eq.\,\eqref{sourcetermdiffe} can be written as
\begin{align}\label{eq:formalsol}
    F_{n,a}(\veck_1,\dots,\veck_n;\eta)&=\etaint {G_a}^b(\eta,\eta')S_b^{(n)}(\veck_1,\dots,\veck_n,\eta'),
\end{align}
where the lower integration boundary at $-\infty$ corresponds to neglecting transient contributions, analogous as in SPT~\cite{Scoccimarro:1997gr}, and ${G_a}^b(\eta,\eta')$ is the linear propagator given by (see App.\,\ref{propderiv} for a derivation)
\begin{align}\label{crudeprop}
    G(\eta,\eta')=
    \begin{pmatrix}
        (x(\eta')-1-\partial_{\eta'})g(\eta,\eta') & g(\eta,\eta')\\
        (x(\eta')-1-\partial_{\eta'})\partial_\eta g(\eta,\eta') & \partial_\eta g(\eta,\eta')
    \end{pmatrix},
\end{align}
with
\begin{align}\label{getaetaint}
    &g(\eta,\eta')\equiv\int^\eta_{\eta'}\text{d}\eta''e^{\eta-\eta''} \, \exp\left[ -\int_{\eta'}^{\eta''}\text{d}\eta'''x(\eta''') \right].
\end{align}
It satisfies the differential equation 
\be
  (\partial_\eta + x(\eta))(\partial_\eta-1)g(\eta,\eta') = 0\,,
\ee
with boundary conditions
\be\label{eq:gboundary}
  g(\eta',\eta')=0,\quad \partial_\eta g(\eta,\eta')|_{\eta=\eta'} = 1\,.
\ee
A further useful relation is
\be\label{eq:detaprimeg}
  (x(\eta')-\partial_{\eta'})g(\eta,\eta')=e^{\eta-\eta'}\,,
\ee
which allows us to write the propagator in the equivalent form
\be\label{eq:propresult}
  G(\eta,\eta') = e^{\eta-\eta'}\left(\begin{array}{cc} 1 & 0 \\ 1 & 0 \end{array}\right)
  + g(\eta,\eta')\left(\begin{array}{cc} -1 & 1 \\ 0 & 0 \end{array}\right)
  + \partial_\eta g(\eta,\eta') \left(\begin{array}{cc} 0 & 0 \\ -1 & 1 \end{array}\right)\,.
\ee

In the EdS limit $x(\eta)\stackrel{\mbox{\scriptsize EdS}}{=}3/2$,  Eq.\,\eqref{getaetaint} becomes  $g(\eta,\eta')\stackrel{\mbox{\scriptsize EdS}}{=}\frac25(e^{\eta-\eta'}-e^{-3/2(\eta-\eta')})$. Inserting this into Eq.\,\eqref{crudeprop} yields the usual EdS-SPT linear propagator
\begin{align}\label{eq:propEdS}
    G(\eta,\eta')\stackrel{\mbox{\scriptsize EdS}}{=}
    \frac15\begin{pmatrix}
        3 & 2\\
        3 & 2
    \end{pmatrix}e^{\eta-\eta'} +
    \frac15\begin{pmatrix}
        2 & -2\\
        -3 & 3
    \end{pmatrix}e^{-3/2(\eta-\eta')}.
\end{align}

\subsection{Limit of constant $\Omega_m/f^2$}\label{sec:constantx}

A slight generalization of the EdS limit involves the assumption that
\be
  x(\eta)\equiv\frac{3}{2}\frac{\Omega_m(\eta)}{f^2(\eta)} \stackrel{\dot x=0}{=} x_0\,,
\ee
 is constant in time. While this is in practice not a useful approximation for $\Lambda$CDM,
it provides a generalization of the EdS limit while at the same time retaining most of its simplicity~\cite{Scoccimarro:1997st, Joyce:2022oal}.
We do not further employ this approximation in this work, but provide expressions
valid in this limit for illustration and give the limit of some of our results in that case for reference (see Sec.\,\ref{sec:naivebasis} and App.\,\ref{applindep}). 

We note that a constant $x(\eta)$ is for example realized in a matter-dominated universe in which a constant fraction $f_\nu$ of the total matter does not cluster. In this case the \mesza equation Eq.\,\eqref{meszaroslcdm} is modified as $\Omega_m\mapsto \Omega_m^{\text{clustering}}=(1-f_\nu)\Omega_m$. Assuming matter domination for the background (i.e. $\Omega_m=1$) yields $D_1=e^\eta=e^{f\ln a}$ with constant growth rate 
\be
  f=(\sqrt{25-24f_\nu}-1)/4\simeq 1-3f_\nu/5\,,
\ee
 and 
\be
  x_0=3\Omega_m^{\text{clustering}}/(2f^2)=3(1-f_\nu)/(2f^2)\simeq \frac32(1+f_\nu/5)\,,
\ee
where the approximate expressions hold for $f_\nu\ll 1$. This limit has been argued to provide an approximate description for neutrino cosmologies if the wavenumbers of all relevant perturbation modes are much larger than the neutrino free-streaming scale~\cite{Joyce:2022oal}.

Assuming constant $x\stackrel{\dot x=0}{=}x_0$ in Eq.\,\eqref{getaetaint}, one has 
$  g(\eta,\eta')\stackrel{\dot x=0}{=}\frac{1}{1+x_0}(e^{\eta-\eta'}-e^{-x_0\, (\eta-\eta')})\,.$
Using Eq.\,\eqref{eq:propresult} this yields the propagator
\be
  G(\eta,\eta') \stackrel{\dot x=0}{=} \frac{1}{1+x_0}\left(\begin{array}{cc} x_0 & 1 \\ x_0 & 1 \end{array}\right)e^{\eta-\eta'}
  + \frac{1}{1+x_0}\left(\begin{array}{cc} 1 & -1 \\ -x_0 & x_0 \end{array}\right)e^{-x_0\, (\eta-\eta')}\,,
\ee
which reduces to the EdS result~Eq.\,\eqref{eq:propEdS} for $x_0= 3/2$.
It is easy to see that for constant  $x\stackrel{\dot x=0}{=}x_0$ the time- and scale-dependence factorizes, similarly as for the EdS approximation,
\bea
  F_n(\veck_1,\dots,\veck_n;\eta) & \stackrel{\dot x=0}{=} & e^{n\eta}F_n^{\dot x=0}(\veck_1,\dots,\veck_n),\nn\\
  G_n(\veck_1,\dots,\veck_n;\eta) & \stackrel{\dot x=0}{=} & e^{n\eta}G_n^{\dot x=0}(\veck_1,\dots,\veck_n)\,,
\eea
and the equation of motions~Eq.\,\eqref{sourcetermdiffe} and Eq.\,\eqref{sourcetermdeff} reduce to algebraic recursion relations,
\bea
  F_n^{\dot x=0}(\veck_1,\dots,\veck_n) &=& \sum_{m=1}^{n-1}\Big[ g_{11}^{(n)}\alpha(\vecq_{1\dots m},\vecq_{m+1\dots n}) G_m^{\dot x=0}(\vecq_1,\dots,\vecq_m)F_{n-m}^{\dot x=0}(\vecq_{m+1},\dots,\vecq_n) \nn\\
  && {} + g_{12}^{(n)}\beta(\vecq_{1\dots m},\vecq_{m+1\dots n}) G_m^{\dot x=0}(\vecq_1,\dots,\vecq_m)G_{n-m}^{\dot x=0}(\vecq_{m+1},\dots,\vecq_n) \Big]_{sym}\nn\,,\\
  G_n^{\dot x=0}(\veck_1,\dots,\veck_n) &=& \sum_{m=1}^{n-1}\Big[ g_{21}^{(n)}\alpha(\vecq_{1\dots m},\vecq_{m+1\dots n}) G_m^{\dot x=0}(\vecq_1,\dots,\vecq_m)F_{n-m}^{\dot x=0}(\vecq_{m+1},\dots,\vecq_n) \nn\\
  && {} + g_{22}^{(n)}\beta(\vecq_{1\dots m},\vecq_{m+1\dots n}) G_m^{\dot x=0}(\vecq_1,\dots,\vecq_m)G_{n-m}^{\dot x=0}(\vecq_{m+1},\dots,\vecq_n) \Big]_{sym}\nn\,,\\
\eea
with symmetrization defined as in Eq.\,\eqref{sourcetermdeff} above, $F_1^{\dot x=0}=G_1^{\dot x=0}=1$, and coefficients
\be
  \left(\begin{array}{cc} g_{11}^{(n)} & g_{12}^{(n)} \\ g_{21}^{(n)} & g_{22}^{(n)} \end{array}\right) 
  \equiv \int_{-\infty}^\eta d\eta'\, G(\eta,\eta')\Big|_{\dot x=0}\, e^{n(\eta'-\eta)} = \frac{1}{(n+x_0)(n-1)}\left(\begin{array}{cc} n+x_0-1 & 1 \\ x_0 & n \end{array}\right)\,.
\ee
These reduce to the well-known EdS recursion relations~\cite{Bernardeau_2002} for $x_0= 3/2$.

In the following, we use do not consider this approximation (unless stated otherwise). Instead, we use the general time-dependent form Eq.\,\eqref{crudeprop} of the propagator to construct kernels that are valid beyond the EdS approximation, and for $x(\eta)$ with arbitrary time-dependence.

\subsection{Structure of the time-dependent kernels}

Time-dependent kernels for arbitrary cosmologies can be obtained from the formal solution Eq.\,\pref{eq:formalsol} of the differential equation Eq.\,\eqref{sourcetermdiffe}. In order to obtain explicit expressions, we insert the source terms defined in Eq.\,\pref{sourcetermdeff}. They consist of lower order kernels, which means that this procedure has a recursive structure, analogous to the well-known EdS-SPT case. 

The starting point is given by the linear kernels for $n=1$. They result from the growing mode solution of Eq.\,\pref{sourcetermdiffe} with vanishing right-hand side, which is 
\be\label{eq:startingpoint}
  F_1(\veck;\eta)=e^\eta, \qquad G_1(\veck;\eta)=e^\eta\,.
\ee
This is a consequence of $\eta=\ln D_1$ as well as the property that the non-trivial time-dependence captured by $x(\eta)$ affects only the decaying mode. However, starting from $n=2$, the non-linear kernels are also sensitive to contributions of the decaying mode entering the propagator $G(\eta,\eta')$, and thus differ from the EdS limit if $x(\eta)$ depends on time~\cite{Skewness_and_kurtosis}.

The recursive algorithm and our assumption of time-dependent but scale-independent growth implies that each kernel can be decomposed into a sum of terms, with each summand factorizing into a time-dependent generalized growth factor and a wavevector-dependent basis function~\cite{Skewness_and_kurtosis,Takahashi_2008}. The kernels can thus be written as
\bea \label{redbasisdef} 
  F_n(\veck_1,\dots,\veck_n;\eta) &=& \sum_i d_{n,i}^{F}(\eta) h_{n,i}^{F}(\veck_1,\dots,\veck_n)\,,\nn\\
  G_n(\veck_1,\dots,\veck_n;\eta) &=& \sum_i d_{n,i}^{G}(\eta) h_{n,i}^{G}(\veck_1,\dots,\veck_n)\,,
\eea
with generalized growth factors ${d}_{n,i}^{F(G)}$ and a set of basis functions $h_{n,i}^{F(G)}$ (see Sec.\,\ref{sec:red} for more details). In the EdS approximation, one would have 
$d_{n,i}^{F(G)}(\eta) \stackrel{\mbox{\scriptsize EdS}}{=} c_{n,i}^{F(G)}e^{n\eta}$ with some constants $c_{n,i}^{F(G)}$, which depend on the precise definition of the growth factors as given below, and $\sum_i c_{n,i}^F h_{n,i}^{F}=F_n^{\mbox{\scriptsize EdS}}$ as well as $\sum_i c_{n,i}^G h_{n,i}^{G}=G_n^{\mbox{\scriptsize EdS}}$.

In the following, we proceed in two steps to obtain explicit expressions in the time-dependent case:
\begin{enumerate}
\item We first introduce a non-minimal basis, following the lines of~\cite{Takahashi_2008,zvonimir}, which is convenient to obtain in an algorithmic way. We refer to this as ``naive'' basis, see Eq.\,\eqref{naive_basis}, and denote the growth factors ${D}$ and basis functions $H$ by capital letters.
\item In a second step, we use non-trivial relations to reduce the basis to a minimal size (``reduced basis''). This basis has less elements,  and also makes manifest fundamental constraints from underlying symmetries, in particular momentum conservation, as will be shown below. This set of basis functions represents one of the new result of this work, for $n\geq 4$. Due to the manifest scaling imposed by momentum conservation for all individual basis functions, the reduced basis is convenient for the computation of non-linear corrections to power- and bispectra. We use the notation from Eq.\,\eqref{redbasisdef} to denote this basis, i.e. small letters $d$ and $h$ for the growth factors and basis functions, respectively.
\end{enumerate}
The number of terms contributing to the $F_n$ and $G_n$ kernels up to $n=5$ within the naive and reduced bases, respectively, are summarized in Tab.\,\ref{terms_overview}.
In addition, we display the number of relations among the growth factors $D$ and basis functions $H$ that we use to reduce the number of terms when going from the
naive to the reduced basis. We start by defining the naive basis, and then explain our strategy in constructing the reduced basis.

\begin{table}[t]
    \centering
    \begin{tabular}{c|rrrrrrrr}
             & $G_2$ & $F_2$ & $G_3$ & $F_3$ & $G_4$ & $F_4$ & $G_5$ & $F_5$ \\\hline
       Number of  terms, naive basis  & 2 & 2 & 6 & 6 & 27 & 27 & 127 & 127 \\
        $H$-relations  & 0 & 0 & 1 & 1 & 7 & 7 & 46 & 46 \\
        $D$-relations  & 0 & 0 & 0 & 1 & 8 & 11 & 54 & 64 \\
       Number of  terms, reduced basis & 2 & 2 & 5 & 4 & 14 & 11 & 47 & 39 
    \end{tabular}
    \caption{\small Number of summands for the general time-dependent kernels within the ``naive basis'' (first line, see Eq.\,\eqref{naive_basis}) and the ``reduced basis'' (last line, see Eq.\,\eqref{redbasisdef}). The latter basis is one of the main results of this work, for $n\geq 4$, and makes constraints from momentum conservation manifest for all individual basis functions. The second and third line show the number of independent relations among the set of basis functions (second line) and growth factors (third line), respectively, for the naive basis. These relations are valid independently of the assumed expansion history $\m{H}(z)$ and matter density evolution $\Omega_m(z)$, and are used to obtain the reduced basis, which is thus valid for all models characterized by arbitrary time-dependent linear growth factor $D_1(z)\equiv e^\eta$. Note, that while the relations in the second and third line are independent among themselves, respectively, it is still possible that the number in the last line is larger than the difference of the first minus the second and third, see Sec.\,\ref{sec:red} for details. }
    \label{terms_overview}
\end{table}

\subsection{Naive basis}\label{sec:naivebasis}

Let us start with the first step, obtaining the non-minimal or ``naive'' basis. This basis is largely analogous to~\cite{Takahashi_2008}, except that we use the rescaled time variable $\eta=\ln(D_1)$ instead $\ln(a)$, which is in principle equivalent but
makes the equations somewhat simpler.
It is convenient to slightly rewrite the formal solution Eq.\,\eqref{eq:formalsol} using the structure of the propagator Eq.\,\eqref{crudeprop} as
\be\label{gtmaster}
  F_n = \m{F}_n + \m{C}_n,\qquad G_n = \m{G}_n + \m{C}_n\,,
\ee
with 
\bea\label{eq:defFGC}
  \m{F}_n(\veck_1,\dots,\veck_n;\eta)&\equiv& \etaint g(\eta,\eta')S_\Delta^{(n)}(\veck_1,\dots,\veck_n;\eta')\,,\nn\\
  \m{G}_n(\veck_1,\dots,\veck_n;\eta)&\equiv& \etaint \partial_\eta g(\eta,\eta')S_\Delta^{(n)}(\veck_1,\dots,\veck_n;\eta') = \partial_\eta \m{F}_n(\veck_1,\dots,\veck_n;\eta) \,,\nn\\
  \m{C}_n(\veck_1,\dots,\veck_n;\eta)&\equiv& \etaint e^{\eta-\eta'} S_C^{(n)}(\veck_1,\dots,\veck_n;\eta')\,,
\eea
where $S^{(n)}_\Delta\equiv S^{(n)}_2-S^{(n)}_1$ and $S^{(n)}_C \equiv S^{(n)}_{1}$ refer to the source terms $S^{(n)}_{a}$ from Eq.\,\pref{sourcetermdeff}
that contain the lower order kernels $m<n$, and $g(\eta,\eta')$ is defined in Eq.\,\eqref{getaetaint}.
Since the source terms from Eq.\,\pref{sourcetermdeff} are symmetrized with respect to arbitrary permutations of the wavevector arguments, the resulting kernels inherit this property. We use symmetrized quantities throughout this work.

For the {\it naive basis} we write the expansion of the time-dependent kernels in the form
\bea
    \m{F}_n(\veck_1,\dots,\veck_n;\eta) &=& \sum_{i} D^{\m F}_{n,i}(\eta)H^{\m F}_{n,i}(\veck_1,\dots,\veck_n)\,,\nn\\
    \m{G}_n(\veck_1,\dots,\veck_n;\eta) &=& \sum_{i} D^{\m G}_{n,i}(\eta)H^{\m G}_{n,i}(\veck_1,\dots,\veck_n)\,,\nn\\
    \m{C}_n(\veck_1,\dots,\veck_n;\eta) &=& \sum_i D^{\m C}_{n,i}(\eta)H^{\m C}_{n,i}(\veck_1,\dots,\veck_n)\,.
    \label{eq:expansionFGC}
\eea
Here, we introduced three sets of distinct growth factors $D^{X}_{n,i}(\eta)$ for $X=\m{F},\m{G},\m{C}$.
From Eq.\,\eqref{eq:defFGC} we see that
\be
  D^{\m G}_{n,i}(\eta) = \partial_\eta D^{\m F}_{n,i}(\eta)\,,
\ee
and
\be
  H^{\m F}_{n,i} = H^{\m G}_{n,i} \equiv H_{n,i}\,.
\ee
The index $i$ labels the number of terms of each type. For the $\m{F}_n$ and $\m{G}_n$ the index $i$ corresponds to $0,1,3,14,65$ terms for $n=1,2,3,4,5$. 
For $\m{C}_n$ there are $1,1,3,13,62$ terms. 
The total number of terms for $F_n$ and $G_n$ is given by the sum (see first line of Tab.\,\ref{terms_overview} for $n\leq 5$).

At linear order $n=1$, using Eq.\,\eqref{eq:startingpoint} one has trivially 
\be
  \m{F}_{n=1}(\veck;\eta)=\m{G}_{n=1}(\veck;\eta)=0,\qquad \m{C}_{n=1}(\veck;\eta)=e^\eta\,,
\ee
such that only a single term contributes, with
\be
  D^{\m C}_{n=1,i=1}(\eta)=e^\eta,\qquad H^{\m C}_{n=1,i=1}(\veck)=1\,,
\ee
while the $\m{F}_n$ and $\m{G}_n$ start only at order $n\geq 2$.

To obtain explicit expressions at order $n$, we proceed iteratively and use Eq.\,\eqref{eq:expansionFGC} for all $m<n$ to evaluate the lower order kernels entering the source terms Eq.\,\pref{sourcetermdeff}. The source terms entering the first two and last line of Eq.\,\eqref{eq:defFGC}, respectively,  can be brought into the form
\bea
  S^{(n)}_\Delta(\veck_1,\dots,\veck_n;\eta) &=& \sum_i \m{I}_{n,i}(\eta)H_{n,i}(\veck_1,\dots,\veck_n)\,,\nn\\
  S^{(n)}_C(\veck_1,\dots,\veck_n;\eta) &=& \sum_j \m{I}_{n,j}(\eta)H^{\m C}_{n,j}(\veck_1,\dots,\veck_n)\,,
\eea
where the $\m{I}_{n,i}(\eta')$ are given by a product of two lower order growth functions of the form $D^X_{m,k}(\eta')D^Y_{n-m,l}(\eta')$ for some $1\leq m\leq n-1$, with some indices $X,Y=\m{F},\m{G},\m{C}$ and $k,l$ selecting two growth functions out of the (naive) basis set at order $m$ and $n-m$, respectively. Explicit expressions are given below for $n\leq 4$ and are summarized in App.\,\ref{applindep}, where we also give our numbering convention for these functions up to order $n=5$. The growth factors at order $n$ are then given by
\bea\label{gfintrep}
    D^{\m F}_{n,i}(\eta) &=& \getaint \m{I}_{n,i}(\eta')\,,\nn\\
    D^{\m G}_{n,i}(\eta) &=& \etaint \pgeta \m{I}_{n,i}(\eta') = \partial_\eta D^{\m F}_{n,i}(\eta) \,,\nn\\
    D^{\m C}_{n,i}(\eta) &=& \eetaint \m{I}_{n,i}(\eta')\,.
\eea
For numerical evaluation it is more convenient to obtain the growth factors from a set of differential equations instead of using the integral representation.
For this, it is advantageous to define also 
\be\label{eq:Ddeltadef}
  D^{\Delta}_{n,i}(\eta) \equiv D^{\m G}_{n,i}(\eta)-D^{\m F}_{n,i}(\eta)\,,
\ee
and first solve the two independent first-order equations,
\bea\label{eq:ODEforDdeltaandDC}
  (\partial_\eta+x(\eta))D^{\Delta}_{n,i}(\eta) &=& \m{I}_{n,i}(\eta)\,,\nn\\
  (\partial_\eta-1)D^{\m C}_{n,i}(\eta) &=& \m{I}_{n,i}(\eta)\,,
\eea
and then determine $D^{\m F}_{n,i}(\eta)$ by solving 
\be\label{eq:ODEforDF}
  (\partial_\eta-1)D^{\m F}_{n,i}(\eta) = D^{\Delta}_{n,i}(\eta)\,,
\ee
and $D^{\m G}_{n,i}(\eta)$ from Eq.\,\eqref{eq:Ddeltadef}. In practice, we find it most convenient to solve the differential equations for
the rescaled growth functions $\tilde D_{n,i}^X(\eta) \equiv e^{-n\eta}D_{n,i}^X(\eta)$.

There are a few limits where analytical solutions can be obtained. In particular, in the limit where $x(\eta)\stackrel{\dot x=0}{=} x_0$ is constant 
in time (see Sec.\,\ref{sec:constantx}) all growth factors $D^{X}_{n,i}(\eta)$ are proportional to $e^{n\eta}$. A prominent example is the EdS limit that corresponds to $x(\eta)\stackrel{\mbox{\scriptsize EdS}}{=}3/2$.
We can easily proof this result by induction. For $n=1$ it is trivially true due to $D_1\equiv e^\eta$. Assuming it holds for all $n'<n$, this implies that all $\m{I}_{n,i}(\eta)$, which are products of growth factors of order lower than $n$ of the form $D^X_{m,k}(\eta)D^Y_{n-m,l}(\eta)$ with $1\leq m\leq n-1$, are proportional to $e^{n\eta}$. Using either the differential or integral equations and $x(\eta)\stackrel{\dot x=0}{=} x_0$ yields for $n\geq 2$,
\be\label{eq:constx0}
  D^{\Delta}_{n,i}\stackrel{\dot x=0}{=}\frac{1}{n+x_0}\m{I}_{n,i},\quad
  D^{\m C}_{n,i}\stackrel{\dot x=0}{=}\frac{1}{n-1}\m{I}_{n,i},\quad
  D^{\m F}_{n,i}\stackrel{\dot x=0}{=}\frac{1}{n-1}D^{\Delta}_{n,i},\quad
  D^{\m G}_{n,i}\stackrel{\dot x=0}{=} n D^{\m F}_{n,i}\,,
\ee
which shows that the assumption is true also for order $n$. Using these results recursively also yields explicit expressions for the limit of constant $x(\eta)$,
that we provide in App.\,\ref{applindep} for reference, but do not use in this work. The EdS limit corresponds to $x_0=3/2$,
\be
  D^{\Delta}_{n,i}\stackrel{\mbox{\scriptsize EdS}}{=}\frac{2}{2n+3}\m{I}_{n,i},\quad
  D^{\m C}_{n,i}\stackrel{\mbox{\scriptsize EdS}}{=}\frac{1}{n-1}\m{I}_{n,i},\quad
  D^{\m F}_{n,i}\stackrel{\mbox{\scriptsize EdS}}{=}\frac{1}{n-1}D^{\Delta}_{n,i},\quad
  D^{\m G}_{n,i}\stackrel{\mbox{\scriptsize EdS}}{=} n D^{\m F}_{n,i}\,.
\ee

Let us now return to the case of a cosmological model with general time-dependence of $x(\eta)=3\Omega_m/(2f^2)$.
In this case the time-dependence of the growth factors is more complicated, and has to be determined numerically in general.
Nevertheless, they satisfy certain universal, cosmology-independent properties that we discuss in detail in the following and in Sec.\,\ref{sec:red}, before showing numerical examples in Sec.\,\ref{application}.

As an illustrative example, consider $n=2$. Inspecting Eq.\,\pref{sourcetermdeff}, the only contribution to the source term comes from the product of two first-order kernels, which implies there is only a single contribution $\m{I}_{n=2,i=1}(\eta')=(D_1(\eta'))^2=e^{2\eta'}$. Using Eq.\,\eqref{gfintrep}, this yields the three growth factors
\begin{align}\label{eq:D2i}
    D^{\m F}_{2,1}(\eta)= \etaint g(\eta,\eta')e^{2\eta'}, \qquad D^{\m G}_{2,1}(\eta)= \partial_\eta D^{\m F}_{2,1}(\eta), \qquad D^{\m C}_{2,1}(\eta)= e^{2\eta}.
\end{align}
The first (second) growth factor enters in $F_2$ ($G_2$), while the last one is trivial and enters both density and velocity second order kernels.
The wavevector-dependent coefficients also follow directly from the source terms, and are given by
\begin{align}
    H_{2,1}(\veck_1,\veck_2)=\xi(\veck_1,\veck_2) \qquad \text{and} \qquad H^{\m C}_{2,1}=\alpha^s(\veck_1,\veck_2),
\end{align}
where $\xi\equiv \beta-\alpha^s$, $\alpha^s(\veck_1,\veck_2)\equiv\frac{1}{2}(\alpha(\veck_1,\veck_2)+\alpha(\veck_2,\veck_1))$, such that
\bea
  F_2(\veck_1,\veck_2;\eta) &=& {D}^{\m F}_{2,1}H_{2,1}+{D}^{\m C}_{2,1}H^{\m C}_{2,1} = {D}^{\m F}_{2,1}(\eta)\xi(\veck_1,\veck_2) + e^{2\eta}\alpha^s(\veck_1,\veck_2)\,,\nn\\
  G_2(\veck_1,\veck_2;\eta) &=& {D}^{\m G}_{2,1}H_{2,1}+{D}^{\m C}_{2,1}H^{\m C}_{2,1} = \partial_\eta {D}^{\m F}_{2,1}(\eta)\xi(\veck_1,\veck_2) + e^{2\eta}\alpha^s(\veck_1,\veck_2)\,.
\eea
For $n=3$, the source terms contain the three functions
\be\label{eq:In3}
    \m{I}_{3,1}(\eta)= {D}_1(\eta) D^{\m F}_{2,1}(\eta),\qquad \m{I}_{3,2}(\eta)= {D}_1(\eta) D^{\m C}_{2,1}(\eta),\qquad
    \m{I}_{3,3}(\eta) = {D}_1(\eta) D^{\m G}_{2,1}(\eta)\,,
\ee
where ${D}_1=e^\eta$ and $D^{\m C}_{2,1}(\eta)=e^{2\eta}$ are trivial, such that $\m{I}_{3,2}(\eta)=e^{3\eta}$.
Using Eq.\,\eqref{gfintrep} yields nine growth factors ${D}_{3,i}^X(\eta)$ for $i=1,2,3$ and $X=\m{F},\m{G},\m{C}$ at third order.
One of them is trivial, ${D}_{3,2}^{\m C}(\eta)=\frac12 e^{3\eta}$. We will discuss further cosmology-independent relations
among them in Sec.\,\ref{sec:red}.

For $n=4$, we have
\begin{alignat}{8}
    &\m{I}_{4,1-3} &&= {D}_1 {D}^{\m F}_{3,1-3}, &\qquad&
    \m{I}_{4,4-6} &&= {D}_1 {D}^{\m C}_{3,1-3}, &\qquad&
    \m{I}_{4,7-9} &&= {D}_1 {D}^{\m G}_{3,1-3}, &\qquad&
    \m{I}_{4,10} &&={D}^{\m F}_{2,1} {D}^{\m C}_{2,1}, \nonumber\\
    &\m{I}_{4,11} &&={D}^{\m F}_{2,1} {D}^{\m G}_{2,1}, &\qquad&\m{I}_{4,12} &&={D}^{\m G}_{2,1} {D}^{\m C}_{2,1}, &\qquad&
    \m{I}_{4,13} &&={D}^{\m C}_{2,1} {D}^{\m C}_{2,1}, &\qquad&
    \m{I}_{4,14} &&={D}^{\m G}_{2,1}{D}^{\m G}_{2,1} , \label{eq:I4idef}
\end{alignat}
where $\m{I}_{4,5}(\eta)=\frac12 e^{4\eta}$ and $\m{I}_{4,13}(\eta)= e^{4\eta}$ are trivial. This yields fourth order growth factors
${D}_{4,i}^X(\eta)$ for $1\leq i\leq 14$ and $X=\m{F},\m{G},\m{C}$ via Eq.\,\eqref{gfintrep}, with ${D}_{4,5}^{\m C}(\eta)=\frac16 e^{4\eta}$ and ${D}_{4,13}^{\m C}(\eta)=\frac13 e^{4\eta}$ being trivial.
Furthermore, due to the structure of the source term, only the terms with $i\leq 13$ contribute for $X=\m{C}$ \footnote{This is because the corresponding source term $S^{(n)}_C \equiv S^{(n)}_{1}$ entering in Eq.\,\eqref{eq:defFGC} contains the non-linearity obtained from the continuity equation only (related to the $\alpha$-vertex), which couples the density and the velocity divergence, and no contributions from a product of two velocity fields.}. This leads to $14+13=27$ terms contributing to both $F_4$ and $G_4$ within the naive basis. Again, there exist further relations that allow us to reduce the number of independent terms at fourth order, see Sec.\,\ref{sec:red}.

For $n=5$, we find 65 integrands. The naive basis contains $127$ terms contributing to both $F_5$ and $G_5$, respectively.
We collect all equations for $n=5$ in App.\,\ref{applindep}, where we also summarize those for $n\leq 4$ for convenience.

In summary, the {\it naive basis} yields the decomposition
\bea\label{naive_basis}
  F_n(\veck_1,\dots,\veck_n;\eta) &=& \sum_{X=\m{F},\m{C}}\sum_i {D}^{X}_{n,i}(\eta)H^{X}_{n,i}(\veck_1,\dots,\veck_n)\,,\nn\\
  G_n(\veck_1,\dots,\veck_n;\eta) &=& \sum_{X=\m{G},\m{C}}\sum_i {D}^{X}_{n,i}(\eta)H^{X}_{n,i}(\veck_1,\dots,\veck_n)\,,
\eea
with growth functions ${D}^{X}_{n,i}$ at order $n$ and basis functions
$H^{\m F}_{n,i}=H^{\m G}_{n,i}\equiv H_{n,i}$ as well as $H^{\m C}_{n,i}$. Explicit expressions for the corresponding wavevector-dependent $H$-functions are given up to the fifth order in App.\,\ref{appHfunc},
and the growth factors are defined recursively in Eq.\,\eqref{gfintrep}. While the $H$-functions are independent of the cosmological model, the growth factors have to be evaluated numerically for a given expansion history and matter density evolution,
which is most conveniently done by solving the set of differential equations Eqs.\,\eqref{eq:ODEforDdeltaandDC} and~\eqref{eq:ODEforDF}. Before showing examples for specific cosmologies, we turn to our main result, namely the elimination of redundancies that are present within the naive basis.

\section{Reduced basis}
\label{sec:red}

The representation of the time-dependent non-linear kernels $F_n$ and $G_n$ worked out in Sec.\,\ref{sec:naivebasis} is easy to evaluate algorithmically.
However, it contains redundancies, that exist even for an arbitrary expansion history. In this section, we discuss how to find and classify these redundancies. Then, we use them to derive a minimal {\it reduced basis}, as opposed to the {\it naive basis} from Sec.\,\ref{sec:naivebasis}.

Before doing so, one may ask what is the advantage of using such a reduced basis. Apart from lowering the number of summands, that must be included for each kernel (see Tab.\,\ref{terms_overview}),
there is a more important technical advantage of the reduced basis. In particular, it is well-known that due to mass and momentum conservation the non-linear kernels for the density must satisfy\footnote{Note that for the
velocity kernels $G_n$, the same property holds due to the symmetry $\vecq\to -\vecq$. However, while for example $F_3(\veck-\vecp-\vecq,\vecp,\vecq)\propto k^2$ for $k\to 0$ is guaranteed by mass/momentum conservation, 
a linear scaling occurs for the corresponding $G_3$ in general~\cite{Garny:2022kbk}. Nevertheless, within SPT, where vorticity and higher cumulants are neglected, the linear scaling of $G_3$ is absent.}
\be\label{eq:scaling}
  F_n(\veck_1,\dots,\veck_{n-2},\vecq,-\vecq;\eta) \propto \frac{k^2}{q^2}\quad\text{for}\ k^2\equiv\Big(\sum_i\veck_i\Big)^2 \ll q^2\,. 
\ee
Since the class of cosmological models with general time-dependence respect mass and momentum conservation for $\delta$, also the
time-dependent kernels must respect Eq.\,\eqref{eq:scaling}. However, it turns out that for the {\it naive basis} the individual basis functions in general
do not have this scaling, but generically go as $H_{n,i}^X(\veck_1,\dots,\veck_{n-2},\vecq,-\vecq) \propto {\cal O}(q^0)$, i.e. are constant for $k\ll q$ (see Eq.\,\eqref{naive_basis}). The first case where this happens is at $n=4$. The contributions that are constant or go as $1/q$ have to cancel when summing over all contributions labelled by $i$ and $X$ to obtain $F_n$. However, this requires some non-trivial
relations among the corresponding growth factors ${D}_{n,i}^X(\eta)$. A priori, these relations are not obvious, and making them manifest is the objective of this section.

In practice, it is rather advantageous, if every single basis function respects the scaling Eq.\,\eqref{eq:scaling}, when using the kernels to compute higher order loop corrections to the power- and bispectrum. Otherwise, large numerical cancellations would have to occur among the various contributions involving different growth factors, which is rather cumbersome, especially starting from two-loop order. While a workaround has been suggested in Ref.\,\cite{zvonimir}, it is clearly preferred to eliminate the redundancy of the set of basis functions. This reduces the number of required terms {\it and} makes symmetry constraints manifest for arbitrary time-dependent cosmologies. In the following, we show how to do this, and provide explicit results for the non-linear kernels within this {\it reduced basis}, with notation introduced already in Eq.\,\eqref{redbasisdef}. As we will show below, each individual basis functions  within the {\it reduced basis} satisfies 
\be\label{eq:momentumconstraint}
  h_{n,i}^X(\veck_1,\dots,\veck_{n-2},\vecq,-\vecq) \propto \frac{k^2}{q^2}\quad\text{for}\ k^2\equiv\Big(\sum_i\veck_i\Big)^2 \ll q^2\,,
\ee
for all $n>2$, and all basis functions labelled by $i$ and $X=F,G$. 

\subsection{Redundancies of the naive basis}

Starting from the naive basis discussed in Sec.\,\ref{sec:naivebasis}, we are looking for cosmology-independent relations among either the wavevector-dependent $H$-functions or among the time-dependent growth factors ${D}$, that we dub $H$-relations and ${D}$-relations, respectively.

\subsubsection{Relations among wavevector-dependent functions}\label{sec:Hrelations}

Let us start with the simpler case of $H$-relations. We are looking for linear dependencies of the form
\be
  0=\sum_{X}\sum_i e_{n,i}^X H_{n,i}^X(\veck_1,\dots,\veck_n)\,,
\ee
that hold for arbitrary arguments $\veck_1,\dots,\veck_n$ for some fixed constants $e_{n,i}^X$.
Since the set of functions $H_{n,i}\equiv H_{n,i}^{\m F}=H_{n,i}^{\m G}$ as well as $H_{n,i}^{\m C}$ within the naive basis
enters both the density and velocity divergence kernels, $H$-relations apply to both of them. It is thus sufficient to sum over $X=\m{F},\m{C}$ or equivalently $X=\m{G},\m{C}$ here, see Eqs.\,\eqref{gtmaster} and \eqref{naive_basis}.
Starting from the explicit expressions given in App.\,\ref{appHfunc}, it is straightforward to identify such relations using standard methods in linear algebra.
The first order where a non-trivial relation occurs is for $n=3$, being 
\be\label{eq:H3rel}
  0=H_{3,1}(\veck_1,\veck_2,\veck_3)+H^{\m C}_{3,1}(\veck_1,\veck_2,\veck_3)\,.
\ee
At $n=4$ we find the seven relations (omitting the common wavevector arguments $\veck_1,\dots,\veck_4$)
\begin{align}\label{eq:H4rel}
    0 &= H^{\m C}_{4,1-3}+H_{4,1-3},  && 0=H^{\m C}_{4,10-11}+H_{4,10-11},\nn\\
    0 &= H_{4,1} + H_{4,4} + H_{4,7}, && 0 = H^{\m C}_{4,1} + H^{\m C}_{4,4} + H^{\m C}_{4,7}.
\end{align}
We checked that these relations are all linearly independent from each other.

At $n=5$ we find 46 linearly independent relations, see Eq.\,\eqref{eq:H5rel}.

Overall, the number of independent $H$-relations is $0,0,1,7,46$ at orders $n=1,2,3,4,5$ (see also Tab.\,\ref{terms_overview}). While the precise form of those relations depends on the
convention for the basis, their number is universal. All relations are summarized in App.\,\ref{appHfunc} for convenience.

\subsubsection{Relations among growth functions}\label{sec:Drelations}

The ${D}$-relations among the growth factors that we are looking for are of the form
\bea
  0&=&\sum_{X=\m{F},\m{C}}\sum_i f_{n,i}^X {D}_{n,i}^X(\eta)\,,\nn\\
  0&=&\sum_{X=\m{G},\m{C}}\sum_i g_{n,i}^X {D}_{n,i}^X(\eta)\,,
\eea
that are valid at all times $\eta$ and for arbitrary time-dependence of $x(\eta)=3\Omega_m(\eta)/(2f(\eta)^2)$, with some fixed universal (i.e. cosmology-independent) constants $f_{n,i}^X$ and $g_{n,i}^X$.
The relations of the form of the first line can be used to simplify the time-dependent $F_n$ kernel, and those from the second line for the $G_n$ kernel, see Eqs.\,\eqref{gtmaster} and~\eqref{naive_basis}.

Finding such relations is involved since the propagator function $g(\eta,\eta')$ defined in Eq.\,\eqref{getaetaint}, that enters the definition of the growth factors in Eq.\,\eqref{gfintrep}, is not known explicitly for arbitrary $x(\eta)$. Nevertheless, using partial integration identities, it is possible to rewrite some of the growth factors. The first example occurs at $n=3$. From Eqs.\,\eqref{gfintrep} and~\eqref{eq:In3}, it follows that
\be
    {D}^{\m F}_{3,3}(\eta)-{D}^{\m F}_{3,1}(\eta) = \getaint D_1(\eta') {D}_{2,1}^\Delta(\eta')\,,
\ee
where we set ${D}_{2,1}^\Delta(\eta)\equiv {D}_{2,1}^{\m G}-{D}_{2,1}^{\m F}=(\partial_\eta-1){D}_{2,1}^{\m F}(\eta)$.
We can write $D_1(\eta')=e^{\eta'}=\partial_{\eta'}e^{\eta'}$ and use partial integration, giving
\be\label{calcO3relation}
    {D}^{\m F}_{3,3}(\eta)-{D}^{\m F}_{3,1}(\eta) = \left[ e^{\eta'}\geta {D}_{2,1}^\Delta(\eta') \right]^{\eta'=\eta}_{\eta'\to-\infty} -\etaint e^{\eta'}\partial_{\eta'}\left[ \geta {D}_{2,1}^\Delta(\eta') \right]\,.
\ee
The integrand at $\eta'=\eta$ vanishes since $g(\eta,\eta)=0$ (see Eq.\,\ref{getaetaint}) and also for $\eta'\to-\infty$ due to the exponential suppression of $e^{\eta'}{D}_{2,1}^\Delta(\eta')$. 
Next, we use Eq.\,\eqref{eq:detaprimeg} and the differential equation Eq.\,\eqref{eq:ODEforDdeltaandDC} for ${D}_{2,1}^\Delta(\eta')$ to evaluate the $\partial_{\eta'}$ derivative,
\be
  \partial_{\eta'}\left[ \geta {D}_{2,1}^\Delta(\eta') \right] = \left(x(\eta')g(\eta,\eta')-e^{\eta-\eta'}\right){D}_{2,1}^\Delta(\eta') + g(\eta,\eta')\left(\m{I}_{2,1}(\eta') - x(\eta'){D}_{2,1}^\Delta(\eta')\right)\,.
\ee
Note, that the terms involving $x(\eta')$ cancel and that $\m{I}_{2,1}(\eta')=e^{2\eta'}={D}_{2,1}^{\m C}(\eta')$ yields
\be
    {D}^{\m F}_{3,3}(\eta)-{D}^{\m F}_{3,1}(\eta)  =\etaint e^{\eta'}\left[e^{\eta-\eta'} {D}_{2,1}^\Delta(\eta')-\geta {D}_{2,1}^{\m C}(\eta) \right].
\ee
Substituting ${D}_{2,1}^\Delta={D}_{2,1}^{\m G}-{D}_{2,1}^{\m F}$ and using the definition Eq.\,\eqref{gfintrep} as well as Eq.\,\eqref{eq:In3} and $e^{\eta'}=D_1(\eta')$ leads to the relation 
\be\label{eq:D3relation}
  0 = {D}^{\m F}_{3,3}(\eta)-{D}^{\m F}_{3,1}(\eta) - {D}^{\m C}_{3,3}(\eta) + {D}^{\m C}_{3,1}(\eta) + {D}^{\m F}_{3,2}(\eta)\,.
\ee
This relations holds independently of the cosmological model, and for all times. It involves the growth factors that enter in the third order density kernel $F_3$, and therefore this relation can be used to reduce the basis for this kernel. We did not find any further relations involving exclusively growth factors with $n=3$.

However, we did find additional relations involving growth factors for $n=3$ as well as products of the form $D_1D_{2,i}^X$.
While these relations are not immediately useful to reduce the basis at third order, they yield relations among the integrands $\m{I}_{4,i}$ at fourth order,
and corresponding relations among the fourth-order growth factors, see below. In addition, they are instrumental to check constraints on the
kernels imposed by Galilean symmetry, see Sec.\,\ref{sec:softlimit}. At $n=3$ we found two such relations, 
\bea
 0 &=& {D}^{\Delta}_{3,3}(\eta)-{D}^{\Delta}_{3,1}(\eta)-D_1(\eta){D}^{\Delta}_{2,1}(\eta)+{D}^{\Delta}_{3,2}(\eta)\,, \label{eq:D3D2rel1}\\
 0 &=& {D}^{\m C}_{3,3}(\eta) - D_1(\eta){D}^{\m F}_{2,1}(\eta)\,. \label{eq:D3D2rel2}
\eea
The first relation follows from applying $\partial_\eta-1$ to Eq.\,\eqref{eq:D3relation}
and using $D^\Delta_{n,i}\equiv D^{\m G}_{n,i}-D^{\m F}_{n,i}=(\partial_\eta-1)D^{\m F}_{n,i}$, as well as $(\partial_\eta-1)({D}^{\m C}_{3,3}(\eta) - {D}^{\m C}_{3,1}(\eta))=\m{I}_{3,3}-\m{I}_{3,1}=D_1(D_{2,1}^{\m G}-D_{2,1}^{\m F})$, see Eq.\,\eqref{eq:ODEforDdeltaandDC} and Eq.\,\eqref{eq:In3}.
The second relation follows from 
\be
  {D}^{\m C}_{3,3}(\eta) = \int_{-\infty}^\eta d \eta' e^{\eta-\eta'}D_1(\eta')D_{2,1}^{\m G}(\eta') = e^{\eta}\int_{-\infty}^\eta d \eta' D_{2,1}^{\m G}(\eta')\,,
\ee
and using $D_{2,1}^{\m G}(\eta')=\partial_{\eta'}D_{2,1}^{\m F}(\eta')$ as well as $D_1(\eta)=e^\eta$.

At order $n=4$ in perturbation theory, we find the following cosmology-independent relations,
\bea
 0 &=& {D}^{\m F}_{4,1}  - {D}^{\m F}_{4,7} -D^{\m F}_{4,10} -{D}^{\m C}_{4,1}  + {D}^{\m C}_{4,7}\,, \label{eq:D4relextra1} \\
 0 &=& {D}^{\m F}_{4,2}  - {D}^{\m F}_{4,8} -D^{\m F}_{4,13} -{D}^{\m C}_{4,2}  + {D}^{\m C}_{4,8}\,, \label{eq:D4relextra2} \\
  0 &=& {D}^{\m F}_{4,3}  - {D}^{\m F}_{4,9} -D^{\m F}_{4,12} -{D}^{\m C}_{4,3}  + {D}^{\m C}_{4,9}\,, \label{eq:D4relextra3} \\
 0 &=& D^X_{4,1}-D^X_{4,2}-D^X_{4,3}-D^X_{4,4}+D^X_{4,6}\,, \label{eq:D4rel2}\\
 0 &=& D^X_{4,1}-D^X_{4,2}-D^X_{4,3}-D^X_{4,7}+D^X_{4,8}+D^X_{4,9}+D^X_{4,10}-D^X_{4,12}\,, \label{d3erelation}\\
 0 &=& D^X_{4,6}-D^X_{4,10}\,, \label{eq:D4rel4}\\
 0 &=& 2D^X_{4,5}-D^X_{4,13}\,, \label{eq:D4rel5}
\eea
where the last four lines hold for each choice $X=\m{F},\m{G},\m{C}$, respectively, making in total $3+3\times 4=15$ relations.
Out of these, the first three lines, as well as the relations in the last four lines for $X=\m{F},\m{C}$, can be used to reduce the basis
for the $F_4$ kernel of the density contrast, making $3+2\times 4=11$ relations. For the $G_n$ kernel of the velocity divergence,
only the relations for $X=\m{G},\m{C}$ can be used, making $2\times 4=8$.

The first three relations Eqs.\,\eqref{eq:D4relextra1}--\eqref{eq:D4relextra3} follow from partial integration similarly to Eq.\,\eqref{eq:D3relation}, see App.\,\ref{applindep} for a derivation.
The other relations follow from corresponding relations among the integrands $\m{I}_{n,i}$. In particular, Eq.\,\eqref{eq:D4rel2}, Eq.\,\eqref{d3erelation}
and Eq.\,\eqref{eq:D4rel4} are inherited from the $n\leq 3$ relations Eq.\,\eqref{eq:D3relation}, Eq.\,\eqref{eq:D3D2rel1} and  Eq.\,\eqref{eq:D3D2rel2}, respectively,
together with the definitions Eq.\,\eqref{eq:I4idef} of the integrands for the fourth order growth functions, as well as the property that $D_1(\eta)=e^\eta$ and $D^{\m C}_{2,1}(\eta)=e^{2\eta}$
have a trivial time-dependence, see  Eq.\,\eqref{eq:D2i}. The last relation Eq.\,\eqref{eq:D4rel5} follows from using
$D^{\m C}_{2,1}(\eta)=e^{2\eta}$ again, and in addition ${D}^{\m C}_{3,2}(\eta)=\frac12 e^{2\eta}$, see Eq.\,\eqref{eq:In3}.

In general, we find that the cosmology-independent relations can be classified in two categories:
\begin{itemize}
\item Relations that follow from using partial integration identities, similarly to Eq.\,\eqref{eq:D3relation} for $n=3$ from above. In contrast to the second type, these relations contain growth factors $D_{n,i}^X(\eta)$ involving both $X=\m{F}$ and $X=\m{C}$.
\item Relations among growth factors, that are inherited from relations among the integrands $\m{I}_{n,i}$ that enter the growth factors, see Eq.\,\eqref{gfintrep}. In particular,
if $\sum_i j_{n,i}\m{I}_{n,i}(\eta)=0$ for all $\eta$ and some fixed constants $j_{n,i}$, this immediately implies the three relations $\sum_i j_{n,i} D_{n,i}^X(\eta)=0$ for $X=\m{F},\m{G},\m{C}$, respectively. These relations involve growth factors with a given type $X$ only.
\end{itemize}
At $n=4$, the first three relations Eqs.\,\eqref{eq:D4relextra1}--\eqref{eq:D4relextra3} are of the first kind, and Eqs.\,\eqref{eq:D4rel2}--\eqref{eq:D4rel5} of the second kind.
At $n=5$ we find ten linearly independent relations of the first category, and $27$ of the second, making in total $10+3\times 27=91$ relations.
These relations are reported and derived in App.\,\ref{applindep}.
Out of them, the first ten as well as $2\times 27$ of the others (for $X=\m{F},\m{C}$) can be used to simplify $F_5$, and  $2\times 27$ (for $X=\m{G},\m{C}$) for $G_5$.
This makes in total $10+2\times 27=64$ relations for $F_5$, and $2\times 27=54$ for $G_5$.

\subsection{Reduced basis of the kernels}\label{sec:redbasis}

We can use the cosmology-independent relations among the growth factors and basis functions discussed above
to obtain a reduced basis for the density and velocity divergence kernels. We denote the growth factors in
this basis by $d_{n,i}^X(\eta)$ and the basis functions by $h_{n,i}^X(\veck_1,\dots,\veck_n)$, with $X=F$
for the $F_n$ and $X=G$ for the $G_n$, 
\bea 
  F_n(\veck_1,\dots,\veck_n;\eta) &=& \sum_i d_{n,i}^{F}(\eta) h_{n,i}^{F}(\veck_1,\dots,\veck_n)\,,\nn\\
  G_n(\veck_1,\dots,\veck_n;\eta) &=& \sum_i d_{n,i}^{G}(\eta) h_{n,i}^{G}(\veck_1,\dots,\veck_n)\,,
\eea
as already pointed out in Eq.\,\eqref{redbasisdef}. We note that due to the basis reduction, we find it more convenient to use a labelling, which is slightly
different from the naive basis Eq.\,\eqref{naive_basis}, where we introduced different labels for the contributions to the kernels that
are in common (superscript $\m{C}$) and specific (superscript $\m{F}$ or $\m{G}$) for the $F_n$ and $G_n$ kernels, respectively.
Here, we introduce separate basis sets for the density contrast and velocity divergence, since there is a different number of relations among the corresponding
growth functions entering $F_n$ and $G_n$, respectively.

After using all relations discussed in Secs.\,\ref{sec:Hrelations} and~\ref{sec:Drelations}, we find that the number of terms labelled by the index $i$ is
$1,2,4,11,39$ for $n=1,2,3,4,5$ for $F_n$, and $1,2,5,14,47$ for $G_n$, see Tab.\,\ref{terms_overview}.
For comparison, the naive basis contains $1,2,6,27,127$ elements for both $F_n$ and $G_n$.

At first and second order, there is no difference between the naive and reduced basis, and we can simply set $d_{1,1}^F(\eta)=d_{1,1}^G(\eta)\equiv e^\eta$, $h_{1,1}^F(\veck)=h_{1,1}^G(\veck)\equiv 1$,
and $h_{2,1}^X\equiv H_{2,1}$, $h_{2,2}^X\equiv  H^{\m C}_{2,1}$ for $X=F,G$, as well as $d_{2,1}^F\equiv D_{2,1}^{\m F}$,  $d_{2,1}^G\equiv D_{2,1}^{\m G}$
and $d_{2,2}^X\equiv D_{2,1}^{\m C}$ for $X=F,G$, such that (see Eq.\,\ref{eq:D2i})
\bea\label{eq:F2G2red}
  F_2(\veck_1,\veck_2;\eta) &=& {d}^{F}_{2,1}h_{2,1}+{d}^{F}_{2,2}h^{F}_{2,2} = {d}^{ F}_{2,1}(\eta)\xi(\veck_1,\veck_2) + e^{2\eta}\alpha^s(\veck_1,\veck_2)\,,\nn\\
  G_2(\veck_1,\veck_2;\eta) &=& {d}^{G}_{2,1}h_{2,1}+{d}^{G}_{2,2}h^{G}_{2,2} = \partial_\eta {d}^{ F}_{2,1}(\eta)\xi(\veck_1,\veck_2) + e^{2\eta}\alpha^s(\veck_1,\veck_2)\,,
\eea
with
\begin{align}\label{eq:d2i}
    d^{F}_{2,1}(\eta)= \etaint g(\eta,\eta')e^{2\eta'}, \qquad d^{G}_{2,1}(\eta)= \partial_\eta d^{F}_{2,1}(\eta), \qquad d^{F}_{2,2}(\eta)=d^{G}_{2,2}(\eta)= e^{2\eta}.
\end{align}
At $n=3$, there is one relation among the naive basis $H$-functions, see Eq.\,\eqref{eq:H3rel}, that can be used to eliminate one basis function in both $F_3$ and $G_3$.
In addition, there is one growth factor relation among the $D$-functions, see Eq.\,\eqref{eq:D3relation}. Some of the growth functions entering this relation appear only in $F_3$
but not in $G_3$ (specifically being the $D^{\m F}_{3,i}$). This means, this relation can be used to further reduce the number of terms contributing to $F_3$ by one.
Specifically, we choose the freedom to eliminate terms to arrive at the reduced basis expression
\bea\label{g3kernel}
  F_3(\veck_1,\veck_2,\veck_3;\eta) &=& \sum_{i=1}^4 {d}^{F}_{3,i}(\eta) h^F_{3,i}(\veck_1,\veck_2,\veck_3),\nn\\
  G_3(\veck_1,\veck_2,\veck_3;\eta) &=& \sum_{i=1}^5 {d}^{G}_{3,i}(\eta) h^G_{3,i}(\veck_1,\veck_2,\veck_3)\,.
\eea
For the $F_3$ kernel the relevant growth functions, in terms of those defined for the naive basis, are
\be\label{eq:d3iF}
  d_{3,1}^F\equiv D_{3,2}^{\m C}=\frac12 e^{3\eta},\quad 
  d_{3,2}^F\equiv D_{3,2}^{\m F},\quad 
  d_{3,3}^F\equiv D_{3,3}^{\m F},\quad 
  d_{3,4}^F\equiv D_{3,3}^{\m C}\,.
\ee
Note, that due to Eq.\,\eqref{eq:D3D2rel2} $d_{3,4}^F=D_1d_{2,1}^F$ is related to a second-order growth factor,
and $d_{3,1}^F=D_1^3/2$ to the one at first-order, such that effectively only two genuinely new third-order growth functions appear~\cite{DAmico:2021rdb,Joyce:2022oal}.
The relevant basis functions in terms of those of the naive basis are
\be\label{eq:h3iF}
  h_{3,1}^F\equiv H_{3,2}^{\m C},\quad 
  h_{3,2}^F\equiv H_{3,2}-H_{3,1}^{\m C},\quad 
  h_{3,3}^F\equiv H_{3,3}-H_{3,1}^{\m C},\quad 
  h_{3,4}^F\equiv H_{3,3}^{\m C}+H_{3,1}^{\m C}\,.
\ee
Explicit expressions for the $h_{3,i}^F$, as well as the $d_{3,i}^G$ and $h_{3,i}^G$ can be found in App.\,\ref{sec:redbasisthirdorder}.
For reference, we also give the values of the non-trivial growth functions in the EdS approximation, as well as in the approximation of constant $x(\eta)\stackrel{\dot x=0}{=} x_0$
considered in Sec.\,\ref{sec:constantx}, 
\begin{align}\label{eq:d3EdS}
   d_{3,2}^F &\stackrel{\mbox{\scriptsize EdS}}{=} \frac{1}{9}e^{3\eta},\
  &d_{3,3}^F &\stackrel{\mbox{\scriptsize EdS}}{=} \frac{4}{63}e^{3\eta},\
  &d_{3,4}^F &\stackrel{\mbox{\scriptsize EdS}}{=} \frac{2}{7}e^{3\eta},\nn\\
   d_{3,2}^F &\stackrel{\dot x=0}{=} \frac{1}{2 (x_0+3)}e^{3\eta},\
  &d_{3,3}^F &\stackrel{\dot x=0}{=} \frac{1}{(x_0+2) (x_0+3)}e^{3\eta},\
  &d_{3,4}^F &\stackrel{\dot x=0}{=} \frac{1}{x_0+2}e^{3\eta}\,.
\end{align}
We note, that the representation of the $F_3$ kernel with four basis functions and three non-trivial growth functions matches
the form found in Ref.~\cite{Skewness_and_kurtosis}\footnote{The relation to the form given in Ref.~\cite{Skewness_and_kurtosis} is
\begin{align}\label{f3kernel}
    F_3(\vecq_1, \vecq_2, \vecq_3; \eta) &= e^{3\eta}\left(\m{R}^s_1(\vecq_1, \vecq_2, \vecq_3) + \nu_2(\eta)\m{R}^s_2(\vecq_1, \vecq_2, \vecq_3) + \nu_3(\eta)\m{R}^s_3(\vecq_1, \vecq_2, \vecq_3)+ \lambda_3(\eta)\m{R}^s_4(\vecq_1, \vecq_2, \vecq_3)\right)\,,
\end{align}
with
\begin{align*}
    &\m{R}^s_1=\frac{1}{2}h^F_{3,1}+\frac{1}{6}h^F_{3,2}-\frac{5}{4}h^F_{3,3}+\frac{3}{2}h^{F}_{3,4},
    &&\m{R}^s_2=-\frac{4}{3}h^F_{3,4}-2h^F_{3,3}, 
    &&\m{R}^s_3=\frac{3}{16}h^F_{3,3},
    &&\m{R}^s_4=-\frac{1}{3}h^F_{3,2}-\frac{1}{4}h^F_{3,3}\,,
\end{align*}
\begin{align*}
    &\nu_2(\eta)=\frac{3}{4}(3 d^F_{3,1} - d^F_{3,4})e^{-3\eta}, && 
    \nu_3(\eta)=\frac{4}{3}(29 d^F_{3,1}-3d^F_{3,2}+4d^F_{3,3}-6d^F_{3,4})e^{-3\eta}, && \lambda_3(\eta)=(d^F_{3,1}-3d^F_{3,2})e^{-3\eta}.
\end{align*}}.
For convenience, we also give an analytical result for the kernel $F_3(\veck,\vecq,-\vecq;\eta)$ averaged over the angle of $\vecq$,
that enters the one-loop correction to the density power spectrum,
\begin{align}
    \int \frac{d\Omega_q}{4\pi} F_3(\veck,\vecq,-\vecq;\eta) =\hspace{0.1cm}& \frac{1}{36 r^2}\bigg[\left(r^2+1\right) \left(3 r^4-14 r^2+3\right)(d^F_{3,2}-d^F_{3,3}) -\big[ \left(r^4-8 r^2+3\right)d^F_{3,1} \nn\\
    &+\left(3 r^4+8 r^2-3\right)d^F_{3,4} \big]\nonumber\\
    &-\frac{3 \left(r^2-1\right)^3 }{2r} \text{ artanh}\left[\frac{2 r}{r^2+1}\right]\left[ \left(r^2-1\right)(d^F_{3,2}-d^F_{3,3})+d^F_{3,1}-d^F_{3,4}\right]\bigg]\,, \label{f3bare}
\end{align}
where $r\equiv |\veck|/|\vecq|$.

It is worth to note a subtlety in the construction of the reduced basis that
appears for $n\geq 4$. For example, the number of terms in the naive basis for $F_4$ is $27$. We found $7$ relations among the wavevector-dependent $H$-functions
and $11$ among the time-dependent $D$-functions entering in $F_4$. This could lead to the expectation that, after using these relations, the $F_4$ kernel can be
expressed as a sum of $27-7-11=9$ terms. However, the minimal reduced basis for $F_4$ actually contains $11$ terms. The reason is that it is not possible to use all
$7$ $H$-relations and all $11$ $D$-relations simultaneously to eliminate terms. To see this, we write the naive basis expression for $F_4$ in the compact notation
\be
  F_4(\veck_1,\dots,\veck_4;\eta)=\sum_{i=1}^{14} D^{\m F}_{4,i}(\eta)H_{4,i}+\sum_{i=1}^{13} D^{\m C}_{4,i}(\eta)H_{4,i}^{\m C}\equiv \sum_{\alpha=1}^{27} D_\alpha(\eta)H_\alpha(\veck_1,\dots,\veck_4)\,,
\ee
 where we defined a representation where the index $\alpha$ runs over all $14+13=27$ contributions within the naive basis, see Sec.\,\ref{sec:naivebasis}. Using the $7$ linearly independent relations among the set of $27$ $H$-functions at fourth order
means that we can write 
\be
  H_\alpha(\veck_1,\dots,\veck_4)=\sum_{\beta=1}^{20} R_{\alpha\beta} h_\beta(\veck_1,\dots,\veck_4)\qquad \mbox{for}\ 1\leq \alpha\leq 27\,,
\ee
 with some $27\times 20$ matrix $R_{\alpha\beta}$
and a set of $27-7=20$ functions $h_\beta$ that potentially enter in the reduced basis for $F_4$. Similarly, we can use the $11$ linearly independent relations among the $27$ $D$-functions
entering in $F_4$ to write 
\be
  D_\alpha(\eta)=\sum_{\gamma=1}^{16}d_\gamma(\eta)L_{\gamma\alpha}\qquad \mbox{for}\ 1\leq \alpha\leq 27\,,
\ee
 with some $16\times 27$ matrix $L_{\gamma\alpha}$ and a set of $27-11=16$ growth functions
$d_\gamma$ that potentially enter in the reduced basis for $F_4$. This leads to the representation 
\be
  F_4(\veck_1,\dots,\veck_4;\eta)=\sum_{\gamma\leq 16,\beta\leq 19} d_\gamma(\eta) M_{\gamma\beta} h_\beta(\veck_1,\dots,\veck_4)\,,
\ee
with a $16\times 20$ coefficient matrix given by $M_{\gamma\beta}=\sum_{\alpha\leq 27} L_{\gamma\alpha}R_{\alpha\beta}$, that encodes all the information about the redundancies in the naive basis.
In matrix-vector notation, 
\be
  F_4(\veck_1,\dots,\veck_4;\eta)=d^TMh,\qquad \mbox{and}\qquad M=LR\,.
\ee
However, this result is not yet of the desired form. To achieve this, we need to bring the matrix $M_{\gamma\beta}$ into a form where only terms on the diagonal are non-zero, i.e. for $\beta=\gamma$. This can be done by realizing that the choice of the $h_\beta$ and $d_\gamma$ functions is not unique so far. Indeed, an equally admissible choice is obtained by forming arbitrary linear combinations of the $h_\beta$, which corresponds to writing $h=Vh'$ with an arbitrary invertible $20\times 20$ matrix $V$. Similarly, there is a freedom to build linear combinations of the $d_\gamma$, by writing $d=Ud'$ with an arbitrary invertible $16\times 16$ matrix $U$. This leads to 
\be
  F_4(\veck_1,\dots,\veck_4;\eta)=d^TMh=(d')^TM'h'\,,
\ee
 with a modified $16\times 20$ matrix $M'\equiv U^TMV$.
We can therefore use the freedom to choose unitary (and therefore invertible) matrices $U$ and $V$ to perform a singular value decomposition of $M$, such that  $M'_{\beta\beta}$ are real and non-zero for $1\leq\beta\leq r$ where $r$ is the rank of the $16\times 20$ matrix $M$, and all other $M'_{\gamma\beta}=0$. As for any rectangular matrix, we know that $r\leq \mbox{min}(16,20)=16$. For the chosen example of the $F_4$ kernel, we find $r=11$, meaning that the reduced basis has $11$ summands. To bring it in our standard form, we can use the freedom to redefine the basis again,
with $h'=V'h''$, $d'=U'd''$ and diagonal, invertible matrices $U'$ and $V'$ such that $F_4=(d'')^TM''h''$ and $M''\equiv (U')^TM'V'=(U')^TUMVV'$. The entries on the diagonals of $U'$ and $V'$ can be chosen such that the non-zero entries on the diagonal are normalized, $M''_{\beta\beta}=1$ for $1\leq\beta\leq r$, while all other entries of $M''$ remain zero, as for $M'$. 
Identifying the components $d''_\gamma$ for $1\leq\gamma\leq r$ and $h''_\beta$ for $1\leq\beta\leq r$ with the set of fourth order growth functions and basis functions for the $F_4$ kernel, respectively, we arrive at the decomposition 
\be\label{eq:F4redbasis}
  F_4(\veck_1,\dots,\veck_4;\eta)=\sum_{i=1}^{11} d^{F}_{4,i}(\eta)h^F_{4,i}(\veck_1,\dots,\veck_4)\,.
\ee
This procedure yields the minimal number of terms, which is given by the rank of the matrix $M$, being $11$ for the $F_4$ kernel. It also yields a possible choice for the set of reduced basis functions. We note, that there is still a remaining freedom to redefine $d^F_{4,i}$ and $h^F_{4,i}$ by multiplying both with an identical, arbitrary unitary $11\times 11$ matrix.

The algorithm outlined above can be generalized in a straightforward way to all $F_n$ and $G_n$, respectively, and used to derive the mapping between the reduced and naive basis.
Furthermore, it automatically yields an optimal choice for the reduced basis, showing that it features the minimal amount of terms possible given the $H$- and $D$-relations that
were identified for the underlying naive basis.

Explicit expressions for the growth functions $d_{n,i}^F(\eta)$ and $d_{n,i}^G(\eta)$ as well as basis functions $h_{n,i}^F(\veck_1,\dots,\veck_n)$ and $h_{n,i}^G(\veck_1,\dots,\veck_n)$ of the reduced basis at order $n=4$ and at order $n=5$ can be found in App.\,\ref{sec:redbasisfourthorder} 
and App.\,\ref{sec:redbasisfifthorder}, respectively. The number of terms are summarized in Tab.\,\ref{terms_overview}.

\subsection{Constraints from momentum conservation and Galilean symmetry}\label{sec:softlimit}

The system of continuity and Euler equation is covariant under generalized Galilean transformations~\cite{Kehagias:2013yd,Peloso:2013zw}
and respects mass and momentum conservation. These properties hold irrespective of the time-dependence
of ${\cal H}(a)$ and $\Omega_m(a)$, and therefore apply also to the time-dependent kernels discussed in
this work.

As discussed already at the beginning of this section, momentum conservation requires $F_n(\veck_1,\dots,\veck_{n-2},\vecq,-\vecq;\eta)\propto k^2/q^2$
for $k^2\equiv(\sum_i\veck_i)^2\ll q^2$. We checked (up to $n=5$) that within the reduced basis, this scaling is respected for \emph{every basis function}
$h_{n,i}^X$ individually, see Eq.\,\eqref{eq:momentumconstraint}. This makes manifest that the corresponding scaling property holds for all kernels $F_n$ 
\emph{independently of the expansion history and matter density evolution}, i.e. for an arbitrary time-dependence of $x(\eta)=3\Omega_m(\eta)/2f(\eta)^2$.
As stressed in the beginning, the property that the scaling is respected by all basis functions is useful for the numerical evaluation of loop corrections.
In contrast, the functions $H_{n,i}^X$ within the naive basis scale generically as $q^0$, and the $q^{-2}$ scaling is only obtained after summing over all terms. 
Furthermore, the manifest $1/q^2$ scaling allows us to match the UV limit described by large wavenumbers $q$ to the EFT description, see Sec.\,\ref{sec:eft}.

Generalized Galilean invariance implies a relation between kernels of different order in the infrared (IR) limit, i.e. when one of the wavenumber arguments becomes small~\cite{softLimRed},
\begin{align}\label{eq:Galilei}
    {F}_n(\veck_1,\dots,\veck_{n-1},\vecq;\eta)\to\frac{1}{n}\frac{\veck \cdot \vecq e^\eta}{\vecq^2}{F}_{n-1}(\veck_1,\dots,\veck_{n-1};\eta)\quad \mbox{for}\ |\vecq|\to 0\,,
\end{align}
with $\veck=\sum\veck_i$ fixed. A similar relation holds for the $G_n$ and $G_{n-1}$ kernel.
This relation underlies the cancellation of IR sensitive terms when computing loop corrections to power- and bispectra~\cite{Jain:1995kx,Scoccimarro:1995if,Blas_2013,Blas:2013aba}. Additionally, it is also the basis for the commonly used IR-resummation
of loop corrections, that receive large corrections from bulk flows in the context of baryon acoustic oscillations (BAO)~\cite{Eisenstein:2006nj,Crocce:2007dt,Baldauf:2015xfa,Senatore:2014via,Blas_2016}. Indeed, the IR-resummation formula for the power spectrum can be derived
using only Eq.\,\eqref{eq:Galilei}, see App.\,F in~\cite{Blas_2016}. While Eq.\,\eqref{eq:Galilei} has been derived in~\cite{softLimRed} within the EdS approximation of SPT, it has been shown to hold in general in App.\,C of~\cite{Garny:2022kbk} (see also~\cite{DAmico:2021rdb}). This proof also applies to the class of cosmological models with arbitrary expansion history and matter density evolution considered here.
Therefore, Eq.\,\eqref{eq:Galilei} has to be valid also for the time-dependent kernels, implying the cancellation of IR sensitive terms as well as the applicability of the standard IR-resummation formula~\cite{Blas_2016}. We checked that Eq.\,\eqref{eq:Galilei} is valid for $n=2,3,4$, using the relations among growth functions of different orders from Eqs.\,\eqref{eq:D3D2rel1},~\eqref{eq:D3D2rel2} and Eqs.\,\eqref{eq:D4D3D2rel1}--\eqref{eq:D4D3D2relextra3}. 

In addition, by applying Eq.\,\eqref{eq:Galilei} twice it follows that
\begin{align}\label{eq:Galilei2}
    \int\frac{d\Omega_q}{4\pi}\,{F}_n(\veck_1,\dots,\veck_{n-2},\vecq,-\vecq;\eta)
    \to -\frac{1}{3n(n-1)}\frac{\veck^2 e^{2\eta}}{\vecq^2}{F}_{n-2}(\veck_1,\dots,\veck_{n-2};\eta)\quad \mbox{for}\ |\vecq|\to 0\,.
\end{align}
Explicitly, using Eq.\,\eqref{f3bare} we find $\int\frac{d\Omega_q}{4\pi}\,{F}_3(\veck,\vecq,-\vecq;\eta)\to -\veck^2/(9\vecq^2) d_{3,1}^F(\eta)=-\veck^2/(18\vecq^2)e^{3\eta}$, where the last step
follows from Eq.\,\eqref{eq:d3iF} and confirms Eq.\,\eqref{eq:Galilei2} for $n=3$.
For $n=4$ we find 
\bea
    \int\frac{d\Omega_q}{4\pi}\,{F}_4(\veck_1,\veck_{2},\vecq,-\vecq;\eta)& \to&
    -\frac{\left(\veck_1+\veck_2\right)^2}{36q^2}\frac{1}{2 k_1 k_2} \bigg[2 k_1 k_2 \left(\mu ^2-1\right) \left(d^F_{4,10}+d^F_{4,11}\right) \nn\\
    && {} +3 \left(\left(k_1^2+k_2^2\right) \mu +2 k_1 k_2\right) d^F_{4,3}\bigg]\,,
\eea
where $k_i=|\veck_i|$ and $\mu=\veck_1\cdot\veck_2/k_1k_2$. Using the definitions of the fourth order growth functions from App.~\ref{sec:redbasisfourthorder} as well as Eqs.\,\eqref{eq:D4D3D2rel1}--\eqref{eq:D4D3D2relextra3} one finds 
$d^{F}_{4,10}+d^{F}_{4,11}=e^{2\eta}d^F_{2,1}$ and $3d^{F}_{4,3}=e^{2\eta}d^F_{2,2}$.  Using the explicit expression of the second order kernel
in Eq.\,\eqref{eq:F2G2red}, we confirm by explicit computation that Eq.\,\eqref{eq:Galilei2} is satisfied also for $n=4$. 

Presenting every individual angle-averaged $h$-function is out of scope due to their length, but we provide $\int\frac{d\Omega_q}{4\pi}\, {F}_4^{\text{EdS}}(\veck_1,\veck_{2},\vecq,-\vecq)$
for general $q$ in App.\,\ref{F4AAappendix} for reference.

We note that an ``inverse'' strategy to construct generalized kernels has been followed in~\cite{DAmico:2021rdb}, where symmetry constraints have been used as starting point.
Up to third order, the number of terms and free coefficients is identical to the one found in our analysis. 
For example, for $F_3$ we find four terms, see Eq.\,\eqref{eq:d3iF}, out of which only two have growth functions, that are not trivially related to those known already from
lower order. It would also be interesting to investigate, whether symmetry
constraints from Galilei invariance can be related to the structure of generalized kernels within Lagrangian perturbation theory~\cite{Matsubara:2015ipa,Schmidt:2020ovm}.

\section{Effective field theory with time-dependent kernels}
\label{sec:eft}

\subsection{Effective field theory setup}

The framework of standard perturbation theory approximates matter as an ideal fluid, taking only the density contrast and velocity into account,
and neglecting the stress tensor $\sigma^{ij}$ contributing to the Euler equation for the velocity field. Even if cold dark matter is assumed to possess an
initially vanishing velocity dispersion, a non-zero stress tensor is generated non-linearly by shell (or orbit) crossing~\cite{Pueblas}.
Apart from the possibility to take velocity dispersion and potentially higher cumulants of the matter distribution function
into account explicitly within the perturbative expansion~\cite{McDonald:2009hs,Erschfeld:2018zqg,Garny:2022tlk,Garny:2022kbk,Erschfeld:2023aqr}, its impact can also be described within
an effective field theory (EFT) description~\cite{Baumann:2010tm}. In this approach, an effective stress tensor is added to the Euler equations, that is in turn
parameterized by a sum of ``operators'' ${\cal O}$ allowed by the underlying symmetries (Galilean invariance and mass and momentum conservation), each of them multiplied with a free
parameter, a so-called ``Wilson coefficient'' $c_{\cal O}$ (also frequently dubbed counterterm in this context), schematically
\be
  \sigma^{ij} = \sum_{\cal O} c_{\cal O}\, {\cal O}\,.
\ee
The operators can be further classified by the order in perturbation theory in which they contribute, the number of spatial derivatives acting on the density and velocity fields composing the operator, and whether they contribute to the EFT corrections of an elementary field (viz. the density contrast and velocity) or a composite field (i.e. the product of two density fields). The latter are often represented by stochastic ``noise terms'', and the former are referred to as ``deterministic corrections''~\cite{Assassi:2014fva,Mirbabayi:2014zca,Abolhasani:2015mra}. 

Following the treatment of~\cite{baldauf2015bispectrum,baldauf2021twoloop,Steele_2021}, the deterministic EFT corrections can at each order in perturbation theory be represented by adding an additional contribution $\tilde F_n$ to the kernels,
\be
  F_n^{\rm EFT}(\veck_1,\dots,\veck_n;\eta) = F_n(\veck_1,\dots,\veck_n;\eta) + \tilde F_n(\veck_1,\dots,\veck_n;\eta) \,.
\ee
For $n\leq 2$ and at lowest order in the derivative expansion, the \emph{most general} deterministic EFT corrections for the total matter density field can in Fourier space be parameterized as~\cite{baldauf2015bispectrum}
\bea
  \tilde F_1(\veck;\eta) &=& -c_s^2(\eta)k^2\,,\label{f1tilde} \\
  \tilde F_2(\veck_1,\veck_2;\eta) &=& \sum_{j=1}^3\epsilon_j(\eta) E_j (\veck_1,\veck_2) +\gamma(\eta)\Gamma(\veck_1,\veck_2)\,, \label{counterkernelf4}
\eea
with Wilson coefficients $c_s^2(\eta)$ and $\epsilon_j(\eta), \gamma(\eta)$, respectively, and shape functions derived from the EFT operators at second order given by~\cite{baldauf2021twoloop}
\begin{align}
    E_1(\veck_1,\veck_2)&\equiv (\veck_1+\veck_2)^2,\nn\\
    E_2(\veck_1,\veck_2)&\equiv (\veck_1+\veck_2)^2\left[\frac{(\veck_1\cdot\veck_2)^2}{k_1^2k_2^2}-\frac{1}{3}\right],\nn\\
    E_3(\veck_1,\veck_2)&\equiv  -\frac{(\veck_1+\veck_2)^2}{6}+\kkscalar + \frac{\kkscalar^2(k_1^2+k_2^2)}{2k_1^2k_2^2},\nn\\
    \Gamma(\veck_1,\veck_2)&\equiv (\veck_1+\veck_2)^2 \left(\frac{\veck_1\cdot\veck_2}{2k_1^2}+\frac{\veck_1\cdot\veck_2}{2k_2^2}\right) \nn\\
    & {} +\frac{1}{11}\left( \frac{207}{21}E_1(\veck_1,\veck_2) +\frac{12}{7}E_2(\veck_1,\veck_2) -6E_3(\veck_1,\veck_2)\right)\,. \label{shapefkt}
\end{align}
Here $c_s^2$ as well as $\epsilon_j$, $\gamma$ are Wilson coefficients, related to the operators $\partial^k\partial^l\phi$ as well as various index contractions
of $ \partial^k\partial^l\phi\partial^m\partial^n\phi$ at first and second order, respectively, where $\phi$ is the rescaled gravitational potential with $\partial^2\phi=\delta$.

\subsection{Application to general time-dependent kernels}

In the following, we show that, as expected, the EFT corrections as given above are sufficient to ``renormalize'' the leading UV dependence of the one-loop power- and bispectrum 
also for the case of cosmological models with general time-dependence as considered in this work. In Sec.\,\ref{application} we then use these results to discuss in how far the deviations from the EdS limit are degenerate with the Wilson coefficients. 

\subsubsection{Power spectrum}

For illustration, we first consider the density power spectrum, which up to one-loop order is given by
\bea
  P_{1L}(k,\eta) &=& e^{2\eta}P_{11}(k) -2c_s^2(\eta)k^2e^{\eta}P_{11}(k) + \int_{\vecq}  \Big(6F_3(\veck,\vecq,-\vecq;\eta)e^\eta P_{11}(q)P_{11}(k)\nn\\
  && {} +2F_2(\veck-\vecq,\vecq;\eta)^2P_{11}(q)P_{11}(|\veck-\vecq|) \Big)\,,
\eea
where $P_{11}(k)$ is the usual linear matter power spectrum, $D_1=e^\eta$ is the growth function, $F_n$ are the time-dependent kernels as discussed in the previous sections, and $c_s^2(\eta)$ is the Wilson coefficient associated to the first-order EFT correction. 
Due to the scaling Eq.\,\eqref{eq:scaling} imposed by momentum conservation, we can \emph{equivalently} rewrite the one-loop spectrum as
\bea\label{eq:P1Lren}
  P_{1L}(k,\eta) &=& e^{2\eta}P_{11}(k) -2c_{s,{\rm ren}}^2(\eta)k^2e^{\eta}P_{11}(k) + \int_{\vecq}  \Big(6F_3^{\rm ren}(\veck,\vecq,-\vecq,\eta)e^\eta P_{11}(q)P_{11}(k)\nn\\
  && {} +2F_2(\veck-\vecq,\vecq,\eta)^2P_{11}(q)P_{11}(|\veck-\vecq|) \Big)\,,
\eea
where we defined a ``renormalized kernel'' (applied for $n=3$ in the equation above)
\be\label{eq:Fnren}
  F_n^{\rm ren}(\veck_1,\dots,\veck_{n-2},\vecq,-\vecq;\eta) \equiv F_n(\veck_1,\dots,\veck_{n-2},\vecq,-\vecq;\eta) - \frac{1}{q^2} F_n^\infty(\veck_1,\dots,\veck_{n-2},\eta)W_\Lambda(q)\,,
\ee
and split the Wilson coefficient into two parts,
\be\label{eq:csren}
  c_s^2(\eta) = c_{s,{\rm ren}}^2(\eta) + \delta c_s^2(\eta)\,,
\ee
with $\delta c_s^2(\eta)\equiv -9F_3^\infty(\veck,\eta)\sigma_\Lambda^2/k^2$ and 
\be
  \sigma_\Lambda^2\equiv \frac13\int_{\vecq}  \, \frac{P_{11}(q)}{q^2}W_\Lambda(q)\,.
\ee
Furthermore, $W_\Lambda(q)$ is a high-pass filtering function that approaches unity for $q\gg \Lambda$, e.g. chosen as $W_\Lambda(q)=\Theta(q-\Lambda)$, and we introduced the \emph{hard limit}
\be\label{eq:Fnhard}
  F_n^\infty(\veck_1,\dots,\veck_{n-2},\eta) \equiv  \lim_{q\to\infty} \int \frac{d\Omega_q}{4\pi} q^2\, F_n(\veck_1,\dots,\veck_{n-2},\vecq,-\vecq,\eta)\,.
\ee
The rationale behind this rewriting is that the leading UV behaviour for large loop wavenumbers $q=|\vecq|$ is subtracted for the renormalized kernel, and instead absorbed into the (a priori) free choice of the Wilson coefficient. The EFT setup guarantees that this reshuffling is possible. For $F_3$, this is ensured by the property that $F_3^\infty(\veck,\eta)\propto k^2$,
such that $\delta c_s^2(\eta)$ as defined above is indeed independent of $k$, and only depends on time. At this order it is easy to see that the required scaling is a direct consequence of Eq.\,\eqref{eq:scaling}, and using the general time-dependent $F_3$ kernel we find 
\begin{align}
    F_3^\infty(\veck,\eta) = \sum_i h^{F,\infty}_{3,i}(\veck)d^F_{3,1}(\eta)
    = \left(\frac{7}{45} d^F_{3,1}(\eta)-\frac{32}{45} d^F_{3,2}(\eta)+\frac{32}{45} d^F_{3,3}(\eta)-\frac{4}{15} d^F_{3,4}(\eta)\right)k^2\,, \label{f3hard}
\end{align}
where $d^F_{3,i}(\eta)$ are the third-order growth functions defined in Eq.\,\eqref{eq:d3iF} and the coefficients follow from the hard limits 
\be\label{eq:hnhard}
  h_{n,i}^{F,\infty}(\veck_1,\dots,\veck_{n-2})\equiv\lim_{q\to\infty} \int \frac{d\Omega_q}{4\pi} q^2\, h^{F}_{n,i}(\veck_1,\dots,\veck_{n-2},\vecq,-\vecq)\,,
\ee
of the individual basis functions, defined in Eq.\,\eqref{eq:d3iF} for the case $n=3$. Following the discussion from above, within the reduced basis it is guaranteed that all
basis functions individually respect the $k^2$ scaling imposed by momentum conservation. This property is in particular instrumental to extend the EFT description to the bispectrum, see below.
In the EdS approximation all growth functions appearing in Eq.\,\eqref{f3hard} are proportional to $e^{3\eta}$, and inserting their values from Eq.\,\eqref{eq:d3EdS} one recovers the familiar result
$F_3^\infty(\veck,\eta)  \stackrel{\mbox{\scriptsize EdS}}{=} e^{3\eta}F_3^{{\rm EdS},\infty}= -\frac{61}{1890} k^2 e^{3\eta}$.
In the approximation of constant $x(\eta) \stackrel{\dot x=0}{=} x_0$  (see Sec.\,\ref{sec:constantx} and Eq.\,\ref{eq:d3EdS}) we find\footnote{We checked that the value of $x_0$, for which $F_3^\infty$ vanishes formally, corresponds to a growth rate $f\simeq 0.1635$ as pointed out in~\cite{Joyce:2022oal}.} $F_3^\infty(\veck,\eta)  \stackrel{\dot x=0}{=} \frac{\left(7 x_0^2-21 x_0-30\right)}{90  (x_0+2) (x_0+3)} k^2 e^{3\eta}$.

The renormalization procedure introduces a dependence on the arbitrary smoothing scale $\Lambda$, which by construction cancels in the sum of the loop integral
and the EFT correction containing the renormalized Wilson coefficient in Eq.\,\eqref{eq:P1Lren}. The required $\Lambda$-dependence of $c_{s,{\rm ren}}^2(\eta)$
can be obtained from the property that the ``bare'' Wilson coefficient is independent of $\Lambda$, $0=dc_s^2/d\Lambda$. Using Eq.\,\eqref{eq:csren}, this leads
to the renormalization group equation
\be\label{eq:rgecs}
  \frac{d}{d\Lambda}c_{s,{\rm ren}}^2(\eta) = - \beta_{c_s^2}(\eta) \frac{d\sigma_\Lambda^2}{d\Lambda}\,,
\ee
with $\beta_{c_s^2}d\sigma_\Lambda^2/d\Lambda \equiv d\delta c_s^2/d\Lambda=-9(F_3^\infty(\veck,\eta)/k^2)d\sigma_\Lambda^2/d\Lambda$ given by
\be\label{eq:betacs}
  \beta_{c_s^2}(\eta) = -9\left(\frac{7}{45} d^F_{3,1}(\eta)-\frac{32}{45} d^F_{3,2}(\eta)+\frac{32}{45} d^F_{3,3}(\eta)-\frac{4}{15} d^F_{3,4}(\eta)\right)\,.
\ee
Correspondingly, in the EdS limit one has $\beta_{c_s^2}\stackrel{\mbox{\scriptsize EdS}}{=} \frac{61}{210}e^{3\eta}$ and in the limit of
constant $x(\eta) \stackrel{\dot x=0}{=} x_0$, we find $\beta_{c_s^2}\stackrel{\dot x=0}{=}- \frac{\left(7 x_0^2-21 x_0-30\right)}{10  (x_0+2) (x_0+3)} e^{3\eta}$.
Let us now generalize this procedure to the bispectrum.

\subsubsection{Bispectrum}

The bispectrum up to one-loop order~\cite{Scoccimarro:1997st} and including the leading EFT corrections~\cite{baldauf2015bispectrum} is given by
\bea
  \lefteqn{ B_{1L}(k_1,k_2,k_3,\eta) = \Big[ 2P_{11}(k_2)P_{11}(k_3)e^{2\eta}F_2(\veck_2,\veck_3,\eta)+ 2\ \text{perm.}\Big] }\nn\\
  && + \left[2P_{11}(k_2)P_{11}(k_3)e^{2\eta}\left(\tilde F_2(\veck_2,\veck_3,\eta) +6\int_{\vecq} \, F_4(\veck_2,\veck_3,\vecq,-\vecq,\eta)P_{11}(q)\right) + 2\ \text{perm.}\right]\nn\\
  && + \left[2P_{11}(k_2)P_{11}(k_3)F_2(\veck_2,\veck_3,\eta)e^{3\eta} \left(-c_s^2k_3^2  +3\int_{\vecq} \, F_3(\veck_3,\vecq,-\vecq,\eta)P_{11}(q)\right) + 5\ \text{perm.}\right]\nn\\
  && + \left[B_{321}^I(k_1,k_2,k_3,\eta)+ 5\ \text{perm.}\right]+B_{222}(k_1,k_2,k_3,\eta)\,,
\eea
where the last two terms contain no leading UV sensitivity (see e.g. \cite{baldauf2021twoloop}), and are therefore irrelevant for the present discussion.
Similarly to the power spectrum, we can equivalently rewrite the bispectrum as
\bea
  \lefteqn{ B_{1L}(k_1,k_2,k_3,\eta) = \Big[ 2P_{11}(k_2)P_{11}(k_3)e^{2\eta}F_2(\veck_2,\veck_3,\eta)+ 2\ \text{perm.}\Big] }\nn\\
  && + \left[2P_{11}(k_2)P_{11}(k_3)e^{2\eta}\left(\tilde F_2^{\text{ren}}(\veck_2,\veck_3,\eta) +6\int_{\vecq} \, F_4^{\text{ren}}(\veck_2,\veck_3,\vecq,-\vecq,\eta)P_{11}(q)\right) + 2\ \text{perm.}\right]\nn\\
  && + \left[2P_{11}(k_2)P_{11}(k_3)F_2(\veck_2,\veck_3,\eta)e^{3\eta} \left(-c_{s,\text{ren}}^2k_3^2  +3\int_{\vecq} \, F_3^{\text{ren}}(\veck_3,\vecq,-\vecq,\eta)P_{11}(q)\right) + 5\ \text{perm.}\right]\nn\\
  && + \left[B_{321}^I(k_1,k_2,k_3,\eta)+ 5\ \text{perm.}\right]+B_{222}(k_1,k_2,k_3,\eta)\,,
\eea
where the third line contains the renormalized $c_s^2$ and $F_3$ kernels as for the power spectrum,
and the second line contains the renormalized $F_4$ kernel as defined in Eq.\,\eqref{eq:Fnren} for $n=4$.
Furthermore, we split the second-order EFT correction into a renormalized part and a part that accounts for the subtraction term introduced in $F_4^{\text{ren}}$,
\be
  \tilde F_2(\veck_1,\veck_2,\eta) = \tilde F_2^{\text{ren}}(\veck_1,\veck_2,\eta) + \delta \tilde F_2(\veck_1,\veck_2,\eta)\,,
\ee
with
\be
  \delta \tilde F_2(\veck_1,\veck_2,\eta) \equiv 18\, F_4^{\infty}(\veck_1,\veck_2,\eta)\sigma_\Lambda^2\,,
\ee
containing the hard limit $F_4^{\infty}$ of the fourth order kernel as defined in Eq.\,\eqref{eq:Fnhard} for $n=4$.
For this rewriting to be possible, it is a necessary condition that $F_4^{\infty}$ can be represented as a linear
combination of the four shape functions from Eq.\,\eqref{shapefkt}, composing the second-order EFT correction.
We find that, using the reduced basis representation Eq.\,\eqref{eq:F4redbasis} for the general time-dependent $F_4$ kernels, this is indeed the case.
More precisely, each of the basis functions $h_{4,i}^F(\veck_1,\veck_2,\vecq,-\vecq)$ has a hard limit Eq.\,\eqref{eq:hnhard}, that can be matched
onto a linear combination of shape functions,
\be
  h_{4,i}^{F,\infty}(\veck_1,\veck_2) = \sum_{j=1}^3 e_{j,i} E_j (\veck_1,\veck_2) +g_i\Gamma(\veck_1,\veck_2)\qquad \text{for}\ 1\leq i\leq 11\,.
\ee
The corresponding matching coefficients $e_{j,i}$ and $g_i$ are given in Tab.\,\ref{tab:h4coeff}. We emphasize that this matching is possible term-by-term only within the \emph{reduced basis} of the $F_4$ kernel introduced above, since the naive
basis would contain extra terms with a different structure that cancel only when summing over all terms, with the cancellation hinging on a precise
numerical evaluation of the growth functions. The reduced basis instead allows for a manifest and direct EFT matching, independently of the expansion
history and matter density evolution with time, and without relying on implicit cancellations.

\begin{table}[t]
    \centering
    \setlength{\extrarowheight}{0.3cm}
    \addtolength{\tabcolsep}{-1pt}
    \begin{tabular}{l|rrrrrrrrrrr}
        &$ h_{4,1}^{F,\infty} $ & $ h_{4,2}^{F,\infty} $ & $ h_{4,3}^{F,\infty} $&$ h_{4,4}^{F,\infty} $&$ h_{4,5}^{F,\infty} $&$ h_{4,6}^{F,\infty} $&$ h_{4,7}^{F,\infty} $&$ h_{4,8}^{F,\infty} $&$ h_{4,9}^{F,\infty} $&$ h_{4,10}^{F,\infty} $&$ h_{4,11}^{F,\infty}$ \\\hline
        $\epsilon_1 $&$ -\frac{4}{27} $&$ \frac{4}{45} $&$ \frac{89}{990} $&$ \frac{104}{135} $&$ -\frac{128}{135} $ & $ \frac{4}{3} $&$ -\frac{62}{135} $&$ -\frac{1231}{10395} $&$ \frac{1924}{10395}$ & $\frac{2869}{20790}$ & $-\frac{7603}{20790}$ \\
        $\epsilon _2 $&$ -\frac{2}{45} $&$ \frac{2}{315} $&$ -\frac{373}{13860} $&$ \frac{4}{315} $&$ \frac{16}{63} $&$ -\frac{46}{315} $&$ -\frac{1}{15} $&$ \frac{527}{6930} $&$ -\frac{122}{693} $&$ -\frac{757}{13860} $&$ \frac{481}{4620}$ \\
        $\epsilon _3 $&$ 0 $&$ \frac{32}{315} $&$ -\frac{527}{6930} $&$ -\frac{32}{63} $&$ \frac{32}{63} $&$ \frac{104}{315} $&$ -\frac{16}{45} $&$ \frac{1088}{3465} $&$ -\frac{1088}{3465} $&$ -\frac{8}{231} $&$ \frac{8}{495}$ \\
        $\gamma  $&$ 0 $&$ 0 $&$ \frac{7}{60} $&$ 0 $&$ 0 $&$ 0 $&$ 0 $&$ -\frac{16}{45} $&$ \frac{16}{45} $&$ -\frac{2}{15} $&$ -\frac{2}{15}$ \\
    \end{tabular}
    \caption{Matching coefficients for the basis functions $h_{4,i}^{F}$ composing the kernel $F_4$. Specifically, the entries show the matching coefficients of the hard limits $h_{4,i}^{F,\infty}(\veck_1,\veck_2) = \lim_{q\to\infty} \int \frac{d\Omega_q}{4\pi} q^2\,  h_{4,i}^F(\veck_1,\veck_2,\vecq,-\vecq)$
    onto the four EFT operators at second order corresponding to the shape functions $E_{1-3}(\veck_1,\veck_2)$ (row one to three) and $\Gamma(\veck_1,\veck_2)$ (row four) from Eq.\,\eqref{shapefkt}. The results are valid for cosmologies with arbitrary time-dependence of $H(a)$ and $\Omega_m(a)$.}
    \label{tab:h4coeff}
\end{table}

This allow us to split the second-order EFT Wilson coefficients into two parts analogously to $c_s^2$,
\be
  \epsilon_j(\eta) = \epsilon_j^{\text{ren}}(\eta)+\delta\epsilon_j(\eta),\qquad \gamma(\eta) = \gamma^{\text{ren}}(\eta) + \delta\gamma(\eta)\,,
\ee
with 
\be
  \tilde F_2^{\text{ren}}(\veck_1,\veck_2,\eta)= \sum_{j=1}^3 \epsilon_j^{\text{ren}}(\eta) E_j (\veck_1,\veck_2) +\gamma^{\text{ren}}(\eta)\Gamma(\veck_1,\veck_2)\,,
\ee
and
\be
  \delta\epsilon_j(\eta) = 18\,\sum_{i=1}^{11} e_{j,i} d_{4,i}^F(\eta) \sigma_\Lambda^2,\quad
  \delta\gamma(\eta) = 18\,\sum_{i=1}^{11} g_{i} d_{4,i}^F(\eta) \sigma_\Lambda^2\,.
\ee
Furthermore, we can derive renormalization group equations for the renormalized second-order Wilson coefficients similarly as for $c_s^2$,
\be\label{eq:rge2nd}
  \frac{d}{d\Lambda}\epsilon_j^{\text{ren}}(\eta) = - \beta_{\epsilon_j}(\eta) \frac{d\sigma_\Lambda^2}{d\Lambda},\qquad
  \frac{d}{d\Lambda}\gamma^{\text{ren}}(\eta) = - \beta_{\gamma}(\eta) \frac{d\sigma_\Lambda^2}{d\Lambda}\,,
\ee
with $\beta$-functions
\be\label{eq:beta2ndorder}
  \beta_{\epsilon_j}(\eta) = 18\,\sum_{i=1}^{11} e_{j,i} d_{4,i}^F(\eta),\quad
  \beta_{\gamma}(\eta) = 18\,\sum_{i=1}^{11} g_{i} d_{4,i}^F(\eta) \,,
\ee
given in terms of the fourth order growth functions $d_{4,i}^F(\eta)$.
This describes the dependence of the second order EFT coefficients $\epsilon_j$ and $\gamma$
on the smoothing scale $\Lambda$ for an arbitrary time-dependent cosmology, with coefficients from Tab.\,\ref{tab:h4coeff}.

For convenience, we also give the result when assuming a constant $x=3\Omega_m/(2f^2)\mapsto x_0$, 
\bea\label{eq:beta2ndorderEdS}
  &&\beta_{\epsilon_1}\stackrel{\dot x=0}{=} \frac{e^{4\eta} \left(1869 x_0^4-5019 x_0^3-30473 x_0^2-12066 x_0+19080\right)}{3465 (x_0+2)^2 (x_0+3) (x_0+4)},\nn\\
  &&\beta_{\epsilon_2}\stackrel{\dot x=0}{=}-\frac{e^{4\eta} \left(373 x_0^4+1317 x_0^3+1999 x_0^2+10110 x_0+19080\right)}{2310 (x_0+2)^2 (x_0+3) (x_0+4)},\nn\\
  &&\beta_{\epsilon_3}\stackrel{\dot x=0}{=} -\frac{e^{4\eta} \left(527 x_0^4+5013 x_0^3+20292 x_0^2+33012 x_0+15120\right)}{1155 (x_0+2)^2 (x_0+3) (x_0+4)},\nn\\
  &&\beta_{\gamma}\stackrel{\dot x=0}{=} \frac{e^{4\eta} \left(7 x_0^2-21 x_0-30\right)}{10 (x_0+2) (x_0+3)}\,.
\eea
The EdS limit can easily be recovered by setting $x_0=3/2$.

We note that, using our results for $F_5$, an analogous expansion could be made when including the third-order EFT operators contributing to $\tilde F_3$,
extending previous work in this direction within the EdS approximation~\cite{TriEFT1, TriEFT2}. This would be relevant for the renormalization of the
one-loop trispectrum~\cite{trispectrum}, and also enters the most general EFT description of the two-loop power spectrum~\cite{Zaldarriaga:2015jrj}.

In summary, the renormalization procedure subtracts the most UV sensitive part of the loop integrand, and the renormalized loop correction can be used
to quantify in how far the time-dependent kernels lead to an irreducible dependence of the one-loop power- and bispectrum on the cosmological model, see Sec.\,\ref{application}.

\section{Application to \lcdm and \ocdm cosmologies}
\label{application}

In this section, we provide some numerical examples of the impact of taking the precise time-dependence in the $F_n$ kernels into account, as compared to the conventional EdS approximation. In addition, we discuss in how far these deviations are degenerate with the EFT parameters when considering the one-loop density power- and bispectrum.

For illustration, we consider two cosmological models, \lcdm and \ocdm  (also known as CPL model~\cite{LofCPL}) with dark energy equation of state given by $w(a)=w_0+w_a(1-a)$, with two free parameters $w_0$ and $w_a$. We assume a spatially flat geometry in both cases. The relevant quantities for the kernels are the expansion history and matter density evolution within the late Universe, given by
\bea
  {\cal H}(a) &=& aH(a) = \left\{\begin{array}{lcl}
  aH_0\sqrt{\Omega_{m0}a^{-3}+1-\Omega_{m0}} && \text{\lcdm},\\ 
  aH_0\sqrt{\Omega_{m0}a^{-3}+(1-\Omega_{m0})a^{-3(1+w_0+w_a)}e^{-3w_a(1-a)}} && \text{\ocdm},
  \end{array}\right.\nn\\
  \Omega_m(a) &=& \left\{\begin{array}{lcl}
  \Omega_{m0}a^{-3}/(\Omega_{m0}a^{-3}+1-\Omega_{m0}) && \text{\lcdm},\\ 
  \Omega_{m0}a^{-3}/(\Omega_{m0}a^{-3}+(1-\Omega_{m0})a^{-3(1+w_0+w_a)}e^{-3w_a(1-a)}) && \text{\ocdm}\,.
  \end{array}\right.
\eea
For the present matter density parameter we use $\Omega_{m0}=0.31$~\cite{Planck_2018}. In both cases $\Omega_m(a)\to 1$ for $a\ll 1$
and $\Omega_m(a)\to\Omega_{m0}$ for $a\to 1$. We determine the growth function $D_1(a)$ and rate $f=d\ln D_1/d\ln a$ by numerically solving
the \mesza equation Eq.\,\eqref{meszaroslcdm}, using a normalization $D_1(1)=1$ and requiring $f\to 1$ for $a\ll 1$ in order to obtain the growing mode solution.
Note that the value of $H_0$ is not required, since only $d\ln{\cal H}/d\ln a$ enters in Eq.\,\eqref{meszaroslcdm}.

Subsequently, we use $\eta=\ln D_1$ as time-variable instead of the scale-factor $a$ for all computations, and show results for $\eta=0$ (i.e. today, $a=1$) unless
stated otherwise. The only input required to determine the growth functions is $x(\eta)=3\Omega_m/(2f^2)$, which can conveniently be obtained by computing a table
of values $\{\ln D_1(a),3\Omega_m(a)/(2f(a)^2)\}$ for a grid in $a$, and creating an interpolation that yields $x$ as function of $\eta$. We then iteratively solve the differential
equations Eqs.\,\eqref{eq:ODEforDdeltaandDC} and~\eqref{eq:ODEforDF} for the $n$th order growth functions $D^X_{n,i}(\eta)$ within the naive basis, and use the mapping from App.\,\ref{redbasis} to determine the $d_{n,i}^F(\eta)$ and $d_{n,i}^G(\eta)$ within the reduced basis for the $F_n$ and $G_n$ kernels, respectively. In practice, as mentioned above, we find it more convenient to numerically solve for the rescaled functions $\tilde D^X_{n,i}(\eta)\equiv e^{-n\eta}D^X_{n,i}(\eta)$, that approach constant values for $a\ll 1$, i.e. $\eta\ll 0$. Since $\Omega_m(a)\to 1$ for $a\ll 1$, we initialize all growth functions with their EdS values. We use $\eta_{\text{ini}}\simeq -50$, and checked that our results do not depend on the precise choice of the initialization time.

\subsection{Growth factors}

\begin{figure}[t]
    \centering
    \includegraphics[width=\textwidth]{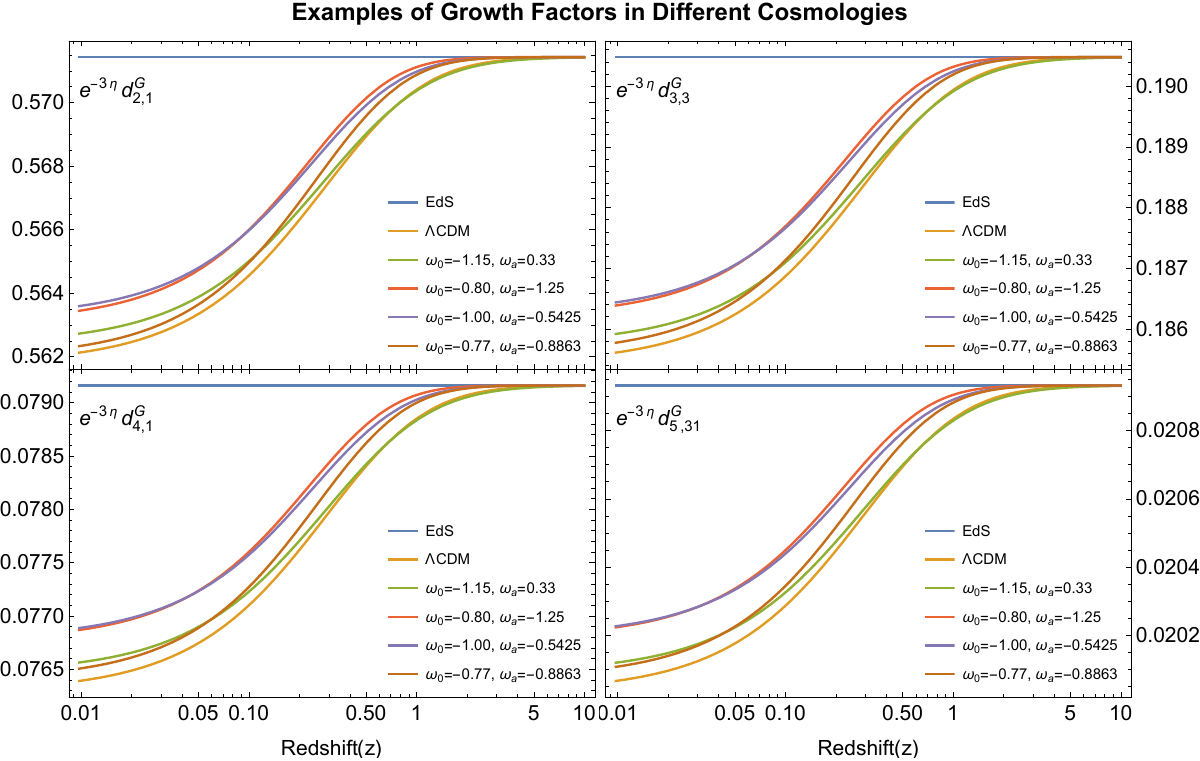}
    \caption{Redshift evolution of rescaled $n$th order growth factors $d_{n,i}^X(z)/D_1(z)^n$ within \lcdm and \ocdm for various values of $w_0$ and $w_a$ (see legends) compatible with Planck and DES-Y3 data~\cite{cpl_ellipse}. The panels correspond to $n=2,3,4,5$, respectively. At early times $z\gg 1$ all growth factors approach the (constant) value obtained for an EdS universe.}
    \label{fig:gfs}
\end{figure}

In Fig.\,\ref{fig:gfs}, we show the dependence of some exemplary $n$th order growth factors $d_{n,i}^X(\eta)$ on redshift $z$ for $n=2,3,4,5$. We show the ratio $d_{n,i}^X(\eta)/D_1(\eta)^n=e^{-n\eta}d_{n,i}^X(\eta)$ to the $n$th power of the linear growth factor $D_1=e^\eta$, that approaches a constant value within the matter dominated epoch at $z\gtrsim 1$, and agrees with the corresponding value obtained within the EdS approximation at high redshift. For $z\lesssim 1$, the growth functions start to deviate from the EdS value and feature a non-trivial time evolution that depends on the cosmological model. We show some examples for \lcdm and various sets of $w_0$ and $w_a$ parameters for \ocdm compatible with Planck and DES-Y3 data~\cite{cpl_ellipse} at $95\%$C.L. For all models, the deviations at $z=0$ are at most at the level of about $2\%, 3\%, 4\%, 5\%$ 
for $n=2,3,4,5$, respectively.
In order to bracket the impact of time-dependence on the non-linear kernels, we chose the growth function $d_{n,i}^X(\eta)$ with largest deviation from EdS in Fig.\,\ref{fig:gfs} at each order.
This generally turns out to be a growth function $d_{n,i}^G$ entering the velocity divergence kernels, whose corresponding integrand ${\cal I}_{n,i}$ also contains lower-order growth functions relevant for the velocity divergence. Among the sets of $d_{n,i}^F$ entering the $n$th order density kernels $F_n$, we find that the largest deviation from EdS at $z=0$ is about
$0.8\%, 1.5\%, 2.3\%, 3\%$ 
for $n=2,3,4,5$, respectively.

\subsection{Cosmology-dependence of $F_3$, $F_4$ and $F_5$}

\begin{table}[t]
    \centering
    \setlength{\extrarowheight}{0.2cm}
    \addtolength{\tabcolsep}{-3pt}
    \resizebox{\textwidth}{!}{
    \begin{tabular}{l|rrrrrrr}
         & $F_3$ & $F_4^{\text{equilateral}}$ & $F_4^{\text{isosceles}}$ & $F_4^{\text{squeezed}}$ & $F_5^{\text{square}}$ & $F_5^{2-\text{loop}} (p\gg k)$ & $F_5^{2-\text{loop}} (p\ll k)$\\\hline
        EdS (hard)$[r^2]$ & $-\dfrac{61}{1890}$ & $-\dfrac{1219}{246960}$ & $\dfrac{909}{19317760}$ & $\dfrac{1544}{509355}\epsilon $ & $-\dfrac{112457}{132432300}$ & $-\dfrac{120424}{45147375}$ & $\dfrac{61}{113400}s^2$ \\
        \lcdm (hard)$[r^2]$ & $-0.03165$ & $-0.004694$ & $0.0000934$ & $0.003371\epsilon$ & $-0.0007740$ & $-0.002643$ & $0.0005275s^2$\\
        Dev. from EdS $[\%]$ & $1.933$ & $4.897$ & $-98.50$ & $-11.19$ & $8.849$ & $0.9092$ & $1.933$ \\\hline
        EdS (soft) $[r^2]$& $-\dfrac{1}{18}$ & $-\dfrac{1}{126}$ & $-\dfrac{13}{32256}$ & $-\dfrac{1}{42}\epsilon^2$ & $-\dfrac{1}{1512}$ & $ \dfrac{61}{113400}s^2$ & $ \dfrac{1}{1080}s^2$\\
        \lcdm (soft) $[r^2]$& $-\frac{1}{18}$ & $-0.007976$ & $-0.0004061$ & $-0.02402\epsilon^2$ & $-0.0006737$ & $ 0.0005275s^2$ & $ \dfrac{1}{1080}s^2$ \\
        Dev. from EdS $[\%]$ & $0$ & $-0.4955$ & $-0.7623$ & $-0.8809$ & $-1.870$ & $1.933$ & $0$
    \end{tabular}}
    \caption{Comparison of the $F_3(\veck,\vecq,-\vecq,\eta)$, $F_4(\veck_1,\veck_2,\vecq,-\vecq,\eta)$, $F_5(\veck_1,\veck_2,\veck_3,\vecq,-\vecq,\eta)$ and $F_5(\veck,\vecp,-\vecp,\vecq,-\vecq,\eta)$ kernels with exact time-dependence for a \lcdm cosmology with $\Omega_{m0}=0.31$ at redshift $z=0$ to the corresponding values within the conventional EdS approximation, and for various configurations of wavenumbers (angle-averaged for $\vecq$ and $\vecp$). Here $r=k/q$ and $s=k/p$, with $k=|\veck|,p=|\vecp|$. For the rows, ``hard'' refers to $q\gg k,p$ and ``soft'' to $q\ll k,p$ (note that $p$ appears only in the last two columns). All entries in the table are understood to be multiplied by $r^2=k^2/q^2$, where either $k=|\veck|$ (columns 1, 6, 7) or $k=|\veck_1|$ (columns 2-5), see text for details. 
    }
    \label{tab:hardsoftlim}
\end{table}

\begin{figure}[t]
    \centering
    \includegraphics[width=\textwidth]{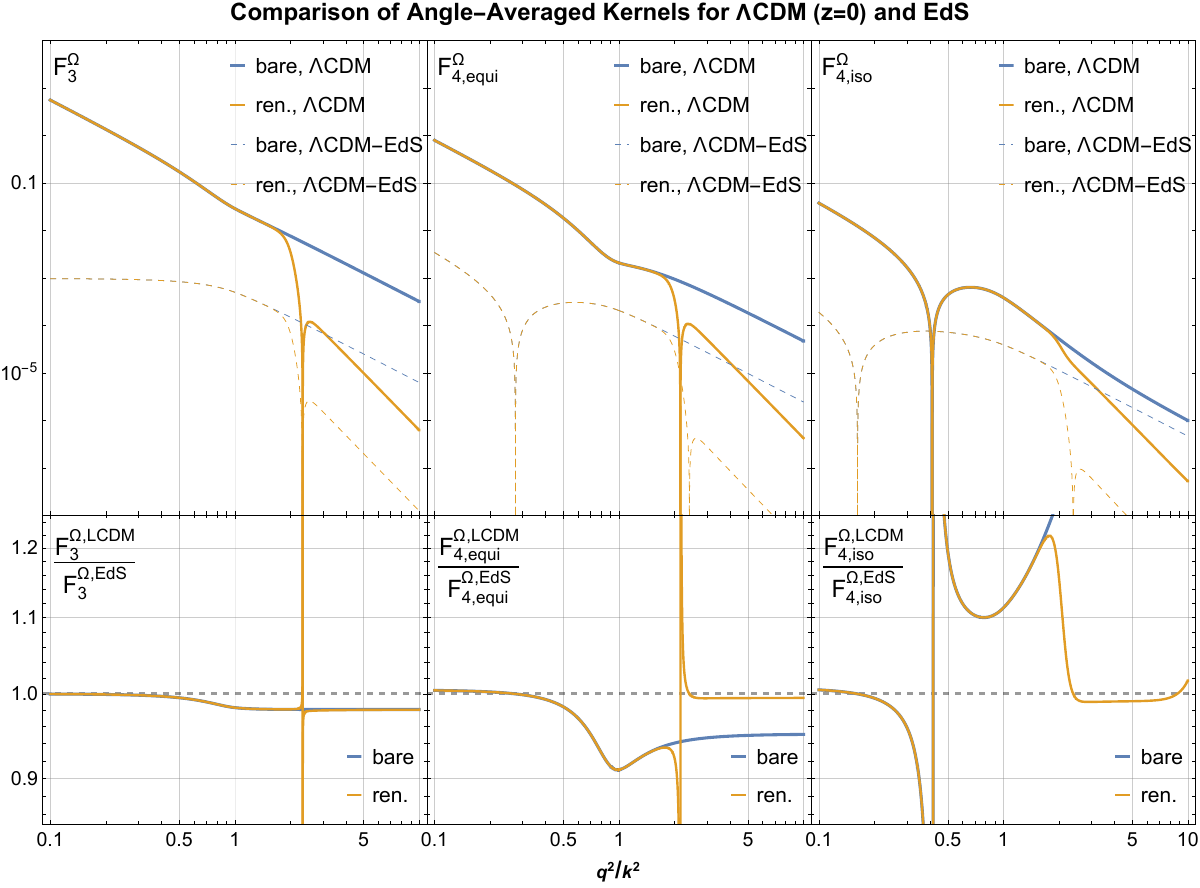}
    \caption{Angle-averaged kernels $\int\frac{d\Omega_q}{4\pi}F_3(\veck,\vecq,-\vecq,\eta)$ (left panels) and $\int\frac{d\Omega_q}{4\pi}F_4(\veck_1,\veck_2,\vecq,-\vecq,\eta)$
    for $(k_1,k_2,k_3)=(k,k,k)$ (middle panels) and $(k,k,k/2)$ (right panels), where $k_3=|\veck_1+\veck_2|$. We show the dependence on the ratio $q^2/k^2$, for redshift $z=0$ (i.e. $\eta=0$).
    The blue solid lines show the time-dependent kernels for a \lcdm cosmology with $\Omega_{m0}=0.31$. The blue dashed lines show the difference to the EdS approximation.
    In addition, we show the corresponding ``renormalized kernels'' defined in Eq.\,\eqref{eq:Fnren}, for which the leading UV dependence in the limit $q\gg k$ is subtracted, such that they decrease more steeply in the limit of large $q$. }
    \label{fig:ren}
\end{figure}

Let us next discuss the cosmology-dependence of the $n$th order kernels $F_n$ that enter in loop corrections to the matter density power-, bi- and trispectrum.
For concreteness we focus on the kernels $F_3(\veck,\vecq,-\vecq,\eta)$, $F_4(\veck_1,\veck_2,\vecq,-\vecq,\eta)$, $F_5(\veck_1,\veck_2,\veck_3,\vecq,-\vecq,\eta)$ and $F_5(\veck,\vecp,-\vecp,\vecq,-\vecq,\eta)$, averaged over the direction
of $\vecq$ (and $\vecp$ for the last case). We show these kernels in various limits in Tab.\,\ref{tab:hardsoftlim}, for the \lcdm model at $z=0$ and within the EdS approximation.
Generally, we find that the relative deviation depends highly on the specific configuration, and can range from (sub-)percent level to order one.

More specifically, the various columns in Tab.\,\ref{tab:hardsoftlim} correspond to the following cases:
\begin{itemize}
\item The first column of Tab.\,\ref{tab:hardsoftlim} shows $\int\frac{d\Omega_q}{4\pi}F_3(\veck,\vecq,-\vecq,\eta)$ entering in the one-loop power spectrum, in the limits $r\equiv k/q\ll 1$ (hard) and $r\gg 1$ (soft). The explicit result for general time-dependence is given in Eq.\,\eqref{f3bare}. Note that the result is proportional to $r^2$ in both limiting cases (the number in the table gives the proportionality factor). The soft limit of $F_3$ is cosmology-independent due to the constraint from Galilean invariance, see Eq.\,\eqref{eq:Galilei2}. The hard limit is related to the EFT corrections, see below.
\item Columns $2, 3$ and $4$ show $\int\frac{d\Omega_q}{4\pi}F_4(\veck_1,\veck_2,\vecq,-\vecq,\eta)$ for configurations given by $(k_1,k_2,k_3)=(k,k,k), (k,k,k/2), (k,k,\epsilon k)$, respectively, where $k_3\equiv|\veck_1+\veck_2|$ and $\epsilon\ll 1$, as well as $r=k/q\ll 1$ (hard) and $r\gg 1$ (soft). These kernels enter the one-loop correction to the bispectrum in equilateral, isosceles and squeezed configuration, respectively. 
\item    Column $5$ shows $\int\frac{d\Omega_q}{4\pi}F_5(\veck_1,\veck_2,\veck_3,\vecq,-\vecq,\eta)$ for $k_1=k_2=k_3\equiv k$ and $\veck_1\cdot\veck_2=\veck_2\cdot\veck_3=0$ as well as $\veck_3=-\veck_1$, that enter the one-loop correction to the trispectrum, in a square configuration.
\item    The last two columns show $\int\frac{d\Omega_q}{4\pi}\int\frac{d\Omega_p}{4\pi}F_5(\veck,\vecp,-\vecp,\vecq,-\vecq,\eta)$ that enters the two-loop power spectrum for various limits of the magnitudes $k,p,q$. The column marked $p\gg k$ corresponds to $q\gg p\gg k$ (row 1-3) or $q\ll k\ll p$ (row 4-6), and the last column marked $p\ll k$ to $q\gg k \gg p$ (row 1-3) or $q\ll p\ll k$ (row 4-6). Since we display all results rescaled by $r^2=k^2/q^2$, the entries involving a factor $s^2=k^2/p^2$ correspond to a scaling of the kernel $\propto s^2r^2=k^2/(p^2q^2)$, in accordance with Galilean invariance and momentum conservation constraints, see Eqs.\,\eqref{eq:Galilei2} and \eqref{eq:scaling}. The last three rows of the last column correspond to the case where both loop wavenumbers are soft. Consequently, Galilean invariance requires the kernel to be proportional to $F_1$ and it has no non-trivial cosmology dependence. Note that the cases $q\gg k \gg p$ and $p\gg k \gg q$ are symmetric under exchange of $q$ and $p$; this results in the same entries in the corresponding rows. We checked agreement with~\cite{baldauf2021twoloop} in the EdS case.
\end{itemize}
Note, that all soft limits are constrained by Galilean invariance, see Eq.\,\eqref{eq:Galilei2}, and therefore only sensitive to growth factors up to order $d^F_{n-2,i}$ for $F_n$.  

In addition, we show the dependence of angular-averaged $F_3$ and $F_4$ kernels on the ratio of loop and external wavenumber arguments by the blue lines in Fig.\,\ref{fig:ren}.
The blue solid lines show the time-dependent kernels within the \lcdm model at $z=0$, and the dashed-blue lines the differences to the corresponding EdS kernels. Both feature the expected scaling proportional to $k^2/q^2$ in both the soft ($q\ll k$) and hard ($q\gg k$) limits, in accordance with Tab.\,\ref{tab:hardsoftlim}. The lower panels show the ratio of the \lcdm and EdS results. The cosmology-dependence is  strongly dependent on the precise wavevector configuration, and generally stronger for $F_4$ than for $F_3$. The wavevector configuration can also determine the sign of the kernel, which can change, as shown in the isosceles case. Furthermore the cosmology-dependence is weaker in the soft than in the hard limit for the $F_n$ kernels given by the blue lines, since the soft limit is constrained by Galilean invariance. Fig.\,\ref{fig:ren} also shows the ``renormalized kernels'' defined in Eq.\,\eqref{eq:Fnren} in orange, to which we come back below. 

\subsection{Wilson coefficients}

\begin{figure}[t]
    \centering
    \includegraphics[width=\textwidth]{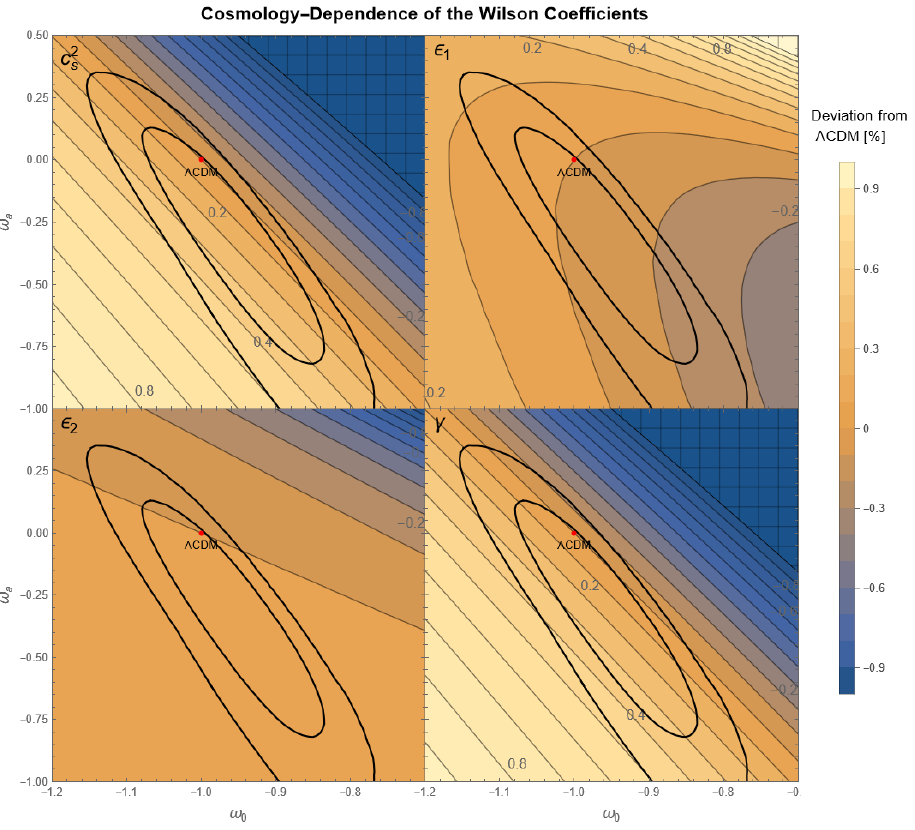}
    \caption{Cosmology-dependence of the UV (large-$q$) limits of the kernels $F_3(\veck,\vecq,-\vecq,\eta)$ and $F_4(\veck_1,\veck_2,\vecq,-\vecq,\eta)$ within the \ocdm model, as parameterized by the $\beta$-functions for the EFT parameters $c_s^2$, $\epsilon_1$, $\epsilon_2$ and $\gamma$, see  Eq.\,\eqref{eq:F34hard}. The panels show the deviation of $\beta_{c_s^2}$, $\beta_{\epsilon_1}$, $\beta_{\epsilon_2}$ and $\beta_{\gamma}$ at redshift $z=0$ relative to their \lcdm values (see text), in the plane spanned by $w_0$ and $w_a$. For comparison, the ellipses show the $68\%$ and $95\%$C.L. allowed regions from Planck and DES-Y3 data~\cite{cpl_ellipse}. We do not show a panel for the parameter $\epsilon_3$ since the difference between \lcdm and \ocdm turns out to be negligibly small for this parameter.}
    \label{fig:wilson}
\end{figure}

Let us now discuss in more detail the UV limit of $F_n(\veck_1,\dots,\veck_{n-2},\vecq,-\vecq,\eta)$ for large $q$. By construction, this part of the kernel is degenerate with EFT corrections, and therefore this type of cosmology-dependence
can be absorbed by the choice of the EFT parameters ($c_s^2$ and $\epsilon_j,\gamma$ for $n=3$ and $n=4$, respectively, see above). We nevertheless consider this hard limit, since $(i)$ it is used below to assess the irreducible cosmology dependence that remains even after taking the freedom in the choice of the EFT parameters into account, and $(ii)$ it has been argued that the ratio of second-order EFT terms tends to unity at first order, i.e. $\epsilon_j/c_s^2$ and $\gamma/c_s^2$ could be fixed to the value predicted by the UV limit of the $F_4$ kernel in order to reduce the amount of free EFT parameters in analyses of the bispectrum, see e.g.~\cite{Steele_2021,baldauf2021twoloop}. In this case these ratios do depend on the cosmological model when considering time-dependent kernels.

Within the notation introduced in Sec.\,\ref{sec:eft}, the UV limits of $F_n(\veck_1,\dots,\veck_{n-2},\vecq,-\vecq,\eta)$ as defined in Eq.\,\eqref{eq:Fnhard}
are parameterized by the functions $\beta_{c_s^2}(\eta)$ for $n=3$, see Eq.\,\eqref{eq:betacs}, as well as $\beta_{\epsilon_j}(\eta), \beta_\gamma(\eta)$ for $n=4$, see Eq.\,\eqref{eq:beta2ndorder}.
They directly determine the cosmology-dependence of the UV limits,
\bea\label{eq:F34hard}
\int\frac{d\Omega_q}{4\pi}F_3(\veck,\vecq,-\vecq,\eta) &\to& -\beta_{c_s^2}(\eta)\frac{k^2}{9q^2}\,,\nn\\
\int\frac{d\Omega_q}{4\pi}F_4(\veck_1,\veck_2,\vecq,-\vecq,\eta) &\to& \left[\sum_{j=1}^3\beta_{\epsilon_j}(\eta)E_j(\veck_1,\veck_2)+\beta_{\gamma}(\eta)\Gamma(\veck_1,\veck_2)\right]\frac{1}{18q^2}\,,
\eea
for large $q$, see Eq.\,\eqref{shapefkt} for the definition of $E_j$ and $\Gamma$.
In addition, the $\beta$-functions describe how the corresponding ``renormalized'' EFT coefficients depend on the smoothing scale $\Lambda$, see Eqs.\,\eqref{eq:rge2nd} and \eqref{eq:rgecs}.

Within the EdS approximation, $\beta_{c_s^2}\stackrel{\mbox{\scriptsize EdS}}{=} \frac{61}{210}e^{3\eta}$, and $\beta_{\epsilon_j},\beta_{\gamma}$ are obtained from Eq.\,\eqref{eq:beta2ndorderEdS}.
Solving for the growth functions numerically for the \lcdm model considered here we find at redshift $z=0$ (i.e. $\eta=0$)
$\beta_{c_s^2}^{\text{\lcdm}}/\beta_{c_s^2}^{\text{EdS}} \simeq 0.981$, and 
$\beta_{\epsilon_1}^{\text{\lcdm}}=-0.07068$, $\beta_{\epsilon_2}^{\text{\lcdm}}=-0.06437$, $\beta_{\epsilon_3}^{\text{\lcdm}}=-0.3713$,  $\beta_{\gamma}^{\text{\lcdm}}=-0.2849$,
deviating by about $-0.1\%$ for $\beta_{\epsilon_2}$ and $\beta_{\epsilon_3}$ from the EdS values, while $\beta_{\epsilon_1}$ is about $1.4\%$ larger and $\beta_\gamma$ is $2.1\%$ larger than for the EdS approximation.

Next, we investigate the cosmology-dependence within \ocdm. In Fig.\,\ref{fig:wilson} we show the relative deviation of $\beta_{c_s^2}, \beta_{\epsilon_1}, \beta_{\epsilon_2}$ and $\beta_{\gamma}$
at $z=0$ between the \ocdm and \lcdm models. As may already be expected from Fig.\,\ref{fig:gfs}, the deviation between \ocdm and \lcdm is smaller than the one between \lcdm and the EdS approximation. Correspondingly, we find a variation of the relative deviation between \ocdm and \lcdm that is at the sub-percent level for all of the EFT $\beta$-functions.

As stated above, the cosmology-dependence related to EFT parameters is actually irrelevant if they are treated as free and determined e.g. by fitting to data or simulations.
Therefore, we next discuss the irreducible dependence that remains even when marginalizing over EFT coefficients.

\subsection{Cosmology-dependence of power and bispectrum and degeneracy with EFT corrections}
\begin{figure}[t]
    \centering
    \includegraphics[width=\textwidth]{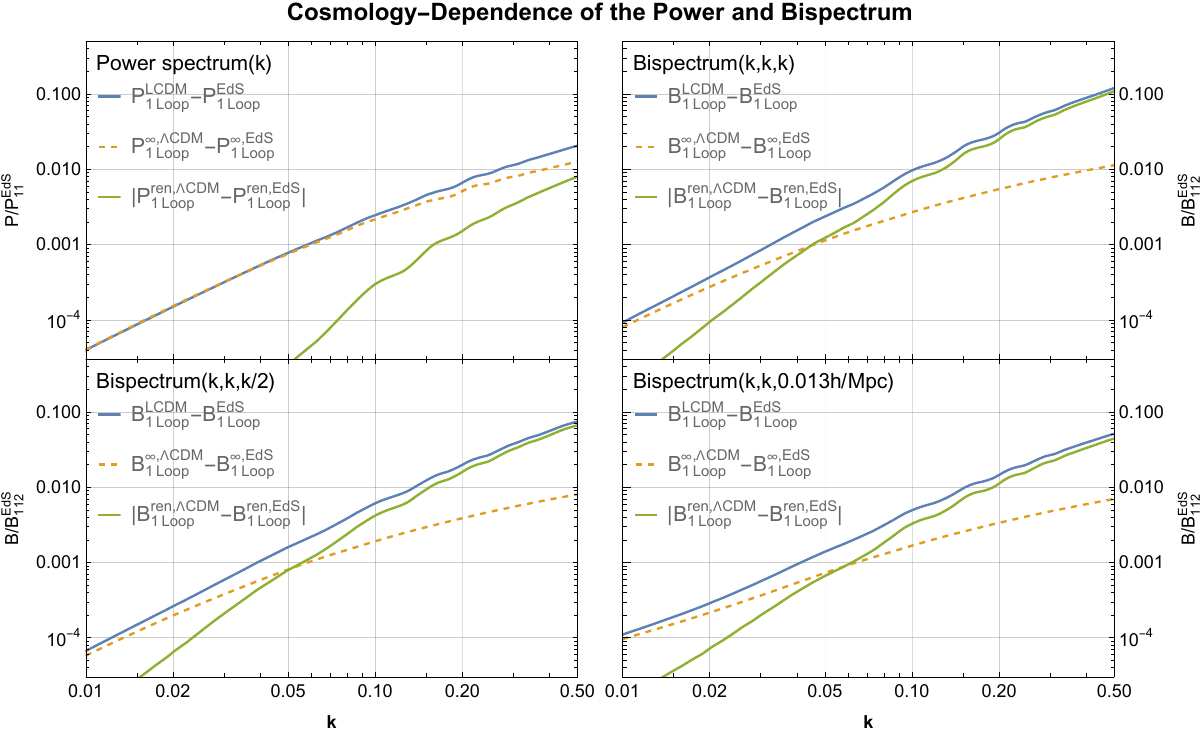}
    \caption{Impact of time-dependent kernels on the one-loop contributions to the power spectrum (upper left panel) and bispectrum (other panels for three configurations as indicated).
    We show the difference between the result using kernels for a \lcdm cosmology at $z=0$ and in the EdS approximation. The curves are normalized to the linear power spectrum or the tree-level bispectrum within EdS, respectively. 
    Blue lines show the result obtained without any EFT corrections. Dashed orange lines correspond to the part of the result that is degenerate with EFT corrections, and solid green lines show the irreducible cosmology-dependence that remains after projecting out contributions that are degenerate with EFT corrections. 
    }
    \label{fig:renplot}
\end{figure}
In Fig.\,\ref{fig:renplot}, we show the size of the corrections due to time-dependent kernels for the one-loop contributions to the power spectrum (upper left panel) and various configurations of the bispectrum, particularly
for the equilateral shape $(k_1,k_2,k_3)=(k,k,k)$ (upper right panel), isosceles triangles $(k,k,k/2)$ (lower left panel) and squeezed triangles $(k,k,0.013h/$Mpc$)$ (lower right panel). In each panel we show the difference between the result using time-dependent kernels $F_n$ for a
\lcdm cosmology at redshift $z=0$, and the EdS approximation. We note that these results have already been obtained in~\cite{baldauf2021twoloop}, using numerically computed $F_3$ and $F_4$ kernels (see also~\cite{taule, Garny:2022fsh, zvonimir} for the two-loop power spectrum). The main difference in this work is rather on the technical side. Specifically, we use the analytical expressions for time-dependent kernels derived above, within the reduced basis featuring manifest $k^2$ scaling of all basis functions in the UV limit (see Eq.\,\ref{eq:scaling}).

This technical difference allows us to study the interplay of time-dependent kernels and EFT corrections in an analytical manner. In particular, the blue lines in Fig.\,\ref{fig:renplot} show the (unrenormalized) results without any EFT corrections. For the unrenormalized power spectrum, the difference of the one-loop correction for \lcdm and EdS kernels grows above $1\%$ of the linear power spectrum for $k \gtrsim 0.25h/$Mpc at $z=0$ (for the bispectrum we find that this threshold occurs for smaller wavenumbers, see below). In addition, we show  the contribution to the loop from the UV region, that is by construction degenerate with EFT corrections, by the dashed orange lines (obtained from replacing the kernels by their hard limits $F_n^\infty$ as defined in Eq.\,\eqref{eq:Fnhard}, and integrating only over the hard region $q>\Lambda$; we set $\Lambda=k$ for the figure, see below). For small $k$ the blue and orange lines approach each other, since most of the loop momentum is associated to wavenumbers within the integration region $q\gg k$ in that limit. This implies that the cosmology-dependence can in that case be absorbed into a small re-adjustment of the EFT correction. 

For larger $k$, the dashed-orange and blue-solid lines start to differ. Their difference corresponds to the irreducible cosmology dependence, and is precisely given by the loop integrals involving the \emph{renormalized} kernels defined in Eq.\,\eqref{eq:Fnren}, shown by the green lines in Fig.\,\ref{fig:renplot}. For illustration, we chose $W_\Lambda(q)=\Theta(q-\Lambda)$ and $\Lambda=k$  to obtain the dashed-orange and solid-green lines, and set the renormalized part of the EFT parameters to zero. In practice, the renormalized EFT parameters would be determined by fitting to observation or simulation data. Here, the main point is that the residual, irreducible cosmology dependence of the one-loop correction to the power spectrum, that cannot be absorbed in the freedom to choose the EFT parameter, is represented by the result with renormalized kernels, shown by the green lines. In this case, the wavenumber, where the correction due to time-dependent kernels is larger than $1\%$ of the linear power, is shifted to $k>0.5h/$Mpc for the power spectrum (we find a much smaller degeneracy of exact time-dependence and EFT corrections
for the bispectrum, see below).

We also show the renormalized kernels in Fig.\,\ref{fig:ren} above. They agree with the original kernels for $q\ll k$ and decrease more strongly for large $q\gg k$, leading to a smaller sensitivity of the renormalized loop integral towards the region with large wavenumbers.
Note that for Fig.\,\ref{fig:ren} we chose a filter function with a smooth transition, $W_\Lambda(q)=(1+\tanh[(q-\Lambda)/ c\Lambda])/2$ with $\Lambda=2k$ and $c=0.1$ for illustration, but none of our conclusions depend on the precise shape. Since the EFT corrections dominate in the hard limit and they have opposite signs, there is a sign change in the renormalized kernels. Since the zero-crossing point is cosmology dependent, it is slightly shifted between the EdS and \lcdm kernels, which also leads to a pole at the ratios in the lower panel of Fig.\,\ref{fig:ren}.

Let us now discuss the cosmology-dependence of the bispectrum.
For the equilateral bispectrum, the difference between the one-loop correction for \lcdm and EdS kernels becomes larger than $1\%$ of the tree-level bispectrum for $k\gtrsim 0.1h/$Mpc (blue solid line in upper right panel in Fig.\,\ref{fig:renplot}). When taking the freedom to choose EFT parameters into account, this wavenumber is shifted to $k\approx 0.15h/$Mpc (green line). Similarly, for isosceles triangles (lower left panel) the $1\%$ threshold is crossed for $k\gtrsim 0.13h/$Mpc for the case without EFT corrections (blue line), and for $k \gtrsim 0.15h/$Mpc when taking them into account (green line). For the squeezed bispectrum (lower right panel), these numbers are $k\gtrsim 0.15h/$Mpc and $k \gtrsim 0.2h/$Mpc, respectively. 

In summary, we find that the impact of time-dependent kernels is larger for the one-loop bispectrum than for the one-loop power spectrum. Some part of this cosmology-dependence can be absorbed into the freedom to choose EFT parameters within the EFT setup, and we precisely quantify the remaining irreducible impact of \lcdm kernels by using the ``renormalized'' time-dependent kernels defined in Sec.\,\ref{sec:eft}. For the bispectrum the impact of time-dependent kernels is much less degenerate with the freedom in adjusting counterterms as compared to the power spectrum (blue solid versus green solid lines in Fig.\,\ref{fig:renplot}).

We furthermore demonstrated that the conventional EFT corrections for the one-loop power- and bispectrum apply to an arbitrary cosmological expansion history. Additionally, we computed the matching of the time-dependent kernels in the UV limit onto the EFT parameters  analytically. This required to identify certain cosmology-independent relations, which allowed us to set up a minimal reduced basis for the time-dependent $F_n$ kernels, in which each basis function individually respects the asymptotic scaling required by mass and momentum conservation.

\section{Conclusion}
\label{sec:conclusion}

In this work, we derived non-linear kernels $F_n$ up to $n=5$, that are applicable for cosmological models
with arbitrary time-dependent expansion history $H(a)$, encompassing in particular \lcdm (for massless neutrinos)
as well as \ocdm. We find that the number of terms contributing to each $F_n$ can be considerably reduced
as compared to a conventional, ``naive'' algorithm for obtaining these kernels, leaving $11$ independent contributions to $F_4$ and $39$ to $F_5$.
We provide explicit expressions for the cosmology-independent basis functions $h_{n,i}^F(\veck_1,\dots,\veck_2)$ as well as
an algorithm for numerically generating the cosmology-dependent growth functions $d_{n,i}^F(\eta)$ furnishing the kernels $F_n$.

Most importantly, the ``reduced'' basis derived in this work has the property that each individual basis function satisfies the scaling required by
mass and momentum conservation. This property is non-trivial for $n\geq 4$. It is not satisfied when
constructing the time-dependent kernels in a ``naive'' way, in which case the correct scaling appears only when summing over all contributions $i$, leading
to potentially large cancellations among the various terms. 
The correct scaling greatly facilitates numerical evaluation of loop corrections for the one-loop bispectrum and two-loop power spectrum. 

In addition, the reduced basis allows us to explicitly show that the UV sensitivity of the cosmology-dependent kernel
$F_4(\veck_1,\veck_2,\vecq,-\vecq,\eta)$, that enters the one-loop bispectrum, can be absorbed into the same set
of four EFT coefficients known from the EdS case. As expected, this implies, that when allowing for
the most general set of EFT corrections relevant for the one-loop bispectrum, the EFT description is applicable
for cosmological models with arbitrary time-dependent $H(a)$. However, the ``running'' of the EFT coefficients
when varying the smoothing scale does depend on cosmology, and we quantify this dependence for \lcdm and \ocdm.

Furthermore, we formulate ``renormalized'' expressions for the one-loop power- and bispectrum including EFT
corrections and cosmology-dependent kernels. We use this rewriting to assess in how far the cosmology-dependence
of the kernels is degenerate with the freedom to adjust the EFT parameters. While there is an approximate degeneracy
for the one-loop power spectrum on weakly non-linear scales, we find that EFT corrections and the impact of
precise, cosmology-dependent kernels are largely orthogonal for the one-loop bispectrum.

The quantitative impact of cosmology-dependent versus EdS kernels depends highly on the precise configuration for the
bispectrum. We generally find that the impact is typically at the percent level for $\Lambda$CDM.

As a by-product, we also provide analytical, algebraic recursion relations for the $F_n$ kernels for all $n$
within the ``generalized EdS'' approximation~\cite{Joyce:2022oal}, that is applicable to models with mixed hot and cold dark matter.
\bigskip
\bigskip

\subsection*{Acknowledgements}

We thank Tobias Baldauf, Henrique Rubira, Fabian Schmidt, Roman Scoccimarro, Petter Taule, and Zvonimir Vlah for discussions.
We acknowledge support by the Excellence Cluster ORIGINS,
which is funded by the Deutsche Forschungsgemeinschaft
(DFG, German Research Foundation) under Germany’s
Excellence Strategy - EXC-2094 - 390783311.

\appendix
\allowdisplaybreaks

\section{Derivation of the propagator}\label{propderiv}

In this appendix, we derive the linear propagator Eq.\,\pref{crudeprop} for a general time-dependent cosmology.
We start from the differential equation Eq.\,\eqref{masterequation}. In linear approximation, it can be turned into a single second-order equation for the density contrast,
\begin{equation}\label{eq:delta2ndorderlin}
     \partial_\eta^2\delta - (1-x(\eta))\partial_\eta\delta - x(\eta)\delta=0\,,
\end{equation}
where $\eta=\ln(D_1)$ and $x(\eta)=3\Omega_m(\eta)/(2f(\eta)^2)$. We write the general solution as $\delta(\eta)=Av(\eta)+Bw(\eta)$ with constants $A,B$.
By construction, one of the solutions is $v(\eta)\equiv D_1(\eta)=e^\eta$, as can be readily verified.
A second linearly independent solution $w(\eta)$ can be obtained by considering the ratio $u\equiv \delta/D_1$, for which Eq.\,\eqref{eq:delta2ndorderlin} yields
$\partial_\eta^2u+(1+x)\partial_\eta u=0$. A formal solution for $\partial_\eta u$ can now be easily obtained by separation of variables, which in turn yields $w(\eta)=\int^\eta d\eta' e^{\eta-\eta'-\int^{\eta'}d\eta''x(\eta'')}$.

In the non-linear regime, the right hand side of Eq.\,\pref{masterequation} does not vanish, changing the differential equation to
\begin{equation}
    \partial_\eta^2\delta - (1-x(\eta))\partial_\eta\delta - x(\eta)\delta=\Tilde{S}(\eta)\,,
\end{equation}
with some time-dependent source function $\Tilde{S}$.
This is a Green's problem and solved by
\begin{equation}
    \delta(\eta)= Av(\eta)+Bw(\eta) + \int^\eta d\eta' g(\eta,\eta')\Tilde{S}(\eta'), \label{deltaS}
\end{equation}
with propagator
\be
  g(\eta,\eta') = \frac{v(\eta)w(\eta')-w(\eta)v(\eta')}{(\partial_{\eta'}v(\eta'))w(\eta')-(\partial_{\eta'}w(\eta'))v(\eta')}\,.
\ee
Inserting the two solutions $v$ and $w$ from above yields Eq.\,\pref{getaetaint} for $g(\eta,\eta')$. 

Next, we formulate the equation of motion for the coupled first-order system Eq.\,\eqref{masterequation} with a general two-component source term $\Tilde{S}_a$ as 
\begin{equation}\label{eq:psisource}
    (\partial_\eta + {\Omega_{a}}^b(\eta)) \psi_b(\eta) = \Tilde{S}_a(\eta).
\end{equation}
Rewriting it into a second-order equation for $\delta=\psi_1$ and using Eq.\,\eqref{deltaS} yields
\begin{align}
    \delta(\eta)&= Av(\eta)+Bw(\eta) +\int^\eta g(\eta,\eta')\left[\Tilde{S}_2(\eta')+\pdiff{\Tilde{S}_1(\eta')}{\eta'}+(x(\eta')-1)\Tilde{S}_1(\eta')\right],\label{inhomdelta}\\
    \theta(\eta)&=A\partial_\eta v(\eta)+B\partial_\eta w(\eta)-\Tilde{S}_1(\eta)+\int^\eta [\partial_\eta g(\eta,\eta')]\left[\Tilde{S}_2(\eta')+\pdiff{\Tilde{S}_1(\eta')}{\eta'}+(x(\eta')-1)\Tilde{S}_1(\eta')\right].\label{inhomtheta}
\end{align}
The second line was obtained using Eq.\,\eqref{eq:psisource} with index $a=2$.
The derivatives acting on the source terms can be eliminated via partial integration. Both equations can then be brought into the form
\begin{align}
    \psi_a(\eta)={G_a}^b(\eta,\eta_0)\psi_b(\eta_0) + \int_{\eta_0}^\eta d\eta' {G_a}^b(\eta,\eta')\Tilde{S}_b(\eta')\,,
    \label{psiequation}
\end{align}
with linear propagator $G(\eta,\eta')$ as given in Eq.\,\pref{crudeprop}, and constants $A,B$ traded for $\psi_a(\eta_0)$.

\section{Source terms}\label{sourceterms}

In this appendix, we provide expressions for the source terms $S_a^{(n)}(\veck_1,\dots,\veck_n;\eta)$ defined in Eq.\,\eqref{sourcetermdeff} for $a=1,2$ and up to order $n\leq 5$, which enter the time-evolution equations of the density and velocity kernels $F_n$ and $G_n$, see Eq.\,\eqref{sourcetermdiffe}. We use the decomposition $F_n = \m{F}_n + \m{C}_n,  G_n = \m{G}_n + \m{C}_n$ defined in Eq.\,\eqref{gtmaster}, which involve the combinations of source terms
$S^{(n)}_\Delta\equiv S^{(n)}_2-S^{(n)}_1$ and $S^{(n)}_C \equiv S^{(n)}_{1}$, see Eq.\eqref{eq:defFGC}.
As a short-hand notation, we use $\m{F}_n(\veck_1,\dots,\veck_n;\eta)\equiv \m{F}^{(n)}_{1,\dots,n}$ and similarly for $\m{G}_n$ and $\m{C}_n$.

We also use $\xi(\veck_1,\veck_2)\equiv \beta(\veck_1,\veck_2)-\alpha^s(\veck_1,\veck_2)$, $\alpha^s(\veck_1,\veck_2)\equiv\frac{1}{2}(\alpha(\veck_1,\veck_2)+\alpha(\veck_2,\veck_1))$,
and the shorthand notation $\alpha_{1,23}=\alpha(\veck_1,\veck_2+\veck_3)$, etc., and similarly for $\alpha^s$ and $\xi$.

At first order $S_a^{(1)}=0$.

\subsection{Second order}

\begin{flalign}
    S_1^{(2)}&=D_1\alpha^s_{1,2},&\\
    S_2^{(2)}&=D_1\beta_{1,2},&\\
    S_\Delta^{(2)}&=D_1\xi_{1,2}.
\end{flalign}
\subsection{Third order}

\begin{flalign}
    S_1^{(3)}&=\frac{{D}_1}{3}\left[ \alpha_{3,12}\m{F}^{(2)}_{1,2}+2\alpha^s_{12,3}\m{C}^{(2)}_{1,2}+\alpha_{12,3}\m{G}^{(2)}_{1,2} +\text{2 perm.}\right],&\\
    S_2^{(3)}&=\frac{{D}_1}{3}\left[2\beta_{12,3}\left( \m{G}^{(2)}_{1,2}+\m{C}^{(2)}_{1,2} \right)+\text{2 perm.}\right],&\\
    S_\Delta^{(3)}&=\frac{{D}_1}{3}\left[ -\alpha_{3,12}\m{F}^{(2)}_{1,2}+2\xi_{12,3}\m{C}^{(2)}_{1,2}+\alpha_{12,3}\m{G}^{(2)}_{1,2} +\text{2 perm.}\right].
\end{flalign}

\subsection{Fourth order}

\begin{flalign}
    S_1^{(4)}&=\frac{{D}_1}{4}\left[\alpha_{4,123}\m{F}^{(3)}_{1,2,3}+2\alpha^s_{123,4}\m{C}^{(3)}_{1,2,3}+\alpha_{123,4}\m{G}^{(3)}_{1,2,3}+\text{3 perm.} \right] \nonumber&\\
    &\qquad + \frac{1}{6}\left[\alpha_{12,34}\m{C}^{(2)}_{1,2}\m{F}^{(2)}_{3,4}+\alpha_{12,34}\m{G}^{(2)}_{1,2}\m{F}^{(2)}_{3,4}+\alpha_{12,34}\m{G}^{(2)}_{1,2}\m{C}^{(2)}_{3,4}+\alpha_{12,34}\m{C}^{(2)}_{1,2}\m{C}^{(2)}_{3,4}+\text{5 perm.}\right]&\\
    S_2^{(4)}&=\frac{{D}_1}{4} \left[2\beta_{123,4}(\m{G}^{(3)}_{1,2,3}+\m{C}^{(3)}_{1,2,3})+\text{3 perm.}\right] +\frac{1}{6}\left[\beta_{12,34}\left(\m{G}^{(2)}_{1,2}+\m{C}^{(2)}_{1,2}\right)\left(\m{G}^{(2)}_{3,4}+\m{C}^{(2)}_{3,4}\right)+\text{5 perm.}\right]&\\
    S_\Delta^{(4)}&=\frac{{D}_1}{4} \left[-\alpha_{4,123}\m{F}^{(3)}_{1,2,3}+2\xi_{4,123}\m{C}^{(3)}_{1,2,3}+\left(2\xi_{123,4}+\alpha_{4,123}\right)\m{G}^{(3)}_{1,2,3} +\text{3 perm.} \right] \nonumber &\\ 
    &\qquad +\frac{1}{6}\left[-\alpha_{12,34}\m{C}^{(2)}_{1,2}\m{F}^{(2)}_{3,4}-\alpha_{12,34}\m{G}^{(2)}_{1,2}\m{F}^{(2)}_{3,4}+\left(2\xi_{12,34}+\alpha_{34,12}\right)\m{G}^{(2)}_{1,2}\m{C}^{(2)}_{3,4}+\text{5 perm.}\right] \nonumber &\\
    &\qquad +\frac{1}{3}\left[\xi_{12,34}\m{C}^{(2)}_{1,2}\m{C}^{(2)}_{3,4}+(\xi_{12,34}+\alpha^s_{12,34})\m{G}^{(2)}_{1,2}\m{G}^{(2)}_{3,4}+\text{2 perm.}\right]
\end{flalign}

\subsection{Fifth order}

\begin{flalign}
    S_1^{(5)}
    &=\frac{{D}_1}{5}\left[ \alpha_{5,1234}\m{F}^{(4)}_{1,2,3,4} + 2\alpha^s_{1234,5}\m{C}^{(4)}_{1,2,3,4}+\alpha_{1234,5}\m{G}^{(4)}_{1,2,3,4} +\text{ 4 perm.}  \right]\nonumber &\\
    &\quad+\frac{1}{10}\left[ \alpha_{345,12}\m{F}^{(2)}_{1,2}\m{C}^{(3)}_{3,4,5}+\alpha_{345,12}\m{F}^{(2)}_{1,2}\m{G}^{(3)}_{3,4,5}+\alpha_{12,345}\m{C}^{(2)}_{1,2}\m{F}^{(3)}_{3,4,5}+\alpha_{345,12}\m{C}^{(2)}_{1,2}\m{G}^{(3)}_{3,4,5}\right.\nonumber&\\
    &\qquad +\left.\alpha_{12,345}\m{G}^{(2)}_{1,2}\m{F}^{(3)}_{3,4,5}+\alpha_{12,345}\m{G}^{(2)}_{1,2}\m{C}^{(3)}_{3,4,5}+2\alpha^s_{12,345}\m{C}^{(2)}_{1,2}\m{C}^{(3)}_{3,4,5} +\text{9 perm.}\right]\label{s15}&\\
    S_2^{(5)}
    &=\frac{{D}_1}{5}\left[ 2\beta_{1234,5}\m{G}^{(4)}_{1,2,3,4} + 2\beta_{1234,5}\m{C}^{(4)}_{1,2,3,4} +\text{ 4 perm.} \right] \nonumber&\\
    &\quad+\frac{1}{10}\Big[ 2\beta_{12,345}\m{C}^{(2)}_{1,2}\m{G}^{(3)}_{3,4,5} + 2\beta_{12,345}\m{G}^{(2)}_{1,2}\m{C}^{(3)}_{3,4,5}+2\beta_{12,345}\m{C}^{(2)}_{1,2}\m{C}^{(3)}_{3,4,5} +2\beta_{12,345}\m{G}^{(2)}_{1,2}\m{G}^{(3)}_{3,4,5}  \nonumber&\\
    &\qquad+ \text{ 9 perm.} \Big]\label{s25}&\\
    S_\Delta^{(5)}&=\frac{{D}_1}{5}\left[ -\alpha_{5,1234}\m{F}^{(4)}_{1,2,3,4} +2\xi_{1234,5}\m{C}^{(4)}_{1,2,3,4}+\left(2\beta_{1234,5}-\alpha_{1234,5}\right)\m{G}^{(4)}_{1,2,3,4} +\text{ 4 perm.}\right] \nonumber&\\
    &\quad+\frac{1}{10}\left[-\alpha_{345,12}\m{F}^{(2)}_{1,2}\m{C}^{(3)}_{3,4,5}-\alpha_{345,12}\m{F}^{(2)}_{1,2}\m{G}^{(3)}_{3,4,5}-\alpha_{12,345}\m{C}^{(2)}_{1,2}\m{F}^{(3)}_{3,4,5} \right. \nonumber&\\
    &\qquad+\left.\left(2\beta_{12,345}-\alpha_{345,12}\right)\m{C}^{(2)}_{1,2}\m{G}^{(3)}_{3,4,5} 
    -\alpha_{12,345}\m{G}^{(2)}_{1,2}\m{F}^{(3)}_{3,4,5}+\left(2\beta_{12,345}-\alpha_{12,345}\right)\m{G}^{(2)}_{1,2}\m{C}^{(3)}_{3,4,5}\right. \nonumber&\\
    &\qquad+\left.2\xi_{12,345}\m{C}^{(2)}_{1,2}\m{C}^{(3)}_{3,4,5}+2\beta_{12,345}\m{G}^{(2)}_{1,2}\m{G}^{(3)}_{3,4,5} +\text{ 9 perm.} \right]\label{ss5}
\end{flalign}

\section{Naive basis: $H$-functions}\label{appHfunc}

In this appendix, we provide explicit expressions for the wavevector-dependent basis functions $H_{n,i}^X$ within the ``naive basis'' introduced in Sec.\,\ref{sec:naivebasis}, see Eq.\,\eqref{naive_basis}, for $n\leq 5$.
In addition, we summarize all relations among them, as discussed in Sec.\,\ref{sec:Hrelations}.

Note, that we discriminate two sets, $H_{n,i}\equiv H_{n,i}^{\m F}=H_{n,i}^{\m G}$ and $H_{n,i}^{\m C}$, see Sec.\,\eqref{sec:naivebasis}.
We use the shorthand notation $H_{n,i}^X(\veck_1,\dots,\veck_n)\equiv H_{n,i}^X(\veck_{1,\dots,n})$ here.

At first order $H_{1,1}^{\m C}(\veck)=1$, while $H_{n,i}$ start only for $n\geq 2$.

\subsection{Second order}

\paragraph{Definitions:}

\begin{align}
    H_{2,1}(\veck_1,\veck_2)=\xi(\veck_1,\veck_2),\qquad H^{\m C}_{2,1}(\veck_1,\veck_2)=\alpha^s(\veck_1,\veck_2)\,.
\end{align}

\paragraph{Relations:} None at $n=2$.

\subsection{Third order}

\paragraph{Definitions:}

\begin{align*}
    H_{3,1}(\vecq_{1,2,3})&\equiv\frac{1}{3}\left[-\alpha_{3,12}H_{2,1}(\vecq_{1,2})+ \text{2 perm.}\right], \\
    H_{3,2}(\vecq_{1,2,3})&\equiv\frac{1}{3}\left[2\xi_{12,3}H^{\m C}_{2,1}(\vecq_{1,2})+ \text{2 perm.}\right], \\ 
    H_{3,3}(\vecq_{1,2,3})&\equiv\frac{1}{3}\left[(2\xi_{12,3}+\alpha_{3,12})H_{2,1}(\vecq_{1,2})+\text{2 perm.}\right], \\
    H^{\m C}_{3,1}(\vecq_{1,2,3})&\equiv\frac{1}{3}\left[\alpha_{3,12}H_{2,1}(\vecq_{1,2})+ \text{2 perm.}\right], \\
    H^{\m C}_{3,2}(\vecq_{1,2,3})&\equiv\frac{1}{3}\left[2\alpha^s_{12,3}H^{\m C}_{2,1}(\vecq_{1,2}) + \text{2 perm.} \right], \\
    H^{\m C}_{3,3}(\vecq_{1,2,3})&\equiv\frac{1}{3}\left[\alpha_{12,3}H_{2,1}(\vecq_{1,2})+ \text{2 perm.}\right]. 
    \hspace{\textwidth/2}
\end{align*}

\paragraph{Relations:} At $n=3$ there is one relation, see Eq.\,\eqref{eq:H3rel},
\be
  0=H_{3,1}(\veck_1,\veck_2,\veck_3)+H^{\m C}_{3,1}(\veck_1,\veck_2,\veck_3)\,.
\ee

\subsection{Fourth order}

\paragraph{Definitions:}

\begin{align*}
\setlength{\mathindent}{0pt}
    &H_{4,1-3}(\vecq_{1,2,3,4})\equiv \frac{1}{4}\left[-\alpha_{4,123}H_{3,1-3}(\vecq_{1,2,3})+\text{3 perm.}\right]\\
    &H_{4,4-6}(\vecq_{1,2,3,4})\equiv \frac{1}{4}\left[2\xi_{4,123}H^{\m C}_{3,1-3}(\vecq_{1,2,3})+\text{3 perm.}\right]\\
    &H_{4,7-9}(\vecq_{1,2,3,4})\equiv \frac{1}{4}\left[(2\xi_{4,123}+\alpha_{4,123})H_{3,1-3}(\vecq_{1,2,3})+\text{3 perm.}\right]\\
    &H_{4,10}(\vecq_{1,2,3,4})\equiv \frac{1}{6}\left[-\alpha_{12,34}\alpha^s_{1,2}\xi_{3,4} +\text{5 perm.}\right] \\
    &H_{4,11}(\vecq_{1,2,3,4})\equiv \frac{1}{3}\left[-\alpha^s_{12,34}\xi_{1,2}\xi_{3,4} +\text{2 perm.}\right]\\
    &H_{4,12}(\vecq_{1,2,3,4})\equiv \frac{1}{6}\left[(2\xi_{12,34}+\alpha_{34,12})\xi_{1,2}\alpha^s_{3,4} +\text{5 perm.}\right] \\
    &H_{4,13}(\vecq_{1,2,3,4})\equiv \frac{1}{3}\left[\xi_{12,34}\alpha^s_{1,2}\alpha^s_{3,4} +\text{2 perm.} \right]  \\
    &H_{4,14}(\vecq_{1,2,3,4})\equiv \frac{1}{3}\left[(\xi_{12,34}+\alpha^s_{12,34})\xi_{1,2}\xi_{3,4} +\text{2 perm.} \right]\\
    &H^{\m C}_{4,1-3}(\vecq_{1,2,3,4})\equiv \frac{1}{4}\left[\alpha_{4,123}H_{3,1-3}(\vecq_{1,2,3})+\text{3 perm.}\right]\\
    &H^{\m C}_{4,4-6}(\vecq_{1,2,3,4})\equiv \frac{1}{4}\left[2\alpha^s_{4,123}H^{\m C}_{3,1-3}(\vecq_{1,2,3})+\text{3 perm.}\right]\\
    &H^{\m C}_{4,7-9}(\vecq_{1,2,3,4})\equiv \frac{1}{4}\left[\alpha_{123,4}H_{3,1-3}(\vecq_{1,2,3})+\text{3 perm.}\right]\\
    &H^{\m C}_{4,10}(\vecq_{1,2,3,4})\equiv \frac{1}{6}\left[\alpha_{12,34}\alpha^s_{1,2}\xi_{3,4} +\text{5 perm.}\right]\\
    &H^{\m C}_{4,11}(\vecq_{1,2,3,4})\equiv \frac{1}{3}\left[\alpha^s_{12,34}\xi_{1,2}\xi_{3,4} +\text{2 perm.}\right]\\
    &H^{\m C}_{4,12}(\vecq_{1,2,3,4})\equiv \frac{1}{6}\left[\alpha_{12,34}\xi_{1,2}\alpha^s_{3,4} +\text{5 perm.}\right]\\
    &H^{\m C}_{4,13}(\vecq_{1,2,3,4})\equiv  \frac{1}{3}\left[\alpha^s_{12,34}\alpha^s_{1,2}\alpha^s_{3,4} +\text{2 perm.} \right]   \hspace{\textwidth/2} 
\end{align*}

\paragraph{Relations:} At $n=4$ there are seven relations, see Eq.\,\eqref{eq:H4rel},
\begin{align}
    0 &= H^{\m C}_{4,1-3}+H_{4,1-3},  && 0=H^{\m C}_{4,10-11}+H_{4,10-11},\nn\\
    0 &= H_{4,1} + H_{4,4} + H_{4,7}, && 0 = H^{\m C}_{4,1} + H^{\m C}_{4,4} + H^{\m C}_{4,7}.
\end{align}

\subsection{Fifth order}

\paragraph{Definitions:}

\begin{equation*}
    \begin{alignedat}{2}
        &H_{5,1-14}\qfive &&\equiv\frac{1}{5}\left[-\alpha_{5,1234}H_{4,1-14}(\vecq_{1,2,3,4}) +\text{ 4 perm.}\right]\\
        &H_{5,15-27}\qfive&&\equiv\frac{1}{5}\left[2\xi_{1,2345}H^{\m C}_{4,1-13}(\vecq_{1,2,3,4}) +\text{ 4 perm.}\right]\\
        &H_{5,28-41}\qfive&&\equiv\frac{1}{5}\left[(2\beta_{1234,5}-\alpha_{1234,5})H_{4,1-14}(\vecq_{1,2,3,4}) +\text{ 4 perm.} \right]\\
        &H_{5,42-44}\qfive&&\equiv\frac{1}{10}\left[-\alpha_{345,12}H^{\m C}_{3,1-3}(\vecq_{3,4,5})H_{2,1}(\vecq_{1,2})+\text{ 9 perm.} \right]\\
        &H_{5,45-47}\qfive&&\equiv\frac{1}{10}\left[-\alpha_{345,12}H_{3,1-3}(\vecq_{3,4,5})H_{2,1}(\vecq_{1,2})+\text{ 9 perm.} \right]\\
        &H_{5,48-50}\qfive&&\equiv\frac{1}{10}\left[-\alpha_{12,345}H_{3,1-3}(\vecq_{3,4,5})H^{\m C}_{2,1}(\vecq_{1,2})+\text{ 9 perm.} \right]\\
        &H_{5,51-53}\qfive&&\equiv\frac{1}{10}\left[(2\beta_{12,345}-\alpha_{345,12})H_{3,1-3}(\vecq_{3,4,5})H^{\m C}_{2,1}(\vecq_{1,2})+\text{ 9 perm.} \right]\\
        &H_{5,54-56}\qfive&&\equiv\frac{1}{10}\left[-\alpha_{12,345}H_{3,1-3}(\vecq_{3,4,5})H_{2,1}(\vecq_{1,2})+\text{ 9 perm.} \right]\\
        &H_{5,57-59}\qfive&&\equiv\frac{1}{10}\left[(2\beta_{12,345}-\alpha_{12,345})H^{\m C}_{3,1-3}(\vecq_{3,4,5})H_{2,1}(\vecq_{1,2})+\text{ 9 perm.} \right]\\
        &H_{5,60-62}\qfive&&\equiv\frac{1}{10}\left[2\xi_{12,345}H^{\m C}_{3,1-3}(\vecq_{3,4,5})H^{\m C}_{2,1}(\vecq_{1,2})+\text{ 9 perm.} \right]\\
        &H_{5,63-65}\qfive&&\equiv\frac{1}{10}\left[2\beta_{12,345}H_{3,1-3}(\vecq_{3,4,5})H_{2,1}(\vecq_{1,2})+\text{ 9 perm.} \right]
        \hspace{5cm}\\
    \end{alignedat}
    \label{h5eq}
\end{equation*}
\begin{equation*}
    \begin{alignedat}{2}
        &H^{\m C}_{5,1-14}\qfive  &&\equiv\frac{1}{5}\left[\alpha_{5,1234}H_{4,1-14}(\vecq_{1,2,3,4}) +\text{ 4 perm.}\right]\\
        &H^{\m C}_{5,15-27}\qfive  &&\equiv\frac{1}{5}\left[2\alpha^s_{1,2345}H^{\m C}_{4,1-13}(\vecq_{1,2,3,4}) +\text{ 4 perm.}\right]\\
        &H^{\m C}_{5,28-41}\qfive &&\equiv\frac{1}{5}\left[\alpha_{1234,5}H_{4,1-14}(\vecq_{1,2,3,4}) +\text{ 4 perm.} \right]\\
        &H^{\m C}_{5,42-44}\qfive&&\equiv\frac{1}{10}\left[\alpha_{345,12}H^{\m C}_{3,1-3}(\vecq_{3,4,5})H_{2,1}(\vecq_{1,2})+\text{ 9 perm.} \right]\\
        &H^{\m C}_{5,45-47}\qfive&&\equiv\frac{1}{10}\left[\alpha_{345,12}H_{3,1-3}(\vecq_{3,4,5})H_{2,1}(\vecq_{1,2})+\text{ 9 perm.} \right]\\
        &H^{\m C}_{5,48-50}\qfive&&\equiv\frac{1}{10}\left[\alpha_{12,345}H_{3,1-3}(\vecq_{3,4,5})H^{\m C}_{2,1}(\vecq_{1,2})+\text{ 9 perm.} \right]\\
        &H^{\m C}_{5,51-53}\qfive&&\equiv\frac{1}{10}\left[\alpha_{345,12}H_{3,1-3}(\vecq_{3,4,5})H^{\m C}_{2,1}(\vecq_{1,2})+\text{ 9 perm.} \right]\\
        &H^{\m C}_{5,54-56}\qfive&&\equiv\frac{1}{10}\left[\alpha_{12,345}H_{3,1-3}(\vecq_{3,4,5})H_{2,1}(\vecq_{1,2})+\text{ 9 perm.} \right]\\
        &H^{\m C}_{5,57-59}\qfive&&\equiv\frac{1}{10}\left[\alpha_{12,345}H^{\m C}_{3,1-3}(\vecq_{3,4,5})H_{2,1}(\vecq_{1,2})+\text{ 9 perm.} \right]\\
        &H^{\m C}_{5,60-62}\qfive&&\equiv\frac{1}{10}\left[2\alpha^s_{12,345}H^{\m C}_{3,1-3}(\vecq_{3,4,5})H^{\m C}_{2,1}(\vecq_{1,2})+\text{ 9 perm.} \right]  \hspace{5cm}
    \end{alignedat}
\end{equation*}

\paragraph{Relations:} 

\begin{align}\label{eq:H5rel}
    0 &= H_{5,1-14} + H^{\m C}_{5,1-14}, && 0 = H_{5,42-50} + H^{\m C}_{5,42-50},\nn\\
    0 &= H_{5,54-56} +H^{\m C}_{5,54-56}, && 0 = H_{5,42} +H_{5,45},\nn\\
    0 &= H^{\m C}_{5,42} + H^{\m C}_{5,45}, && 0 = H^{\m C}_{5,54} + H^{\m C}_{5,57},\nn\\
    0 &= H_{5,1} + H_{5,4} + H_{5,7}, && 0 = H^{\m C}_{5,1} + H^{\m C}_{5,4} + H^{\m C}_{5,7} ,\nn\\
    0 &= H_{5,15} + H_{5,18} + H_{5,21}, && 0= H^{\m C}_{5,15} + H^{\m C}_{5,18} + H^{\m C}_{5,21},\nn\\
    0 &= H_{5,28} + H_{5,31} + H_{5,34}, && 0=H^{\m C}_{5,28} + H^{\m C}_{5,31} + H^{\m C}_{5,34}\,\nn\\
    0 &= H_{5,1-3} + H_{5,28-30} + H_{5,15-17} , && 0=H^{\m C}_{5,1-3} + H^{\m C}_{5,28-30} + H^{\m C}_{5,15-17},\nn\\
    0 &= H_{5,10-11} + H_{5,37-38} + H_{5,24-25} , && 0=H^{\m C}_{5,10-11} + H^{\m C}_{5,37-38} + H^{\m C}_{5,24-25},\nn\\
    0 &= H_{5,48} + H_{5,51} + H_{5,60} , && 0 = H^{\m C}_{5,48} + H^{\m C}_{5,51} + H^{\m C}_{5,60} ,\nn\\
    0 &= H_{5,54} + H_{5,63} + H_{5,57} \,.
\end{align}
These make in total 48 relations. However, two of them are redundant and can be omitted, being in particular ${H^{\m C}_{5,42} = -H^{\m C}_{5,45}}$ and ${H^{\m C}_{5,1} + H^{\m C}_{5,4} + H^{\m C}_{5,7}=0}$. This results in 46 linearly independent relations between $H$-functions at fifth order.

\section{Naive basis: $D$-functions}\label{applindep}

In this appendix, we summarize the definitions of all growth factors $D_{n,i}^X(\eta)$  entering the ``naive basis'' introduced in Sec.\,\ref{sec:naivebasis}, see Eq.\,\eqref{naive_basis} and Eq.\,\eqref{gfintrep}, up to $n\leq 5$. The growth factors with $X=\m{F}$ enter only the density kernels $F_n$, and those with $X=\m{G}$ only the velocity divergence kernels $G_n$.
The growth factors with $X=\m{C}$ enter both into $F_n$ and $G_n$. We also use the notation $D^\Delta_{n,i}\equiv D^{\m G}_{n,i}-D^{\m F}_{n,i}$.

In addition, we collect all (cosmology-independent) relations among the growth factors, as discussed in Sec.\,\ref{sec:Drelations}, and provide proofs if not yet given in the main text.

For convenience, we recall the definitions for $n\geq 2$ from Eq.\,\eqref{gfintrep},
\bea\label{gfintrepAPP}
    D^{\m F}_{n,i}(\eta) &=& \getaint \m{I}_{n,i}(\eta')\,,\nn\\
    D^{\m G}_{n,i}(\eta) &=& \etaint \pgeta \m{I}_{n,i}(\eta') = \partial_\eta D^{\m F}_{n,i}(\eta) \,,\nn\\
    D^{\m C}_{n,i}(\eta) &=& \eetaint \m{I}_{n,i}(\eta')\,,
\eea
with $g(\eta,\eta')$ defined in Eq.\,\eqref{getaetaint}. The growth factors are thus completely specified by the integrands $\m{I}_{n,i}(\eta)$.
Equivalently, they can be obtained from the differential equations Eqs.\,\eqref{eq:ODEforDdeltaandDC} and~\eqref{eq:ODEforDF}.
Ultimately, their cosmology dependence is completely determined by  $x(\eta)=3\Omega_m(\eta)/(2f(\eta)^2)$, which can be an arbitrary function of time.

Even for a general cosmology specified by an arbitrary $x(\eta)$, {\it some} of the
integrands and corresponding growth factors $D^{\m C}_{n,i}(\eta)$ have a trivial time-dependence proportional to $e^{n\eta}$,
and we give them below. 

In the special case, where $x(\eta)\stackrel{\dot x=0}{=} x_0$ is constant in time (see end of Sec.\,\ref{sec:relax}),  {\it all} of the growth factors $D^{ X}_{n,i}(\eta)$ are proportional to
$e^{n\eta}$, see Eq.\,\eqref{eq:constx0}. We give the proportionality constants below for reference, even though this approximation is not used in this work. The well-known EdS limit is recovered for $x_0\stackrel{\mbox{\scriptsize EdS}}{=}\frac{3}{2}$. For a matter-dominated universe, in which a fraction $f_\nu\ll 1$ of the total matter does not cluster, $x_0= \frac32(1+f_\nu/5)$. All relations in this work are valid for general time-dependent $x(\eta)$, unless explicitly referring to the EdS limit or the limit of constant $x$, respectively. 

\medskip

At $n=1$, there is only a single growth factor, which in our parameterization is by definition given by $D_{1}(\eta)\equiv e^\eta$.

\subsection{Second order} 

\paragraph{Definitions:} There is one integrand at $n=2$, giving three growth functions via Eq.\,\eqref{gfintrepAPP},

\be
  \m{I}_{2,1}(\eta)=(D_1(\eta))^2=e^{2\eta}\,.
\ee

\paragraph{Growth factors with trivial time-dependence (for any $x(\eta)$):}

The integrand $\m{I}_{2,1}$ has a trivial time-dependence, leading to
\be\label{eq:secondordertrivial}
  D^{\m C}_{2,1}(\eta) = e^{2\eta}\,.
\ee

\paragraph{Limit of constant $x(\eta)=x_0$:}

\be
  D^{\m F}_{2,1}\stackrel{\dot x=0}{=} \frac{1}{2+x_0} e^{2\eta},\qquad
  D^{\m G}_{2,1}\stackrel{\dot x=0}{=} \frac{2}{2+x_0} e^{2\eta},\qquad
  D^{\m C}_{2,1}\stackrel{\dot x=0}{=} D^{\m C}_{2,1} = e^{2\eta}\,.
\ee
The EdS limit corresponds to the special case $x_0=3/2$.

\paragraph{Relations (for any $x(\eta)$):}
 None at $n=2$.

\subsection{Third order} \label{o3calc}

\paragraph{Definitions:} There are three integrands, see Eq.\,\eqref{eq:In3}, giving nine growth functions via Eq.\,\eqref{gfintrepAPP},

\be
    \m{I}_{3,1}(\eta)= {D}_1(\eta) D^{\m F}_{2,1}(\eta),\qquad \m{I}_{3,2}(\eta)= {D}_1(\eta) D^{\m C}_{2,1}(\eta),\qquad
    \m{I}_{3,3}(\eta) = {D}_1(\eta) D^{\m G}_{2,1}(\eta)\,.
\ee

\paragraph{Growth factors with trivial time-dependence (for any $x(\eta)$):}

The integrand $\m{I}_{3,2}(\eta)=e^{3\eta}$ has trivial time-dependence, leading to
\be\label{eq:thirdordertrivial}
  {D}_{3,2}^{\m C}(\eta)=\frac12 e^{3\eta}\,.
\ee

\paragraph{Limit of constant $x(\eta)=x_0$:}

\be
    \m{I}_{3,1}(\eta)\stackrel{\dot x=0}{=} \frac{1}{2+x_0} e^{3\eta},\qquad \m{I}_{3,2}(\eta)\stackrel{\dot x=0}{=} e^{3\eta},\qquad
    \m{I}_{3,3}(\eta) \stackrel{\dot x=0}{=} \frac{2}{2+x_0} e^{3\eta}\,,
\ee
\bea
D_{3,i}^{\m F}\stackrel{\dot x=0}{=}\frac12\frac{1}{3+x_0}\m{I}_{3,i},\quad
D_{3,i}^{\m C}\stackrel{\dot x=0}{=}\frac12\m{I}_{3,i},\quad
D_{3,i}^{\m G}\stackrel{\dot x=0}{=}\frac32\frac{1}{3+x_0}\m{I}_{3,i}\,.
\eea

\paragraph{Relations (for any $x(\eta)$):}

Relations involving only growth factors with $n=3$ (see Eq.\,\ref{eq:D3relation}):
\be
  0 = {D}^{\m F}_{3,3}(\eta)-{D}^{\m F}_{3,1}(\eta) - {D}^{\m C}_{3,3}(\eta) + {D}^{\m C}_{3,1}(\eta) + {D}^{\m F}_{3,2}(\eta)\,.
\ee
Relations involving growth factors $n\leq 3$ (see Eqs.\,\ref{eq:D3D2rel1}, \ref{eq:D3D2rel2}):
\bea
 0 &=& {D}^{\Delta}_{3,3}(\eta)-{D}^{\Delta}_{3,1}(\eta)-D_1(\eta){D}^{\Delta}_{2,1}(\eta)+{D}^{\Delta}_{3,2}(\eta)\,, \\
 0 &=& {D}^{\m C}_{3,3}(\eta) - D_1(\eta){D}^{\m F}_{2,1}(\eta)\,. 
\eea
These relations are all derived in the main text.

\subsection{Fourth order}

\paragraph{Definitions:} There are 14 integrands, see Eq.\,\eqref{eq:I4idef},

\begin{alignat}{8}
    &\m{I}_{4,1-3} &&= {D}_1 {D}^{\m F}_{3,1-3}, &\qquad&
    \m{I}_{4,4-6} &&= {D}_1 {D}^{\m C}_{3,1-3}, &\qquad&
    \m{I}_{4,7-9} &&= {D}_1 {D}^{\m G}_{3,1-3}, &\qquad&
    \m{I}_{4,10} &&={D}^{\m F}_{2,1} {D}^{\m C}_{2,1}, \nonumber\\
    &\m{I}_{4,11} &&={D}^{\m F}_{2,1} {D}^{\m G}_{2,1}, &\qquad&\m{I}_{4,12} &&={D}^{\m G}_{2,1} {D}^{\m C}_{2,1}, &\qquad&
    \m{I}_{4,13} &&={D}^{\m C}_{2,1} {D}^{\m C}_{2,1}, &\qquad&
    \m{I}_{4,14} &&={D}^{\m G}_{2,1}{D}^{\m G}_{2,1} \,. 
\end{alignat}
This yields $3\times 14$ growth factors $D^X_{4,i}$ via Eq.\,\eqref{gfintrepAPP}.
However, we find that for $X=\m{C}$ only the terms with $i\leq 13$ contribute to the fourth order kernels.

\paragraph{Growth factors with trivial time-dependence (for any $x(\eta)$):}

There are two integrands with trivial time-dependence, $\m{I}_{4,5}(\eta)=\frac12 e^{4\eta}$ and $\m{I}_{4,13}(\eta)= e^{4\eta}$, giving
\be\label{eq:fourthordertrivial}
  {D}_{4,5}^{\m C}(\eta)=\frac16 e^{4\eta},\qquad {D}_{4,13}^{\m C}(\eta)=\frac13 e^{4\eta}\,.
\ee

\paragraph{Limit of constant $x(\eta)=x_0$:}

\be
\begin{array}{lllll}
    &\m{I}_{4,1} \stackrel{\dot x=0}{=} \frac12\frac{1}{3+x_0}\frac{1}{2+x_0}e^{4\eta} , 
    &\m{I}_{4,2} \stackrel{\dot x=0}{=} \frac12\frac{1}{3+x_0}e^{4\eta} , 
    &\m{I}_{4,3} \stackrel{\dot x=0}{=} \frac12\frac{1}{3+x_0}\frac{2}{2+x_0}e^{4\eta} , \\
    &\m{I}_{4,4} \stackrel{\dot x=0}{=} \frac12\frac{1}{2+x_0}e^{4\eta} , 
    &\m{I}_{4,5} \stackrel{\dot x=0}{=} \frac12e^{4\eta} , 
    &\m{I}_{4,6} \stackrel{\dot x=0}{=} \frac12\frac{2}{2+x_0}e^{4\eta} , \\
    &\m{I}_{4,7} \stackrel{\dot x=0}{=} \frac32\frac{1}{3+x_0}\frac{1}{2+x_0}e^{4\eta} , 
    &\m{I}_{4,8} \stackrel{\dot x=0}{=} \frac32\frac{1}{3+x_0}e^{4\eta} , 
    &\m{I}_{4,9} \stackrel{\dot x=0}{=} \frac32\frac{1}{3+x_0}\frac{2}{2+x_0}e^{4\eta} , \\
    &\m{I}_{4,10} \stackrel{\dot x=0}{=} \frac{1}{2+x_0}e^{4\eta} ,  
    &\m{I}_{4,11} \stackrel{\dot x=0}{=} \frac{1}{2+x_0}\frac{2}{2+x_0} e^{4\eta} , 
    &\m{I}_{4,12} \stackrel{\dot x=0}{=}\frac{2}{2+x_0}e^{4\eta} ,  \\
    &\m{I}_{4,13} \stackrel{\dot x=0}{=}e^{4\eta} , 
    &\m{I}_{4,14} \stackrel{\dot x=0}{=}\frac{2}{2+x_0}\frac{2}{2+x_0}e^{4\eta} \,, 
\end{array}
\ee
\bea\label{eq:D4constantx}
D_{4,i}^{\m F}\stackrel{\dot x=0}{=}\frac13\frac{1}{4+x_0}\m{I}_{4,i},\quad
D_{4,i}^{\m C}\stackrel{\dot x=0}{=}\frac13\m{I}_{4,i},\quad
D_{4,i}^{\m G}\stackrel{\dot x=0}{=}\frac43\frac{1}{4+x_0}\m{I}_{4,i}\,.
\eea

\paragraph{Relations (for any $x(\eta)$):}

Relations involving only growth factors with $n=4$ (see Eqs.\,\ref{eq:D4relextra1}--\ref{eq:D4rel5}):
\bea
 0 &=& {D}^{\m F}_{4,1}  - {D}^{\m F}_{4,7} -D^{\m F}_{4,10} -{D}^{\m C}_{4,1}  + {D}^{\m C}_{4,7}\,, \label{eq:D4relextra1APP} \\
 0 &=& {D}^{\m F}_{4,2}  - {D}^{\m F}_{4,8} -D^{\m F}_{4,13} -{D}^{\m C}_{4,2}  + {D}^{\m C}_{4,8}\,, \label{eq:D4relextra2APP} \\
 0 &=& {D}^{\m F}_{4,3}  - {D}^{\m F}_{4,9} -D^{\m F}_{4,12} -{D}^{\m C}_{4,3}  + {D}^{\m C}_{4,9}\,, \label{eq:D4relextra3APP} \\
 0 &=& D^X_{4,1}-D^X_{4,2}-D^X_{4,3}-D^X_{4,4}+D^X_{4,6}\,, \\
 0 &=& D^X_{4,1}-D^X_{4,2}-D^X_{4,3}-D^X_{4,7}+D^X_{4,8}+D^X_{4,9}+D^X_{4,10}-D^X_{4,12}\,, \\
 0 &=& D^X_{4,6}-D^X_{4,10}\,, \\
 0 &=& 2D^X_{4,5}-D^X_{4,13}\,. 
\eea

Relations involving growth factors with $n\leq 4$:
\bea
  0 &=& D^{\m C}_{4,10}(\eta)+D^{\m C}_{4,12}(\eta)-D_1(\eta)^2D_{2,1}^{\m F}(\eta)\,, \label{eq:D4D3D2rel1} \\
  0 &=& D^{\m C}_{4,4}(\eta)+D^{\m C}_{4,10}(\eta)-D_1(\eta)D_{3,1}^{\m C}(\eta)\,, \label{eq:D4D3D2rel2} \\
  0 &=& D^{\m C}_{4,6}(\eta)+D^{\m C}_{4,12}(\eta)-D_1(\eta)D_{3,3}^{\m C}(\eta)\,, \label{eq:D4D3D2rel3} \\
  0 &=& D^\Delta_{4,10}-D^\Delta_{4,11}+D^\Delta_{4,14}-D_{2,1}^{\m F}D_{2,1}^\Delta\,, \label{eq:D4D3D2rel5} \\
  0 &=& {D}^{\m G}_{4,1}  - {D}^{\m G}_{4,7} -D^{\m G}_{4,10} -{D}^{\m C}_{4,1}  + {D}^{\m C}_{4,7} - D_1D_{3,1}^{\m F} + D_1D_{3,1}^{\m G} \,, \label{eq:D4D3D2relextra1} \\
  0 &=& {D}^{\m G}_{4,2}  - {D}^{\m G}_{4,8} -D^{\m G}_{4,13} -{D}^{\m C}_{4,2}  + {D}^{\m C}_{4,8} - D_1D_{3,2}^{\m F} + D_1D_{3,2}^{\m G} \,, \label{eq:D4D3D2relextra2} \\
  0 &=& {D}^{\m G}_{4,3}  - {D}^{\m G}_{4,9} -D^{\m G}_{4,12} -{D}^{\m C}_{4,3}  + {D}^{\m C}_{4,9} - D_1D_{3,3}^{\m F} + D_1D_{3,3}^{\m G} \,. \label{eq:D4D3D2relextra3} 
\eea

The $n=4$ relations Eqs.\,\eqref{eq:D4rel2}--\eqref{eq:D4rel5} are shown in the main text.
To derive Eqs.\,\eqref{eq:D4relextra1APP}--\eqref{eq:D4relextra3APP} we compute
\bea
{D}^{\m F}_{4,7-9} - {D}^{\m F}_{4,1-3} &=& \int_{-\infty}^\eta d\eta' g(\eta,\eta')D_1(\eta')D_{3,1-3}^\Delta(\eta') 
= - \int_{-\infty}^\eta d\eta' D_1(\eta') \partial_{\eta'} \left[g(\eta,\eta')D_{3,1-3}^\Delta(\eta')\right] \nn\\
&=& \int_{-\infty}^\eta d\eta' D_1(\eta')  \left[e^{\eta-\eta'}D_{3,1-3}^\Delta(\eta')-g(\eta,\eta')\m{I}_{3,1-3}(\eta')\right]\,.
\eea
In the first step, we used $D_1(\eta')=e^{\eta'}=\partial_{\eta'}e^{\eta'}$ and partial integration, in addition to Eq.\,\eqref{eq:gboundary}, and in the second step
Eq.\,\eqref{eq:detaprimeg} as well as Eq.\,\eqref{eq:ODEforDdeltaandDC}. Inserting the explicit expressions for $\m{I}_{3,i}$ from  Eq.\,\eqref{eq:In3},
using Eq.\,\eqref{eq:secondordertrivial} and the definitions of the fourth order integrands  Eq.\,\eqref{eq:I4idef}, yields the three relations Eqs.\,\eqref{eq:D4relextra1APP}--\eqref{eq:D4relextra3APP}.

Let us now discuss the proofs of the equations involving both fourth and lower order growth factors.
Eq.\,\eqref{eq:D4D3D2rel1} follows from noting $D^{\m C}_{4,10}(\eta)=\int_{-\infty}^\eta e^{\eta+\eta'}D_{2,1}^{\m F}(\eta')$, writing $e^{\eta'}=\partial_{\eta'}e^{\eta'}$
and using partial integration as well as Eq.\,\eqref{eq:secondordertrivial} and $\partial_\eta D_{2,1}^{\m F}(\eta)=D_{2,1}^{\m G}(\eta)$.
To derive Eq.\,\eqref{eq:D4D3D2rel2}, we use
\be
  D_{4,4}^{\m C}(\eta) = \int_{-\infty}^{\eta} d\eta' \,  e^{\eta-\eta'} D_1(\eta')D_{3,1}^{\m C}(\eta') = e^\eta \int_{-\infty}^{\eta} d\eta' \, D_{3,1}^{\m C}(\eta') 
  = e^\eta \int_{-\infty}^{\eta} d\eta' e^{\eta'} \int_{-\infty}^{\eta'} d\eta'' \, D_{2,1}^{\m F}(\eta'')\,.
\ee
Now, we use again $e^{\eta'}=\partial_{\eta'}e^{\eta'}$ and partial integration as well as Eq.\,\eqref{eq:secondordertrivial}.
Eq.\,\eqref{eq:D4D3D2rel3} can be derived in a very similar way.
Eq.\,\eqref{eq:D4D3D2rel5} can be derived by considering
\bea
  D_{4,14}^{\m F}-D_{4,11}^{\m F} &=& \int_{-\infty}^\eta d\eta'\, g(\eta,\eta')(\m{I}_{4,14}(\eta')-\m{I}_{4,11}(\eta'))
  = \int_{-\infty}^\eta d\eta'\, g(\eta,\eta')D_{2,1}^{\m G}(\eta')D_{2,1}^\Delta(\eta')\nn\\
  &=& -\int_{-\infty}^\eta d\eta'\, D_{2,1}^{\m F}(\eta') \partial_{\eta'}\left[g(\eta,\eta')D_{2,1}^\Delta(\eta')\right]\nn\\
  &=& \int_{-\infty}^\eta d\eta'\, D_{2,1}^{\m F}(\eta') \left[e^{\eta-\eta'}D_{2,1}^\Delta(\eta')-g(\eta,\eta')\m{I}_{2,1}(\eta')\right]\nn\\
  &=& D_{4,11}^{\m C} - \int_{-\infty}^\eta d\eta'\, e^{\eta-\eta'} \left[ D_{2,1}^{\m F}(\eta') \right]^2 - D_{4,10}^{\m F} \,.
\eea
Here, we used Eq.\,\eqref{eq:I4idef} for the second line, $D_{2,1}^{\m G}(\eta')=\partial_{\eta'} D_{2,1}^{\m F}(\eta')$ and partial integration for the third line,
Eq.\,\eqref{eq:detaprimeg} as well as Eq.\,\eqref{eq:ODEforDdeltaandDC} for the fourth line, and Eqs.\,\eqref{eq:I4idef} and\,\eqref{eq:secondordertrivial}
for the last line. The resulting equation would yield another fourth order relation, if we had defined a growth factor with integrand $\left[ D_{2,1}^{\m F}(\eta) \right]^2$.
However, this integrand does not appear, since there is no non-linear vertex connecting two density fields. Nevertheless, applying $(\partial_\eta-1)$ to this relation
and using Eq.\,\eqref{eq:ODEforDdeltaandDC}, $D_{n,i}^{\m G}(\eta)=\partial_{\eta} D_{n,i}^{\m F}(\eta)$ and Eq.\,\eqref{eq:gboundary}, yields the relation
Eq.\,\eqref{eq:D4D3D2rel5} involving fourth and second order growth factors.
Taking a $\partial_\eta$ derivative of Eqs.\,\eqref{eq:D4relextra1APP}--\eqref{eq:D4relextra3APP} yields Eqs.\,\eqref{eq:D4D3D2relextra1}--\eqref{eq:D4D3D2relextra3}.

\subsection{Fifth order}

\paragraph{Definitions:} 

For $n=5$ we find 65 integrands, that we enumerate as follows:
\begin{alignat}{8}
    &\m{I}_{5,1-14}&&= {D}_1 {D}^{\m F}_{4,1-14}, &\quad&
    \m{I}_{5,15-27}&&= {D}_1 {D}^{\m C}_{4,1-13}, &\quad&
    \m{I}_{5,28-41}&&= {D}_1 {D}^{\m G}_{4,1-14}, &\quad&
    \m{I}_{5,42-44}&&={D}^{\m F}_{2,1} {D}^{\m C}_{3,1-3}, \nn\\
    &\m{I}_{5,45-47}&&={D}^{\m F}_{2,1} {D}^{\m G}_{3,1-3}, &\quad& 
    \m{I}_{5,48-50}&&={D}^{\m C}_{2,1} {D}^{\m F}_{3,1-3}, &\quad&
    \m{I}_{5,51-53}&&={D}^{\m C}_{2,1} {D}^{\m G}_{3,1-3}, &\quad&
    \m{I}_{5,54-56}&&={D}^{\m G}_{2,1} {D}^{\m F}_{3,1-3},  \nn\\
    &\m{I}_{5,57-59}&&={D}^{\m G}_{2,1} {D}^{\m C}_{3,1-3}, &\quad& 
    \m{I}_{5,60-62}&&={D}^{\m C}_{2,1} {D}^{\m C}_{3,1-3}, &\quad&
    \m{I}_{5,63-65}&&={D}^{\m G}_{2,1} {D}^{\m G}_{3,1-3}.  \label{eq:I5idef}
\end{alignat}
This yields $3\times 65$ growth factors $D^X_{5,i}$ via Eq.\,\eqref{gfintrepAPP}.
However, we find that for $X=\m{C}$ only the terms with $i\leq 62$ contribute to the fifth order kernels. 
This leads to $65+62=127$ terms within the naive basis contributing to both $F_5$ as well as $G_5$, respectively.

\paragraph{Growth factors with trivial time-dependence (for any $x(\eta)$):}

Here $\m{I}_{5,19}(\eta)=\frac16 e^{5\eta}$, $\m{I}_{5,27}(\eta)=\frac13 e^{5\eta}$ and $\m{I}_{5,61}(\eta)=\frac12 e^{5\eta}$ are trivial. This implies,
\be\label{eq:fifthordertrivial}
 {D}_{5,19}^{\m C}(\eta)=\frac1{24} e^{5\eta},\quad {D}^{\m C}_{5,27}(\eta)=\frac1{12} e^{5\eta},\quad {D}^{\m C}_{5,61}(\eta)=\frac18 e^{5\eta}\,,
\ee
are trivial.

\paragraph{Limit of constant $x(\eta)=x_0$:} Follow from Eqs.\,\eqref{eq:D4constantx}, \eqref{eq:I5idef} and
\bea\label{eq:D5constantx}
D_{5,i}^{\m F}\stackrel{\dot x=0}{=}\frac14\frac{1}{5+x_0}\m{I}_{5,i},\quad
D_{5,i}^{\m C}\stackrel{\dot x=0}{=}\frac14\m{I}_{5,i},\quad
D_{5,i}^{\m G}\stackrel{\dot x=0}{=}\frac54\frac{1}{5+x_0}\m{I}_{5,i}\,.
\eea

\paragraph{Relations (for any $x(\eta)$):} 

\bea
  0 &=& \gffive{2-3}^{\m F}-\gffive{29-30}^{\m F}-\gffive{49-50}^{\m F}+\gffive{29-30}^{\m C}-\gffive{2-3}^{\m C}, \label{eq:D5rel1}\\
  0 &=& \gffive{4-6}^{\m F}-\gffive{31-33}^{\m F}-\gffive{60-62}^{\m F}+\gffive{31-33}^{\m C}-\gffive{4-6}^{\m C}, \label{eq:D5rel2}\\
  0 &=& \gffive{7-9}^{\m F}-\gffive{34-36}^{\m F}-\gffive{51-53}^{\m F}+\gffive{34-36}^{\m C}-\gffive{7-9}^{\m C}, \label{eq:D5rel3}\\
  0 &=&\gffive{42}^{\m F}+\gffive{44}^{\m F}-\gffive{57}^{\m F}-\gffive{59}^{\m F}-\gffive{60}^{\m F} -\gffive{42}^{\m C} + \gffive{57}^{\m C}, \label{eq:D5rel4}\\ 
  0 &=&\gffive{44}^{\m F}+\gffive{57}^{\m F}-\gffive{62}^{\m F}+\gffive{63}^{\m F}-\gffive{64}^{\m F}-\gffive{65}^{\m F}-\gffive{44}^{\m C} + \gffive{59}^{\m C}\,. \label{eq:D5rel5}
\eea

\bea
 0 &=& D^X_{5,48}-D^X_{5,49}-D^X_{5,50}-D^X_{5,60}+D^X_{5,62}, \label{eq:IntegrandD5relFromD3rel1} \\
 0 &=& D^X_{5,54}-D^X_{5,55}-D^X_{5,56}-D^X_{5,57}+D^X_{5,59}, \label{eq:IntegrandD5relFromD3rel2} \\
 0 &=& D^X_{5,42}-D^X_{5,45}+D^X_{5,46}+D^X_{5,47}-D^X_{5,59}, \label{eq:IntegrandD5relFromD3D2rel1} \\
 0 &=& 2D^X_{5,43}-D^X_{5,51}+D^X_{5,52}+D^X_{5,53}-2D^X_{5,58}+D^X_{5,60}-D^X_{5,62}, \label{eq:IntegrandD5relFromD3D2rel2} \\
 0 &=& D^X_{5,1}-D^X_{5,2}-D^X_{5,3}-D^X_{5,4}+D^X_{5,6}, \label{eq:IntegrandD5relFromD4rel2} \\
 0 &=& D^X_{5,15}-D^X_{5,16}-D^X_{5,17}-D^X_{5,18}+D^X_{5,20}, \label{eq:IntegrandD5relFromD4rel3} \\
 0 &=& D^X_{5,28}-D^X_{5,29}-D^X_{5,30}-D^X_{5,31}+D^X_{5,33}, \label{eq:IntegrandD5relFromD4rel4} \\
 0 &=& D^X_{5,1}-D^X_{5,2}-D^X_{5,3}-D^X_{5,7}+D^X_{5,8}+D^X_{5,9}+D^X_{5,10}-D^X_{5,12}, \label{eq:IntegrandD5relFromD4rel5} \\
 0 &=& D^X_{5,15}-D^X_{5,16}-D^X_{5,17}-D^X_{5,21}+D^X_{5,22}+D^X_{5,23}+D^X_{5,24}-D^X_{5,26}, \label{eq:IntegrandD5relFromD4rel6} \\
 0 &=& D^X_{5,28}-D^X_{5,29}-D^X_{5,30}-D^X_{5,34}+D^X_{5,35}+D^X_{5,36}+D^X_{5,37}-D^X_{5,39}, \label{eq:IntegrandD5relFromD4rel7} \\
 0 &=& D^X_{5,6}-D^X_{5,10}, \label{eq:IntegrandD5relFromD4rel8} \\
 0 &=& D^X_{5,20}-D^X_{5,24}, \label{eq:IntegrandD5relFromD4rel9} \\
 0 &=& D^X_{5,33}-D^X_{5,37}, \label{eq:IntegrandD5relFromD4rel10} \\
 0 &=& 2D^X_{5,5}-D^X_{5,13}, \label{eq:IntegrandD5relFromD4rel11} \\
 0 &=& 2D^X_{5,19}-D^X_{5,27}, \label{eq:IntegrandD5relFromD4rel12} \\
 0 &=& 2D^X_{5,32}-D^X_{5,40}, \label{eq:IntegrandD5relFromD4rel13} \\
 0 &=& D^X_{5,1}-D^X_{5,7}-D^X_{5,10}-D^X_{5,15}+D^X_{5,21}, \label{eq:IntegrandD5relFromD4relextra1} \\
 0 &=& D^X_{5,2}-D^X_{5,8}-D^X_{5,13}-D^X_{5,16}+D^X_{5,22}, \label{eq:IntegrandD5relFromD4relextra2} \\
 0 &=& D^X_{5,3}-D^X_{5,9}-D^X_{5,12}-D^X_{5,17}+D^X_{5,23}, \label{eq:IntegrandD5relFromD4relextra3} \\
 0 &=& D^X_{5,24}+D^X_{5,26}-2D^X_{5,43}, \label{eq:IntegrandD5relFromD4D3D2rel1} \\
 0 &=& D^X_{5,18}+D^X_{5,24}-D^X_{5,60}, \label{eq:IntegrandD5relFromD4D3D2rel2} \\
 0 &=& D^X_{5,20}+D^X_{5,26}-D^X_{5,62}, \label{eq:IntegrandD5relFromD4D3D2rel3} \\
 0 &=& D^X_{5,10}-D^X_{5,11}+D^X_{5,14}-D^X_{5,37}+D^X_{5,38}-D^X_{5,41}-D^X_{5,44}+D^X_{5,59}, \label{eq:IntegrandD5relFromD4D3D2rel5} \\
 0 &=& D^X_{5,1}-D^X_{5,7}-D^X_{5,10}-D^X_{5,28}+D^X_{5,34}+D^X_{5,37}+D^X_{5,48}-D^X_{5,51}, \label{eq:IntegrandD5relFromD4D3D2relextra1} \\
 0 &=& D^X_{5,2}-D^X_{5,8}-D^X_{5,13}-D^X_{5,29}+D^X_{5,35}+D^X_{5,40}+D^X_{5,49}-D^X_{5,52}, \label{eq:IntegrandD5relFromD4D3D2relextra2} \\
 0 &=& D^X_{5,3}-D^X_{5,9}-D^X_{5,12}-D^X_{5,30}+D^X_{5,36}+D^X_{5,39}+D^X_{5,50}-D^X_{5,53}, \label{eq:IntegrandD5relFromD4D3D2relextra3} \\
 0 &=& 3D^X_{5,27}-2D^X_{5,61}\,. \label{eq:IntegrandD5relFromTrivialTimeDep1} 
\eea

Again, relations with a superscript $X$ hold for each choice $X=\m{F},\m{G},\m{C}$.
We find $27$ such relations, i.e. in total $3\times 27$ for $X=\m{F},\m{G},\m{C}$.
Out of these $2\times 27$ contain growth factors entering in $F_5$ (for  $X=\m{F},\m{C}$), and $2\times 27$ entering in $G_5$ (for  $X=\m{F},\m{C}$).
In addition, we find ten linearly independent relations connecting $D_{5,i}^{\m F}$ and $D_{5,i}^{\m C}$. These relations can be used for the simplification of $F_5$.

Let us first derive the relations inherited from corresponding relations of the integrands $\m{I}_{5,i}$, i.e. Eqs.\,\eqref{eq:IntegrandD5relFromD3rel1}--\eqref{eq:IntegrandD5relFromTrivialTimeDep1}  above.
Eqs.\,\eqref{eq:IntegrandD5relFromD3rel1} to~\eqref{eq:IntegrandD5relFromD3rel2} follow directly from the third-order relation Eq.\,\eqref{eq:D3relation}. To derive Eqs.\,\eqref{eq:IntegrandD5relFromD3D2rel1} and~\eqref{eq:IntegrandD5relFromD3D2rel2}, we take the
sum of relations Eqs.\,\eqref{eq:D3relation} and \eqref{eq:D3D2rel1} involving third and second order growth factors, giving
\bea
  0 &=& D^{\m C}_{3,1}-D^{\m C}_{3,3}-D^{\m G}_{3,1}+D^{\m G}_{3,2}+D^{\m G}_{3,3}-D_1D^\Delta_{2,1}\,.
\eea
Multiplying with either $D_{2,1}^{\m F}$ or $D_{2,1}^{\m C}$ and using that $D_1D^\Delta_{2,1}D_{2,1}^{\m F}=D^\Delta_{2,1}D_{3,3}^{\m C}=\m{I}_{5,59}-\m{I}_{5,44}$ and $D_1D^\Delta_{2,1}D_{2,1}^{\m C}=2D^\Delta_{2,1}D_{3,2}^{\m C}=2(\m{I}_{5,58}-\m{I}_{5,43})$, which follows using Eqs.\,\eqref{eq:D3D2rel2}, \eqref{eq:secondordertrivial} and \eqref{eq:thirdordertrivial}, we find
 Eqs.\,\eqref{eq:IntegrandD5relFromD3D2rel1} and~\eqref{eq:IntegrandD5relFromD3D2rel2}.

Eqs.\,\eqref{eq:IntegrandD5relFromD4rel2}, \eqref{eq:IntegrandD5relFromD4rel3}, \eqref{eq:IntegrandD5relFromD4rel4} follow from the fourth order relation Eq.\,\eqref{eq:D4rel2} and
eqs.\,\eqref{eq:IntegrandD5relFromD4rel5}, \eqref{eq:IntegrandD5relFromD4rel6}, \eqref{eq:IntegrandD5relFromD4rel7} follow from the fourth order relation Eq.\,\eqref{d3erelation}.
Furthermore, Eqs.\,\eqref{eq:IntegrandD5relFromD4rel8}, \eqref{eq:IntegrandD5relFromD4rel9}, \eqref{eq:IntegrandD5relFromD4rel10} follow from the fourth order relation Eq.\,\eqref{eq:D4rel4} and
Eqs.\,\eqref{eq:IntegrandD5relFromD4rel11}, \eqref{eq:IntegrandD5relFromD4rel12}, \eqref{eq:IntegrandD5relFromD4rel13} follow from the fourth order relation Eq.\,\eqref{eq:D4rel5}.
Note, that Eq.\,\eqref{eq:IntegrandD5relFromD4rel12} could also be obtained from the trivial time-dependence of the corresponding integrands $\m{I}_{5,19}(\eta)=\frac16 e^{5\eta}$
and $\m{I}_{5,27}(\eta)=\frac13 e^{5\eta}$.
Eqs.\,\eqref{eq:IntegrandD5relFromD4relextra1}, \eqref{eq:IntegrandD5relFromD4relextra2}, \eqref{eq:IntegrandD5relFromD4relextra3} follow from the fourth order relations Eqs.\,\eqref{eq:D4relextra1APP}, \eqref{eq:D4relextra2APP}, \eqref{eq:D4relextra3APP}, respectively.

Eq.\,\eqref{eq:IntegrandD5relFromD4D3D2rel1} and Eq.\,\eqref{eq:IntegrandD5relFromD4D3D2rel2} follow from the relations Eq.\,\eqref{eq:D4D3D2rel1} and Eq.\,\eqref{eq:D4D3D2rel2} involving fourth and second order growth factors as well as Eq.\,\eqref{eq:thirdordertrivial}.
Eq.\,\eqref{eq:IntegrandD5relFromD4D3D2rel3} follows from the relation Eq.\,\eqref{eq:D4D3D2rel3} involving fourth and third order growth factors as well as Eq.\,\eqref{eq:secondordertrivial}.
Eq.\,\eqref{eq:D4rel5} as well as Eq.\,\eqref{eq:secondordertrivial}.
Eq.\,\eqref{eq:IntegrandD5relFromD4D3D2rel5} follows from Eq.\,\eqref{eq:D4D3D2rel5}  involving fourth and second order growth factors together with $D_1D^\Delta_{2,1}D_{2,1}^{\m F}=D^\Delta_{2,1}D_{3,3}^{\m C}=\m{I}_{5,59}-\m{I}_{5,44}$  (see above).
Eqs.\,\eqref{eq:IntegrandD5relFromD4D3D2relextra1}--\eqref{eq:IntegrandD5relFromD4D3D2relextra3} follow from difference of Eqs.\,\eqref{eq:D4relextra1APP}, \eqref{eq:D4relextra2APP}, \eqref{eq:D4relextra3APP} and Eqs.\,\eqref{eq:D4D3D2relextra1}, \eqref{eq:D4D3D2relextra2}, \eqref{eq:D4D3D2relextra3}, respectively.

Eq.\,\eqref{eq:IntegrandD5relFromTrivialTimeDep1} follows from the trivial time-dependence of the corresponding integrands. 

\medskip

Let us now derive the relations Eqs.\,\eqref{eq:D5rel1}--\eqref{eq:D5rel5}. Eqs.\,\eqref{eq:D5rel1}--\eqref{eq:D5rel3} can be derived starting from
\bea
  \gffive{28-41}^{\m F}(\eta) -\gffive{1-14}^{\m F}(\eta) &=& \getaint e^{\eta'}{D}^\Delta_{4,1-14}(\eta') \nn\\
  &=& \int_{-\infty}^\eta d\eta'\, e^{\eta'}\left[ e^{\eta-\eta'}{D}^\Delta_{4,1-14}(\eta') - g(\eta,\eta')\m{I}_{4,1-14}(\eta')\right]\nn\\
  &=& \gffive{28-41}^{\m C}(\eta) -\gffive{1-14}^{\m C}(\eta) - \int_{-\infty}^\eta d\eta'\, g(\eta,\eta')e^{\eta'}\m{I}_{4,1-14}(\eta')\,,\nn\\
\eea
where Eq.\,\eqref{eq:I5idef}, $D_1(\eta')=e^{\eta'}=\partial_{\eta'}e^{\eta'}$ along with partial integration, and Eqs.\,\eqref{eq:detaprimeg} and~\eqref{eq:ODEforDdeltaandDC}
were used. Inserting the definitions Eq.\,\eqref{eq:I4idef} of the fourth order integrands, the product $e^{\eta'}\m{I}_{4,1-14}(\eta')$ can be rewritten into
fifth order integrands using Eq.\,\eqref{eq:secondordertrivial} and Eq.\,\eqref{eq:thirdordertrivial}, except for the cases $i=11,14$. For all remaining cases, this leads to the
relations Eqs.\,\eqref{eq:D5rel1}--\eqref{eq:D5rel3}. Note, that we omitted some equations obtained in this way, which would be linearly dependent.

Eq.\,\eqref{eq:D5rel4} can be derived starting from
\bea\label{eq:D5rel10derivation}
  \gffive{57}^{\m F}(\eta)-\gffive{42}^{\m F}(\eta) &=& \int_{-\infty}^\eta d\eta'\, g(\eta,\eta')D^\Delta_{2,1}(\eta')D^{\m C}_{3,1}(\eta')\nn\\
  &=& \int_{-\infty}^\eta d\eta'\, g(\eta,\eta')D^\Delta_{2,1}(\eta') \int_{-\infty}^{\eta'}d\eta'' e^{\eta'-\eta''} D_1(\eta'')D_{2,1}^{\m F}(\eta'')\nn\\
  &=& -\int_{-\infty}^\eta d\eta'\, e^{\eta'} \partial_{\eta'} \left[ g(\eta,\eta')D^\Delta_{2,1}(\eta') \int_{-\infty}^{\eta'}d\eta'' D_{2,1}^{\m F}(\eta'')\right]\nn\\
  &=& \int_{-\infty}^\eta d\eta'\, e^{\eta'}  \Bigg[\left(e^{\eta-\eta'} D^\Delta_{2,1}(\eta') -g(\eta,\eta')\m{I}_{2,1}(\eta')\right)\int_{-\infty}^{\eta'}d\eta'' D_{2,1}^{\m F}(\eta'') \nn\\
  && {} - g(\eta,\eta')D^\Delta_{2,1}(\eta')D^{\m F}_{2,1}(\eta')\Bigg]\nn\\
  &=& \int_{-\infty}^\eta d\eta'\,  \Bigg[\left(e^{\eta-\eta'} D^\Delta_{2,1}(\eta') -g(\eta,\eta')e^{2\eta'}\right)D^{\m C}_{3,1}(\eta') \nn\\
  && {} - g(\eta,\eta')e^{\eta'}D^\Delta_{2,1}(\eta')D^{\m F}_{2,1}(\eta')\Bigg]\nn\\
  &=& \gffive{57}^{\m C}(\eta)-\gffive{42}^{\m C}(\eta)-\gffive{60}^{\m F}(\eta) - \int_{-\infty}^\eta d\eta'\, g(\eta,\eta')e^{\eta'}D^\Delta_{2,1}(\eta')D^{\m F}_{2,1}(\eta')\,.\nn\\
\eea
The first two lines follow from the definitions Eq.\,\eqref{eq:I5idef} and Eq.\,\eqref{eq:In3} of the growth factors, and the third from $e^{\eta'}=\partial_{\eta'}e^{\eta'}$ and
partial integration. Next, we use Eqs.\,\eqref{eq:detaprimeg} and~\eqref{eq:ODEforDdeltaandDC} and then $\m{I}_{2,1}(\eta')=e^{2\eta'}=D_{2,1}^{\m C}(\eta')$ as well as Eq.\,\eqref{eq:I5idef}
again to obtain the last step. The integrand of the remaining integral can be rewritten using 
$  D_1D^\Delta_{2,1}D_{2,1}^{\m F}=D^\Delta_{2,1}D_{3,3}^{\m C}=\m{I}_{5,59}-\m{I}_{5,44}\,, $ (see above).
Inserting this relation, we finally obtain Eq.\,\eqref{eq:D5rel4}.
Eq.\,\eqref{eq:D5rel5} can be derived with a computation similar to Eq.\,\eqref{eq:D5rel10derivation}, but starting with $\gffive{59}^{\m F}(\eta)-\gffive{44}^{\m F}(\eta)$.
This gives 
\be
  \gffive{59}^{\m F}(\eta)-\gffive{44}^{\m F}(\eta) = \gffive{59}^{\m C}(\eta)-\gffive{44}^{\m C}(\eta)-\gffive{62}^{\m F}(\eta) - \int_{-\infty}^\eta d\eta'\, g(\eta,\eta')e^{\eta'}D^\Delta_{2,1}(\eta')D^{\m G}_{2,1}(\eta')\,.
\ee
Now the integrand of the remaining integral can be rewritten using Eq.\,\eqref{eq:D3D2rel1}, $D_1D^\Delta_{2,1}=D^\Delta_{3,3}-D^\Delta_{3,1}+D^\Delta_{3,2}$. Afterwards, we rewrite the integrand as a linear combination
of $\m{I}_{5,i}$ (with six terms), yielding then a relation of fifth order growth factors. To obtain Eq.\,\eqref{eq:D5rel5}, we in addition combined this resulting equation with Eq.\,\eqref{eq:IntegrandD5relFromD3rel2}.

\section{Reduced basis}\label{redbasis}

In this appendix, we provide explicit expressions for the time-dependent growth functions $d_{n,i}^X(\eta)$ and the wavevector-dependent basis functions $h_{n,i}^X$ within the ``reduced basis'', introduced in Sec.\,\ref{sec:redbasis}, see Eq.\,\eqref{redbasisdef}, for $n\leq 5$.\footnote{The growth and basis functions are also available in {\sc Mathematica} format for all $n\leq 5$ on request.} They are given in terms of linear combinations of the growth functions and basis functions within the ``naive basis'', see Sec.\,\ref{sec:naivebasis} and Eq.\,\eqref{naive_basis}, respectively.

We summarise equations, that hold for superscript $F$- and $G$ with a comma notation, e.g. $d_{n,i}^{F,G}={D}_{n,i}^{{\m F},{\m G}}$ is equivalent to $d_{n,i}^{F}={D}_{n,i}^{\m F}$ and $d_{n,i}^{G}={D}_{n,i}^{\m G}$, and $d_{n,i}^{F,G}={D}_{n,i}^{{\m C}}$ is equivalent to $d_{n,i}^{F}={D}_{n,i}^{\m C}$ and $d_{n,i}^{G}={D}_{n,i}^{\m C}$.

At first order, there is a single basis element for both density and velocity divergence, with $d_{1,1}^F(\eta)=d_{1,1}^G(\eta)\equiv D_1(\eta)=e^\eta$, $h_{1,1}^F(\veck)=h_{1,1}^G(\veck)\equiv H_{1,1}^{\m C}(\veck)=1$.

\subsection{Second order}\label{sec:redbasissecondorder}

At second order, the reduced basis contains two terms for both $F_2$ and $G_2$.
\be
  h_{2,1}^{F,G}= H_{2,1},\qquad h_{2,2}^{F,G}= H^{\m C}_{2,1}\,,
\ee
\be
  d_{2,1}^{F,G}\equiv D_{2,1}^{{\m F},{\m G}},\qquad d_{2,2}^{F,G}\equiv D_{2,1}^{\m C}\,.
\ee

\subsection{Third order}\label{sec:redbasisthirdorder}

At third order, the reduced basis contains four terms for $F_3$ and five for $G_3$.

\begin{alignat*}{2}
    &h^{F,G}_{3,1}&&=H^{\m C}_{3,2}=\frac{1}{3}\left[2\alpha^s_{12,3}\alpha^s_{1,2} + \text{2 perm.} \right],\\
    &h^F_{3,2}&&=H_{3,2}-H^{\m C}_{3,1}=\frac{1}{3}\left[2\xi_{12,3}\alpha^s_{1,2}-\alpha_{3,12}\xi_{1,2}+ 2 \text{ perm.}\right],\\
    &h^F_{3,3}&&=H_{3,3}-H^{\m C}_{3,1}=\frac{1}{3}\left[2\xi_{12,3}\xi_{1,2}+2 \text{ perm.}\right],\\
    &h^F_{3,4}&&=H^{\m C}_{3,3}+H^{\m C}_{3,1}=\frac{1}{3}\left[2\alpha^s_{12,3}\xi_{1,2}+ 2 \text{ perm.}\right],\\
    &h^G_{3,2}&&=H_{3,2}=\frac{1}{3}\left[2\xi_{12,3}\alpha^s_{1,2}+ \text{2 perm.}\right],\\
    &h^G_{3,3}&&=H_{3,3}=\frac{1}{3}\left[(2\xi_{12,3}+\alpha_{3,12})\xi^s_{1,2}+\text{2 perm.}\right],\\
    &h^G_{3,4}&&=H^{\m C}_{3,1}=\frac{1}{3}\left[\alpha_{3,12}\xi^s_{1,2}+ \text{2 perm.}\right],\\
    &h^G_{3,5}&&=H^{\m C}_{3,3}=\frac{1}{3}\left[\alpha_{12,3}\xi^s_{1,2}+ \text{2 perm.}\right].
\end{alignat*}
The mapping from the naive to the reduced bases is the following:
\begin{alignat*}{8}
	&d_{3,1}^{F,G}&&={D}^{\m C}_{3,2}, &\qquad& d^F_{3,2}&&={D}^{\m F}_{3,2}, &\qquad& d^F_{3,3}&&={D}^{\m F}_{3,3}, &\qquad& d^F_{3,4}&&={D}^{\m C}_{3,3},\\
	&d^G_{3,2}&&={D}^{\m G}_{3,2}, &\qquad& d^G_{3,3}&&={D}^{\m G}_{3,3}, &\qquad& d^G_{3,4}&&={D}^{\m C}_{3,1}-{D}^{\m G}_{3,1}, &\qquad& d^G_{3,5}&&={D}^{\m C}_{3,3}.
\end{alignat*}

\subsection{Fourth order}\label{sec:redbasisfourthorder}

At fourth order, the reduced basis contains $11$ terms for $F_4$ and $14$ for $G_4$.
\begin{alignat*}{2}
    &h^{F,G}_{4,1}(\vecq_{1,2,3,4})&&=H_{4,14}
	=\frac{1}{3}\left[(\xi_{12,34}+ \alpha^s_{12,34}) \xi_{1,2} \xi_{3,4}+2 \text{ perm.}\right]\\
    &h^{F,G}_{4,2}(\vecq_{1,2,3,4})&&=H^{\m C}_{4,11}
	=\frac{1}{3}\left[\alpha^s_{12,34}\xi_{1,2} \xi_{3,4}+2 \text{ perm.}\right]\\
    &h^{F,G}_{4,3}(\vecq_{1,2,3,4})&&=\frac{1}{2}H^{\m C}_{4,5}+H^{\m C}_{4,13}
	=\frac{1}{12}\left[(2\alpha^s_{123,4} \alpha^s_{12,3} +\alpha^s_{12,34}\alpha^s_{3,4}) \alpha^s_{1,2}+11 \text{ perm.}\right]\\
    &h^F_{4,4}(\vecq_{1,2,3,4})&&=H_{4,7}+H_{4,8}-H^{\m C}_{4,2}+H^{\m C}_{4,4}+H^{\m C}_{4,7}
    =\frac{1}{2}\left[\xi_{123,4}h^F_{3,2}(\vecq_{1,2,3})+\text{3 perm.}\right]\\
    &h^F_{4,5}(\vecq_{1,2,3,4})&&=H_{4,7}+H_{4,9}-H^{\m C}_{4,3}+H^{\m C}_{4,4}+H^{\m C}_{4,7}
    =\frac{1}{2}\left[\xi_{123,4}h^F_{3,3}(\vecq_{1,2,3})+\text{3 perm.}\right]\\
    &h^F_{4,6}(\vecq_{1,2,3,4})&&=H_{4,6}-H_{4,7}+H_{4,12}-H^{\m C}_{4,3}-H^{\m C}_{4,10}
    \\&&&=\frac{1}{12}\left[2(2\xi_{123,4}\alpha^s_{12,3}+\xi_{12,34}\alpha^s_{3,4})\xi_{1,2}-\alpha_{4,123}h^F_{3,3}(\vecq_{1,2,3})+\text{11 perm.}\right]\\
    &h^F_{4,7}(\vecq_{1,2,3,4})&&=\frac{1}{2}(H_{4,5}+H_{4,6}+2 H_{4,13}-2 H^{\m C}_{4,2}+H^{\m C}_{4,4}+H^{\m C}_{4,7}-H^{\m C}_{4,10})
    \\&&&=\frac{1}{24}\left[(2\xi_{123,4}\alpha_{12,3}-\alpha_{4,123}\alpha_{3,12}-\alpha_{34,12}\alpha^s_{3,4})\xi_{1,2} + 2(2\xi_{123,4}\alpha^s_{12,3}+\xi_{12,34}\alpha^s_{3,4})\alpha^s_{1,2}
    \right.\\&&&\qquad \left.-2\alpha_{4,123}h^F_{3,2}(\vecq_{1,2,3})+\text{11 perm.}\right]\\
    &h^F_{4,8}(\vecq_{1,2,3,4})&&=H^{\m C}_{4,2}-H^{\m C}_{4,4}+H^{\m C}_{4,8}=\frac{1}{2}\left[\alpha^s_{123,4}h^F_{3,2}(\vecq_{1,2,3})+\text{3 perm.}\right]\\
    &h^F_{4,9}(\vecq_{1,2,3,4})&&=H^{\m C}_{4,3}-H^{\m C}_{4,4}+H^{\m C}_{4,9}=\frac{1}{2}\left[\alpha^s_{123,4}h^F_{3,3}(\vecq_{1,2,3})+\text{3 perm.}\right]\\
    &h^F_{4,10}(\vecq_{1,2,3,4})&&=\frac{1}{2}(H_{4,6}+H^{\m C}_{4,4}+2 H^{\m C}_{4,6}+H^{\m C}_{4,7}+H^{\m C}_{4,10})
    \\&&&=\frac{1}{24}\left[(2(\xi_{123,4}+2\alpha^s_{123,4})\alpha_{12,3} + \alpha_{4,123}\alpha_{3,12}+\alpha_{34,12}\alpha^s_{3,4})\xi_{1,2}+\text{11 perm.}\right]\\
    &h^F_{4,11}(\vecq_{1,2,3,4})&&=\frac{1}{2}(-H_{4,6}+H^{\m C}_{4,4}-H^{\m C}_{4,7}+H^{\m C}_{4,10})+H^{\m C}_{4,12}
    \\&&&=\frac{1}{24}\left[((2\alpha^s_{123,4}+\alpha_{123,4})\alpha_{3,12}-2\xi_{123,4}\alpha_{12,3}+(2\alpha^s_{12,34}+\alpha_{12,34})\alpha^s_{3,4})\xi_{1,2}+\text{11 perm.}\right]\\
	&h^{G}_{4,4}(\vecq_{1,2,3,4})&&=H_{4,7}+H_{4,8}
	=\frac{1}{4}\left[(2\xi_{123,4} + \alpha_{4,123})h^F_{3,2}(\vecq_{1,2,3})+3 \text{ perm.}\right]\\
	&h^{G}_{4,5}(\vecq_{1,2,3,4})&&=H_{4,7}+H_{4,9}
	=\frac{1}{4}\left[(2\xi_{123,4} + \alpha_{4,123})h^F_{3,3}(\vecq_{1,2,3})+3 \text{ perm.}\right]\\
	&h^{G}_{4,6}(\vecq_{1,2,3,4})&&=H_{4,6}-H^{\m C}_{4,4}-H^{\m C}_{4,7}-H^{\m C}_{4,10}
    =\frac{1}{12}\left[(2\xi_{123,4}\alpha_{12,3}-\alpha_{4,123}\alpha_{3,12}-\alpha_{34,12}\alpha^s_{3,4}) \xi_{1,2}+11 \text{ perm.}\right]\\
    &h^G_{4,7}(\vecq_{1,2,3,4})&&=H_{4,12}-H_{4,7}
	=\frac{1}{12}\left[((2\xi_{123,4}+\alpha_{4,123})\alpha_{3,12}+(2\xi_{12,34}+\alpha_{34,12})\alpha^s_{3,4})\xi_{1,2}+11 \text{ perm.}\right]\\
    &h^G_{4,8}(\vecq_{1,2,3,4})&&=\frac{1}{2}H_{4,5}+H_{4,13}
	=\frac{1}{12}\left[(2\xi_{123,4} \alpha^s_{12,3} + \xi_{12,34} \alpha^s_{3,4})\alpha^s_{1,2}+11 \text{ perm.}\right]\\
    &h^{G}_{4,9}(\vecq_{1,2,3,4})&&=H^{\m C}_{4,2}-H^{\m C}_{4,4}-H^{\m C}_{4,7}
	=\frac{1}{4}\left[\alpha_{4,123}h^F_{3,2}(\vecq_{1,2,3})+3 \text{ perm.}\right]\\
	&h^{G}_{4,10}(\vecq_{1,2,3,4})&&=H^{\m C}_{4,3}-H^{\m C}_{4,4}-H^{\m C}_{4,7}
	=\frac{1}{4}\left[\alpha_{4,123}h^F_{3,3}(\vecq_{1,2,3})+3 \text{ perm.}\right]\\
	&h^{G}_{4,11}(\vecq_{1,2,3,4})&&=H^{\m C}_{4,7}+H^{\m C}_{4,8}
	=\frac{1}{4}\left[\alpha_{123,4}h^F_{3,2}(\vecq_{1,2,3})+3 \text{ perm.}\right]\\
	&h^{G}_{4,12}(\vecq_{1,2,3,4})&&=H^{\m C}_{4,7}+H^{\m C}_{4,9}
	=\frac{1}{4}\left[\alpha_{123,4}h^F_{3,3}(\vecq_{1,2,3})+3 \text{ perm.}\right]\\
	&h^G_{4,13}(\vecq_{1,2,3,4})&&=H^{\m C}_{4,4}+H^{\m C}_{4,6}+H^{\m C}_{4,7}+H^{\m C}_{4,10}
	=\frac{1}{12}\left[(\alpha_{4,123}\alpha_{3,12}+2\alpha^s_{123,4}\alpha_{12,3}+\alpha_{34,12}\alpha^s_{3,4})\xi_{1,2}+11 \text{ perm.}\right]\\
	&h^G_{4,14}(\vecq_{1,2,3,4})&&=H^{\m C}_{4,12}-H^{\m C}_{4,7}=\frac{1}{12}\left[(\alpha_{123,4}\alpha_{3,12}+\alpha_{12,34} \alpha^s_{3,4}) \xi_{1,2}+11 \text{ perm.}\right] \hspace{6cm}
\end{alignat*}
The mapping from the naive to the reduced bases is the following:
\begin{alignat*}{8}
	&d^{F,G}_{4,1}&&={D}^{{\m F},{\m G}}_{4,14}, &\qquad& d^{F,G}_{4,2}&&={D}^{\m C}_{4,11}-{D}^{{\m F},{\m G}}_{4,11}, &\qquad& d^{F,G}_{4,3}&&={D}^{{\m C}}_{4,13}, &\qquad& d^{F,G}_{4,4}&&={D}^{{\m F},{\m G}}_{4,8}, \\
    &d^{F,G}_{4,5}&&={D}^{{\m F},{\m G}}_{4,9},	 &\qquad& d^{F}_{4,6}&&={D}^{\m F}_{4,12}, &\qquad& d^{F}_{4,7}&&={D}^{\m F}_{4,13}, &\qquad& d^{F}_{4,8}&&={D}^{\m C}_{4,8}, \\
	&d^{F}_{4,9}&&={D}^{\m C}_{4,9}, &\qquad& d^F_{4,10}&&={D}^{\m C}_{4,10}, &\qquad& d^F_{4,11}&&={D}^{\m C}_{4,12}, &\qquad& d^G_{4,6}&&={D}^{\m G}_{4,10},  \\
    &d^G_{4,7}&&={D}^{\m G}_{4,12}, &\qquad& d^G_{4,8}&&={D}^{\m G}_{4,13}, &\qquad& d^G_{4,9}&&={D}^{\m C}_{4,2} -{D}^{\m G}_{4,2}, &\qquad& d^G_{4,10}&&={D}^{\m C}_{4,3} -{D}^{\m G}_{4,3},  \\
	& d^G_{4,11}&&={D}^{\m C}_{4,8}, &\qquad& d^G_{4,12}&&={D}^{\m C}_{4,9}, &\qquad& d^G_{4,13}&&={D}^{\m C}_{4,10}, &\qquad& d^G_{4,14}&&={D}^{\m C}_{4,12}.\hspace{3cm}
\end{alignat*}

\subsection{Fifth order}\label{sec:redbasisfifthorder}

At fifth order, the reduced basis contains $39$ terms for $F_5$ and $47$ for $G_5$.

In the following, we simplify the expressions as much as possible. If the substitution of the explicit form of the third order $h$-functions does not simplify the expression, we may not write out the terms to save some space. Additionally, the given fifth order terms are \textit{not symmetrized}. It is understood that in each line the expression on the right-hand side
should be replaced by the symmetrized version, i.e. we omit $\frac{1}{120}[\dots + 119 \text{ perm.}]$ in each line.

\small
\begin{alignat*}{2}
	&h_{5,1}^{FG}(\vecq_{1,2,3,4})&&\equiv -H^{FG}_{5,18}+H^{FG}_{5,22}-H^{FG}_{5,29}-H^{C}_{5,16}-H^{C}_{5,29}=4\xi_{1234,5}\alpha^s_{123,4}(2\xi_{12,3}\alpha^s_{1,2} - \alpha_{3,12}\xi_{1,2}), \\ 
	&h_{5,2}^{FG}(\vecq_{1,2,3,4})&&\equiv -H^{FG}_{5,18}+H^{FG}_{5,23}-H^{FG}_{5,30}-H^{C}_{5,17}-H^{C}_{5,30}=8\xi_{1234,5}\alpha^s_{123,4}\xi_{12,3}\xi_{1,2}, \\ 
	&h_{5,3}^{FG}(\vecq_{1,2,3,4})&&\equiv H^{FG}_{5,6}+H^{FG}_{5,20}+H^{FG}_{5,21}-H^{FG}_{5,26}+H^{FG}_{5,33}\\*
    &&&=2\xi_{1234,5}(2(\xi_{123,4} + \alpha^s_{123,4})\alpha_{12,3} - \alpha_{123,4}\alpha_{3,12} - \alpha_{12,34}\alpha^s_{3,4})\xi_{1,2}, \\ 
	&h_{5,4}^{FG}(\vecq_{1,2,3,4})&&\equiv H^{FG}_{5,38}+H^{FG}_{5,11} =-2\xi_{1234,5}\alpha^s_{12,34}\xi_{1,2}\xi_{3,4}, \\ 
	&h_{5,5}^{FG}(\vecq_{1,2,3,4})&&\equiv H^{FG}_{5,41}-H^{FG}_{5,11} =(2\xi_{1234,5}(\xi_{12,34}+\alpha^s_{12,34}) + \alpha_{5,1234}\xi_{12,34})\xi_{1,2}\xi_{3,4}, \\ 
	&h_{5,6}^{FG}(\vecq_{1,2,3,4})&&\equiv H^{FG}_{5,59}-H^{C}_{5,11}+H^{C}_{5,45}-H^{C}_{5,57}\\*
    &&&=(2\xi_{123,45} + \alpha_{123,45})\alpha_{12,3}\xi_{1,2}\xi_{4,5} - 2\alpha^s_{123,45}\alpha_{3,12}\xi_{1,2}\xi_{4,5} + \alpha_{5,1234}\alpha^s_{12,34}\xi_{1,2}\xi_{3,4}, \\ 
	&h_{5,7}^{FG}(\vecq_{1,2,3,4})&&\equiv H^{C}_{5,14}+H^{C}_{5,11}=\alpha_{5,1234}\xi_{12,34}\xi_{1,2}\xi_{3,4}, \\ 
	&h_{5,8}^{FG}(\vecq_{1,2,3,4})&&\equiv H^{C}_{5,16}-H^{C}_{5,18}+H^{C}_{5,22}=4\alpha^s_{1234,5}\alpha^s_{123,4}(2\xi_{12,3}\alpha^s_{1,2} - \alpha_{3,12}\xi_{1,2}), \\ 
	&h_{5,9}^{FG}(\vecq_{1,2,3,4})&&\equiv H^{C}_{5,17}-H^{C}_{5,18}+H^{C}_{5,23}=8\alpha^s_{1234,5}\alpha^s_{123,4}\xi_{12,3}\xi_{1,2}, \\ 
	&h_{5,10}^{FG}(\vecq_{1,2,3,4})&&\equiv H^{C}_{5,6}+H^{C}_{5,20}+H^{C}_{5,21}-H^{C}_{5,26}+H^{C}_{5,33}\\*
    &&&=2\alpha^s_{1234,5}(2(\xi_{123,4} + \alpha^s_{123,4})\alpha_{12,3} - \alpha_{123,4}\alpha_{3,12} - \alpha_{12,34}\alpha^s_{3,4})\xi_{1,2}, \\ 
	&h_{5,11}^{FG}(\vecq_{1,2,3,4})&&\equiv -H^{C}_{5,25}=-2\alpha^s_{1234,5}\alpha^s_{12,34}\xi_{1,2}\xi_{3,4}, \\ 
	&h_{5,12}^{FG}(\vecq_{1,2,3,4})&&\equiv H^{C}_{5,25}+H^{C}_{5,38}+H^{C}_{5,41}=(\alpha_{1234,5}(\xi_{12,34} + \alpha^s_{12,34}) + \alpha_{5,1234}\alpha^s_{12,34})\xi_{1,2}\xi_{3,4}, \\ 
	&h_{5,13}^{FG}(\vecq_{1,2,3,4})&&\equiv H^{C}_{5,45}+H^{C}_{5,46}=\alpha_{123,45}\xi_{4,5}h^F_{3,2}(\vecq_{1,2,3}), \\ 
	&h_{5,14}^{FG}(\vecq_{1,2,3,4})&&\equiv H^{C}_{5,45}+H^{C}_{5,47}=\alpha_{123,45}\xi_{4,5}h^F_{3,3}(\vecq_{1,2,3}), \\ 
	&h_{5,15}^{FG}(\vecq_{1,2,3,4})&&\equiv H^{C}_{5,55}-H^{C}_{5,57}=\alpha_{45,123}\xi_{4,5}h^F_{3,2}(\vecq_{1,2,3}), \\ 
	&h_{5,16}^{FG}(\vecq_{1,2,3,4})&&\equiv H^{C}_{5,56}-H^{C}_{5,57}=\alpha_{45,123}\xi_{4,5}h^F_{3,3}(\vecq_{1,2,3}), \\ 
	&h_{5,17}^{FG}(\vecq_{1,2,3,4})&&\equiv (H^{C}_{5,19}+2H^{C}_{5,27})/3+H^{C}_{5,61}=\frac{4}{3}(\alpha^s_{1234,5}(2\alpha^s_{123,4}\alpha^s_{12,3} + \alpha^s_{12,34}\alpha^s_{3,4}) + 3\alpha^s_{123,45}\alpha^s_{12,3}\alpha^s_{4,5})\alpha^s_{1,2}, \\ 
    &h_{5,18}^{F}(\vecq_{1,2,3,4})&&\equiv H^{FG}_{5,29}-H^{FG}_{5,31}+H^{FG}_{5,35}-H^{C}_{5,1}-H^{C}_{5,7}-H^{C}_{5,8}+H^{C}_{5,16}+H^{C}_{5,29}\\*
    &&&=4\xi_{1234,5}\xi_{123,4}(2\xi_{12,3}\alpha^s_{1,2} - \alpha_{3,12}\xi_{1,2}), \\ 
	&h_{5,19}^{F}(\vecq_{1,2,3,4})&&\equiv H^{FG}_{5,30}-H^{FG}_{5,31}+H^{FG}_{5,36}-H^{C}_{5,1}-H^{C}_{5,7}-H^{C}_{5,9}+H^{C}_{5,17}+H^{C}_{5,30}=8\xi_{1234,5}\xi_{123,4}\xi_{12,3}\xi_{1,2}, \\ 
	&h_{5,20}^{F}(\vecq_{1,2,3,4})&&\equiv -H^{FG}_{5,20}-H^{FG}_{5,21}+H^{FG}_{5,26}+H^{FG}_{5,30}-H^{FG}_{5,34}+H^{FG}_{5,37}+H^{FG}_{5,39}+H^{C}_{5,7}-H^{C}_{5,10}+H^{C}_{5,12}\\*
    &&&\qquad+H^{C}_{5,17}+H^{C}_{5,30}=-2\xi_{1234,5}(-(2\xi_{123,4} + \alpha_{123,4})\alpha_{3,12} + 2\alpha^s_{123,4}\alpha_{12,3} + 2\alpha_{4,123}\xi_{12,3}\\*
    &&&\qquad - (2\xi_{12,34} + \alpha_{12,34})\alpha^s_{3,4})\xi_{1,2}, \\ 
	&h_{5,21}^{F}(\vecq_{1,2,3,4})&&\equiv (H^{FG}_{5,18}-H^{FG}_{5,20}+H^{FG}_{5,26}+2H^{FG}_{5,29}+H^{FG}_{5,32}+H^{FG}_{5,37}+2H^{FG}_{5,40}\\*
    &&&\qquad-H^{C}_{5,5}-H^{C}_{5,10}-2H^{C}_{5,13}+2H^{C}_{5,16}+2H^{C}_{5,29})/2\\*
    &&&=\xi_{1234,5}(4(\xi_{123,4}\alpha^s_{12,3} - \alpha_{4,123}\xi_{12,3})\alpha^s_{1,2} + 2\alpha^s_{123,4}(\alpha_{3,12} - \alpha_{12,3})\xi_{1,2} \\*
    &&&\qquad+ (2\xi_{12,34}\alpha^s_{3,4} + (\alpha_{34,12} - \alpha_{12,34})\xi_{3,4})\alpha^s_{1,2}), \\ 
	&h_{5,22}^{F}(\vecq_{1,2,3,4})&&\equiv H^{FG}_{5,52}-H^{FG}_{5,60}+H^{C}_{5,4}-H^{C}_{5,8}+H^{C}_{5,16}+H^{C}_{5,29}-H^{C}_{5,49}\\*
    &&&=2(-\alpha_{5,1234}\xi_{123,4} + \xi_{123,45}\alpha^s_{4,5})(2\xi_{12,3}\alpha^s_{1,2} - \alpha_{3,12}\xi_{1,2}), \\ 
	&h_{5,23}^{F}(\vecq_{1,2,3,4})&&\equiv H^{FG}_{5,53}-H^{FG}_{5,60}+H^{C}_{5,4}-H^{C}_{5,9}+H^{C}_{5,17}+H^{C}_{5,30}-H^{C}_{5,50}\\*
    &&&=4(-\alpha_{5,1234}\xi_{123,4} + \xi_{123,45}\alpha^s_{4,5})\xi_{12,3}\xi_{1,2}, \\ 
	&h_{5,24}^{F}(\vecq_{1,2,3,4})&&\equiv (H^{FG}_{5,18}+H^{FG}_{5,19}-H^{FG}_{5,20}-H^{FG}_{5,24}+H^{FG}_{5,26}+2H^{FG}_{5,27}+3H^{FG}_{5,29}-H^{FG}_{5,30}-H^{FG}_{5,58}-2H^{FG}_{5,60}+3H^{FG}_{5,61}\\*
    &&&\qquad +2H^{C}_{5,4}-3H^{C}_{5,5}+2H^{C}_{5,12}-6H^{C}_{5,13}+6H^{C}_{5,16}-2H^{C}_{5,17}+6H^{C}_{5,29}-2H^{C}_{5,30}-3H^{C}_{5,49}+H^{C}_{5,50})/3\\*
    &&&=\frac{1}{3}(2\xi_{1234,5}(\alpha_{123,4}(2\alpha^s_{12,3}\alpha^s_{1,2} + (\alpha_{3,12} - \alpha_{12,3})\xi_{1,2}) + \alpha_{4,123}(2(-3\xi_{12,3} + \alpha^s_{12,3})\alpha^s_{1,2} \\*
    &&&\qquad+ (2\xi_{12,3} + 2\alpha_{3,12} - \alpha_{12,3})\xi_{1,2}) + (2\alpha^s_{12,34}\alpha^s_{3,4} + (\alpha_{34,12} - \alpha_{12,34})\xi_{3,4})\alpha^s_{1,2}) \\*
    &&&\qquad+ \alpha_{5,1234}(4\xi_{123,4}(-3\alpha^s_{12,3}\alpha^s_{1,2} + \alpha_{3,12}\xi_{1,2}) + \alpha_{4,123}(2\xi_{12,3}(-\xi_{1,2} + 3\alpha^s_{1,2}) - \alpha_{3,12}\xi_{1,2}) \\*
    &&&\qquad+ 2(\xi_{12,34}(2\xi_{1,2} - 3\alpha^s_{1,2}) + \alpha_{34,12}\xi_{1,2})\alpha^s_{3,4}) - 4\xi_{12,345}(\alpha^s_{45,3}(\xi_{1,2} - 3\alpha^s_{1,2})\alpha^s_{4,5} \\*
    &&& \qquad+ \alpha_{3,45}\alpha^s_{1,2}\xi_{4,5}) - 2\alpha_{345,12}\alpha^s_{45,3} \xi_{1,2}\alpha^s_{4,5} + \alpha_{12,345}(2\xi_{45,3}(\xi_{4,5} - 3\alpha^s_{4,5}) + \alpha_{3,45}\xi_{4,5})\alpha^s_{1,2}), \\ 
	&h_{5,25}^{F}(\vecq_{1,2,3,4})&&\equiv (H^{FG}_{5,18}+H^{FG}_{5,20}+H^{FG}_{5,24}+H^{FG}_{5,26}+H^{FG}_{5,28}+H^{FG}_{5,30}+H^{FG}_{5,58}+2H^{FG}_{5,60}+2H^{FG}_{5,62}\\*
    &&&\qquad-2H^{C}_{5,6}+2H^{C}_{5,7}-2H^{C}_{5,12} +2H^{C}_{5,17}+2H^{C}_{5,30}-H^{C}_{5,43}-2H^{C}_{5,44}-H^{C}_{5,48}-H^{C}_{5,50})/2\\*
    &&&=(2\xi_{1234,5}(\alpha_{123,4}\alpha^s_{12,3} - \alpha_{4,123}(\xi_{12,3} - \alpha^s_{12,3}) + \alpha^s_{12,34}\alpha^s_{3,4}) - \alpha_{5,1234}(4\xi_{123,4}\alpha^s_{12,3}  \\*
    &&&\qquad- \alpha_{4,123}\xi_{12,3}+ 2\xi_{12,34}\alpha^s_{3,4}))\xi_{1,2} + 2\xi_{12,345}\alpha^s_{45,3}(\xi_{1,2}\alpha^s_{4,5}  \\*
    &&&\qquad+ 2\alpha^s_{1,2}\xi_{4,5}) -(\alpha_{345,12}\alpha_{45,3}\xi_{1,2} + \alpha_{12,345}\xi_{45,3}\alpha^s_{1,2})\xi_{4,5}, \\ 
	&h_{5,26}^{F}(\vecq_{1,2,3,4})&&\equiv H^{FG}_{5,63}-H^{C}_{5,11}+H^{C}_{5,44}\\*
    &&&=(\alpha_{5,1234}\alpha^s_{12,34}\xi_{3,4} - (2(\xi_{12,345} + \alpha^s_{12,345})\alpha_{3,45} - \alpha_{345,12}\alpha_{45,3})\xi_{4,5})\xi_{1,2}, \\ 
	&h_{5,27}^{F}(\vecq_{1,2,3,4})&&\equiv H^{FG}_{5,64}+H^{C}_{5,11}-H^{C}_{5,44}\\*
    &&&=(-\alpha_{5,1234}\alpha^s_{12,34}\xi_{3,4} + 4(\xi_{12,345} + \alpha^s_{12,345})\xi_{45,3}\alpha^s_{4,5} - \alpha_{345,12}\alpha_{45,3}\xi_{4,5})\xi_{1,2}, \\ 
	&h_{5,28}^{F}(\vecq_{1,2,3,4})&&\equiv H^{FG}_{5,65}+H^{C}_{5,11}-H^{C}_{5,44}\\*
    &&&=(-\alpha_{5,1234}\alpha^s_{12,34}\xi_{3,4} + (2(\xi_{12,345} + \alpha^s_{12,345})(2\xi_{45,3} + \alpha_{3,45}) - \alpha_{345,12}\alpha_{45,3})\xi_{4,5})\xi_{1,2}, \\ 
	&h_{5,29}^{F}(\vecq_{1,2,3,4})&&\equiv H^{C}_{5,7}+H^{C}_{5,8}-H^{C}_{5,15}-H^{C}_{5,16}+H^{C}_{5,34}+H^{C}_{5,35}=4\alpha^s_{1234,5}\xi_{123,4}(2\xi_{12,3}\alpha^s_{1,2} - \alpha_{3,12}\xi_{1,2}), \\ 
	&h_{5,30}^{F}(\vecq_{1,2,3,4})&&\equiv H^{C}_{5,7}+H^{C}_{5,9}-H^{C}_{5,15}-H^{C}_{5,17}+H^{C}_{5,34}+H^{C}_{5,36}=8\alpha^s_{1234,5}\xi_{123,4}\xi_{12,3}\xi_{1,2}, \\ 
	&h_{5,31}^{F}(\vecq_{1,2,3,4})&&\equiv -H^{C}_{5,7}+H^{C}_{5,12}-H^{C}_{5,17}-H^{C}_{5,20}-H^{C}_{5,21}-H^{C}_{5,24}+H^{C}_{5,26}-H^{C}_{5,34}+H^{C}_{5,39}\\*
    &&&=2\alpha^s_{1234,5}((2\xi_{123,4} + \alpha_{123,4})\alpha_{3,12} - 2\alpha^s_{123,4}\alpha_{12,3} - 2\alpha_{4,123}\xi_{12,3} + (2\xi_{12,34} + \alpha_{12,34})\alpha^s_{3,4})\xi_{1,2}, \\ 
	&h_{5,32}^{F}(\vecq_{1,2,3,4})&&\equiv (H^{C}_{5,5}+2H^{C}_{5,13}-2H^{C}_{5,16}+H^{C}_{5,18}-H^{C}_{5,20}-H^{C}_{5,24}+H^{C}_{5,26}+H^{C}_{5,32}+2H^{C}_{5,40})/2\\*
    &&&=\alpha^s_{1234,5}(4\xi_{123,4}\alpha^s_{12,3}\alpha^s_{1,2} + 2\alpha^s_{123,4}(\alpha_{3,12} - \alpha_{12,3})\xi_{1,2} \\*
    &&&\qquad+ (-4\alpha_{4,123}\xi_{12,3} + 2\xi_{12,34}\alpha^s_{3,4} + (\alpha_{34,12} - \alpha_{12,34})\xi_{3,4})\alpha^s_{1,2}), \\ 
	&h_{5,33}^{F}(\vecq_{1,2,3,4})&&\equiv (H^{FG}_{5,28}+H^{FG}_{5,29}+2H^{C}_{5,28}+2H^{C}_{5,29}+H^{C}_{5,48}+H^{C}_{5,49})/2\\*
    &&&=\frac{1}{2}((2\xi_{1234,5} + 4\alpha^s_{1234,5} - \alpha_{5,1234})\alpha_{4,123}(-2\xi_{12,3}\alpha^s_{1,2} \\*
    &&&\qquad+ \alpha_{3,12}\xi_{1,2}) + \alpha_{12,345}(2\xi_{45,3}\alpha^s_{4,5} - \alpha_{3,45}\xi_{4,5})\alpha^s_{1,2}), \\ 
	&h_{5,34}^{F}(\vecq_{1,2,3,4})&&\equiv (H^{FG}_{5,28}+H^{FG}_{5,30}+2H^{C}_{5,28}+2H^{C}_{5,30}+H^{C}_{5,50}+H^{C}_{5,48})/2\\*
    &&&=(-2\xi_{1234,5} - 4\alpha^s_{1234,5} + \alpha_{5,1234})\alpha_{4,123}\xi_{12,3}\xi_{1,2} + \alpha_{12,345}\xi_{45,3}\alpha^s_{1,2}\xi_{4,5}, \\ 
	&h_{5,35}^{F}(\vecq_{1,2,3,4})&&\equiv (-H^{FG}_{5,28}-H^{FG}_{5,29}-2H^{C}_{5,28}-2H^{C}_{5,29}+H^{C}_{5,49}+H^{C}_{5,51}+2H^{C}_{5,52}-H^{C}_{5,60})/2\\*
    &&&=\frac{1}{2}((2\xi_{1234,5} + 4\alpha^s_{1234,5} - \alpha_{5,1234})\alpha_{4,123}(2\xi_{12,3}\alpha^s_{1,2} - \alpha_{3,12}\xi_{1,2}) \\*
    &&&\qquad+ (2\alpha^s_{12,345} + \alpha_{345,12})(2\xi_{45,3}\alpha^s_{4,5} - \alpha_{3,45}\xi_{4,5})\alpha^s_{1,2}), \\ 
	&h_{5,36}^{F}(\vecq_{1,2,3,4})&&\equiv (-H^{FG}_{5,28}-H^{FG}_{5,30}-2H^{C}_{5,30}-2H^{C}_{5,28}+H^{C}_{5,50}+H^{C}_{5,51}+2H^{C}_{5,53}-H^{C}_{5,60})/2\\*
    &&&=((2\xi_{1234,5} + 4\alpha^s_{1234,5} - \alpha_{5,1234})\alpha_{4,123} + (2\alpha^s_{45,123} + \alpha_{123,45})\alpha^s_{4,5})\xi_{12,3}\xi_{1,2}, \\ 
	&h_{5,37}^{F}(\vecq_{1,2,3,4})&&\equiv (H^{FG}_{5,20}+H^{FG}_{5,21}+H^{FG}_{5,24}-H^{FG}_{5,26}+H^{FG}_{5,30}+2H^{FG}_{5,28}+H^{FG}_{5,58}+2H^{FG}_{5,60}\\*
    &&&\qquad +2H^{C}_{5,7}-2H^{C}_{5,12}+2H^{C}_{5,17}+3H^{C}_{5,20}+3H^{C}_{5,21}+3H^{C}_{5,24}-3H^{C}_{5,26}+3H^{C}_{5,28}+2H^{C}_{5,30}\\*
    &&&\qquad-H^{C}_{5,50}-3H^{C}_{5,51}+3H^{C}_{5,58}+3H^{C}_{5,60})/3\\*
    &&&=\frac{2}{3}((\xi_{1234,5}(\alpha_{123,4}(\alpha_{12,3} - \alpha_{3,12}) + \alpha_{4,123}(-2\xi_{12,3} + \alpha_{12,3} + \alpha_{3,12}) + (\alpha_{34,12} - \alpha_{12,34})\alpha^s_{3,4})\\*
    &&&\qquad - 3\alpha^s_{1234,5}((\alpha_{123,4} - \alpha_{4,123})\alpha_{3,12} - 2\alpha^s_{123,4}\alpha_{12,3} + (\alpha_{12,34} - \alpha_{34,12})\alpha^s_{3,4}) \\*
    &&&\qquad- \alpha_{5,1234}((2\xi_{123,4} + \alpha_{4,123})\alpha_{3,12} - \alpha_{4,123}\xi_{12,3} + 2\xi_{12,34}\alpha^s_{3,4} + \alpha_{34,12}\alpha^s_{3,4}))\xi_{1,2}  \\*
    &&&\qquad+ 2\xi_{12,345}(\alpha^s_{45,3}\xi_{1,2}\alpha^s_{4,5}+ \alpha_{3,45}\alpha^s_{1,2}\xi_{4,5}) + \alpha_{345,12}(\alpha^s_{45,3}\xi_{1,2}\alpha^s_{4,5} + 3\alpha_{3,45}\alpha^s_{1,2}\xi_{4,5})  \\*
    &&&\qquad+ \alpha_{12,345}(3\alpha^s_{45,3}\xi_{1,2}\alpha^s_{4,5}+ (-\xi_{45,3} + \alpha_{3,45})\alpha^s_{1,2}\xi_{4,5})), \\ 
	&h_{5,38}^{F}(\vecq_{1,2,3,4})&&\equiv H^{C}_{5,44}-H^{C}_{5,45}+H^{C}_{5,57}+H^{C}_{5,59}=4\alpha^s_{123,45}\xi_{4,5}\alpha^s_{12,3}\xi_{1,2}, \\ 
	&h_{5,39}^{F}(\vecq_{1,2,3,4})&&\equiv (-H^{FG}_{5,20}-H^{FG}_{5,21}-H^{FG}_{5,24}+H^{FG}_{5,26}-2H^{FG}_{5,28}-H^{FG}_{5,30}-H^{FG}_{5,58}-2H^{FG}_{5,60}\\*
    &&&\qquad+3H^{C}_{5,1} -2H^{C}_{5,7}+2H^{C}_{5,12}-2H^{C}_{5,17}-6H^{C}_{5,21}+6H^{C}_{5,26}-2H^{C}_{5,30}\\*
    &&&\qquad+3H^{C}_{5,43} +H^{C}_{5,50}+3H^{C}_{5,51}+3H^{C}_{5,60}+6H^{C}_{5,62)/6}\\*
    &&&=\frac{1}{3}((\xi_{1234,5}(\alpha_{123,4}(\alpha_{3,12} - \alpha_{12,3}) + 2\alpha_{4,123}(\xi_{12,3} - \alpha^s_{12,3}) + (\alpha_{12,34} - \alpha_{34,12})\alpha^s_{3,4}) \\*
    &&&\qquad+ 6\alpha^s_{1234,5}(\alpha_{123,4}\alpha_{3,12} + \alpha_{12,34}\alpha^s_{3,4}) + \alpha_{5,1234}((2\xi_{123,4} + \alpha_{4,123})\alpha_{3,12} - \alpha_{4,123}\xi_{12,3}\\*
    &&&\qquad + (2\xi_{12,34} + \alpha_{34,12})\alpha^s_{3,4}))\xi_{1,2} + (-2\xi_{12,345}(\alpha^s_{45,3}\xi_{1,2}\alpha^s_{4,5} + \alpha_{3,45}\alpha^s_{1,2}\xi_{4,5})  \\*
    &&&\qquad+ \alpha_{345,12}(2\alpha^s_{45,3}\xi_{1,2}\alpha^s_{4,5}+ 3\alpha_{45,3}\alpha^s_{1,2}\xi_{4,5}) + \alpha_{12,345}(\xi_{45,3} + 4\alpha^s_{3,45} + \alpha_{45,3})\alpha^s_{1,2}\xi_{4,5})), \\ 
	&h_{5,18}^{G}(\vecq_{1,2,3,4})&&\equiv H^{FG}_{5,29}-H^{FG}_{5,31}+H^{FG}_{5,35}=2(2\xi_{1234,5} + \alpha_{5,1234})\xi_{123,4}h^F_{3,2}(\vecq_{1,2,3}), \\ 
	&h_{5,19}^{G}(\vecq_{1,2,3,4})&&\equiv H^{FG}_{5,30}-H^{FG}_{5,31}+H^{FG}_{5,36}=2(2\xi_{1234,5} + \alpha_{5,1234})\xi_{123,4}h^F_{3,3}(\vecq_{1,2,3}), \\ 
	&h_{5,20}^{G}(\vecq_{1,2,3,4})&&\equiv -H^{FG}_{5,6}-H^{FG}_{5,11}-H^{FG}_{5,20}-H^{FG}_{5,21}+H^{FG}_{5,26}+H^{FG}_{5,30}-H^{FG}_{5,34}+H^{FG}_{5,37}+H^{FG}_{5,39}\\*
    &&&=(2\xi_{1234,5}(2\xi_{123,4}\alpha_{3,12} + \alpha_{123,4}(\alpha_{3,12} - \alpha_{12,3}) - \alpha_{4,123}(2\xi_{12,3} + \alpha_{12,3}) + (2\xi_{12,34} + \alpha_{12,34})\alpha^s_{3,4}) \\*
    &&&\qquad+ \alpha_{5,1234}(4\xi_{123,4}\alpha^s_{12,3} - 2\alpha_{4,123}\xi_{12,3} + 2\xi_{12,34}\alpha^s_{3,4} - \alpha^s_{12,34}\xi_{3,4}))\xi_{1,2}, \\ 
	&h_{5,21}^{G}(\vecq_{1,2,3,4})&&\equiv (-H^{FG}_{5,6}-H^{FG}_{5,11}-H^{FG}_{5,20}-H^{FG}_{5,21}+H^{FG}_{5,26}+H^{FG}_{5,28}+2H^{FG}_{5,29}+H^{FG}_{5,32}+H^{FG}_{5,37}+2H^{FG}_{5,40})/2\\*
    &&&=\xi_{1234,5}(4\xi_{123,4}\alpha^s_{12,3}\alpha^s_{1,2} + 2\alpha^s_{123,4}(\alpha_{3,12} - \alpha_{12,3})\xi_{1,2} - 4\alpha_{4,123}\xi_{12,3}\alpha^s_{1,2} \\*
    &&&\qquad+ (2\xi_{12,34}\alpha^s_{3,4} + (\alpha_{34,12} - \alpha_{12,34})\xi_{3,4})\alpha^s_{1,2}) + \alpha_{5,1234}(\xi_{123,4}(2\alpha^s_{12,3}\alpha^s_{1,2} + \alpha_{12,3}\xi_{1,2}) \\*
    &&&\qquad+ \frac{1}{2}\alpha_{4,123}(-4\xi_{12,3}\alpha^s_{1,2} + \alpha_{3,12}\xi_{1,2}) + \xi_{12,34}\alpha^s_{1,2}\alpha^s_{3,4} - \frac{1}{2}\alpha_{12,34}\alpha^s_{1,2}\xi_{3,4} - \frac{1}{2}\alpha^s_{12,34}\xi_{1,2}\xi_{3,4}), \\ 
	&h_{5,22}^{G}(\vecq_{1,2,3,4})&&\equiv H^{FG}_{5,28} + H^{FG}_{5,29}+H^{FG}_{5,48}+H^{FG}_{5,49}\\*
    &&&=(2\xi_{1234,5} + \alpha_{5,1234})\alpha_{4,123}(\alpha_{3,12}\xi_{1,2} - 2\xi_{12,3}\alpha^s_{1,2}) + \alpha_{12,345}(\alpha_{3,45}\xi_{4,5} - 2\xi_{45,3}\alpha^s_{4,5})\alpha^s_{1,2}, \\ 
	&h_{5,23}^{G}(\vecq_{1,2,3,4})&&\equiv H^{FG}_{5,28}+H^{FG}_{5,30}+H^{FG}_{5,48}+H^{FG}_{5,50}\\*
    &&&=-2(2\xi_{1234,5} + \alpha_{5,1234})\alpha_{4,123}\xi_{12,3}\xi_{1,2} - 2\alpha_{12,345}\xi_{45,3}\xi_{4,5}\alpha^s_{1,2}, \\ 
	&h_{5,24}^{G}(\vecq_{1,2,3,4})&&\equiv -H^{FG}_{5,29}-H^{FG}_{5,28}-H^{FG}_{5,48}+H^{FG}_{5,52}-H^{FG}_{5,60}\\*
    &&&=(2\xi_{1234,5} + \alpha_{5,1234})\alpha_{4,123}(2\xi_{12,3}\alpha^s_{1,2} - \alpha_{3,12}\xi_{1,2}) + (2\xi_{12,345} + \alpha_{12,345})(2\xi_{45,3}\alpha^s_{4,5} - \alpha_{3,45}\xi_{4,5})\alpha^s_{1,2}, \\ 
	&h_{5,25}^{G}(\vecq_{1,2,3,4})&&\equiv -H^{FG}_{5,30}-H^{FG}_{5,28}-H^{FG}_{5,48}+H^{FG}_{5,53}-H^{FG}_{5,60}\\*
    &&&=2(2\xi_{1234,5} + \alpha_{5,1234})\alpha_{4,123}\xi_{12,3}\xi_{1,2} + 2(2\xi_{12,345} + \alpha_{12,345})\xi_{45,3}\xi_{4,5}\alpha^s_{1,2}, \\ 
	&h_{5,26}^{G}(\vecq_{1,2,3,4})&&\equiv H^{FG}_{5,6}+H^{FG}_{5,11}+H^{FG}_{5,20}+H^{FG}_{5,21}-H^{FG}_{5,26}+H^{FG}_{5,28}-H^{FG}_{5,37}+2H^{FG}_{5,48}+H^{FG}_{5,58}+2H^{FG}_{5,60}\\*
    &&&=(2\xi_{1234,5}(\alpha_{123,4}(\alpha_{12,3} - \alpha_{3,12}) + 2\alpha_{4,123}\alpha^s_{12,3} + (\alpha_{34,12} - \alpha_{12,34})\alpha^s_{3,4})  \\*
    &&&\qquad+ \alpha_{5,1234}(-2\xi_{123,4}\alpha_{12,3} + \alpha_{4,123}\alpha_{3,12}+ \alpha^s_{12,34}\xi_{3,4} + \alpha_{34,12}\alpha^s_{3,4}))\xi_{1,2}  \\*
    &&&\qquad + 2(2\xi_{12,345}(\alpha^s_{45,3}\alpha^s_{4,5}\xi_{1,2}+ \alpha_{3,45}\alpha^s_{1,2}\xi_{4,5}) +\alpha_{345,12}\alpha^s_{45,3}\alpha^s_{4,5}\xi_{1,2} + \alpha_{12,345}\alpha_{3,45}\alpha^s_{1,2}\xi_{4,5}), \\ 
	&h_{5,27}^{G}(\vecq_{1,2,3,4})&&\equiv (H^{FG}_{5,19}+2H^{FG}_{5,27}+3H^{FG}_{5,61})/3=\frac{4}{3}(\xi_{1234,5}(2\alpha^s_{123,4}\alpha^s_{12,3} + \alpha^s_{12,34}\alpha^s_{3,4}) + 3\xi_{123,45}\alpha^s_{12,3}\alpha^s_{4,5})\alpha^s_{1,2}, \\ 
	&h_{5,28}^{G}(\vecq_{1,2,3,4})&&\equiv (-2H^{FG}_{5,21}+2H^{FG}_{5,26}+2H^{FG}_{5,62}-H^{C}_{5,1}+H^{C}_{5,6}+H^{C}_{5,10}+H^{C}_{5,11}-H^{C}_{5,43}+2H^{C}_{5,48})/2\\*
    &&&=(2\xi_{1234,5}(\alpha_{123,4}\alpha_{3,12} + \alpha_{12,34}\alpha^s_{3,4}) - \frac{1}{2}\alpha_{5,1234}(-2\xi_{123,4}\alpha_{12,3} + \alpha^s_{12,34}\xi_{3,4} + \alpha_{4,123}\alpha_{3,12}\\*
    &&& + \alpha_{34,12}\alpha^s_{3,4}) + (2\xi_{123,45}\alpha_{12,3} - \alpha_{45,123}\alpha_{3,12} - \alpha_{345,12}\alpha^s_{45,3})\alpha^s_{4,5})\xi_{1,2}, \\ 
	&h_{5,29}^{G}(\vecq_{1,2,3,4})&&\equiv H^{FG}_{5,63}=-2(\xi_{12,345} + \alpha^s_{12,345})\alpha_{3,45}\xi_{1,2}\xi_{4,5}, \\ 
	&h_{5,30}^{G}(\vecq_{1,2,3,4})&&\equiv H^{FG}_{5,64}=4(\xi_{12,345} + \alpha^s_{12,345})\xi_{3,45}\xi_{1,2}\alpha^s_{4,5}, \\ 
	&h_{5,31}^{G}(\vecq_{1,2,3,4})&&\equiv H^{FG}_{5,65}=2(\xi_{12,345} + \alpha^s_{12,345})(2\xi_{3,45} + \alpha_{3,45})\xi_{1,2}\xi_{4,5}, \\ 
	&h_{5,32}^{G}(\vecq_{1,2,3,4})&&\equiv -H^{C}_{5,4}+H^{C}_{5,8}-H^{C}_{5,16}-H^{C}_{5,29}=2\alpha_{5,1234}\xi_{123,4}h^F_{3,2}(\vecq_{1,2,3}), \\ 
	&h_{5,33}^{G}(\vecq_{1,2,3,4})&&\equiv -H^{C}_{5,4}+H^{C}_{5,9}-H^{C}_{5,17}-H^{C}_{5,30}=2\alpha_{5,1234}\xi_{123,4}h^F_{3,3}(\vecq_{1,2,3}), \\ 
	&h_{5,34}^{G}(\vecq_{1,2,3,4})&&\equiv H^{C}_{5,6}-H^{C}_{5,7}+H^{C}_{5,10}+H^{C}_{5,11}+H^{C}_{5,12}-H^{C}_{5,17}-H^{C}_{5,30}\\*
    &&&=\alpha_{5,1234}(4\xi_{123,4}\alpha^s_{12,3} - 2\alpha_{4,123}\xi_{12,3} + 2\xi_{12,34}\alpha^s_{3,4} - \alpha^s_{12,34}\xi_{3,4})\xi_{1,2}, \\ 
	&h_{5,35}^{G}(\vecq_{1,2,3,4})&&\equiv (H^{C}_{5,1}+H^{C}_{5,5}+H^{C}_{5,6}+H^{C}_{5,10}+H^{C}_{5,11}+2H^{C}_{5,13}-2H^{C}_{5,16}-2H^{C}_{5,29})/2\\*
    &&&=\frac{1}{2}\alpha_{5,1234}(2\xi_{123,4}(2\alpha^s_{12,3}\alpha^s_{1,2} + \alpha_{12,3}\xi_{1,2}) + \alpha_{4,123}(-4\xi_{12,3}\alpha^s_{1,2} \\*
    &&&\qquad+ \alpha_{3,12}\xi_{1,2}) + 2\xi_{12,34}\alpha^s_{1,2}\alpha^s_{3,4} - \alpha^s_{12,34}\xi_{1,2}\xi_{3,4} - \alpha_{12,34}\alpha^s_{1,2}\xi_{3,4}), \\ 
	&h_{5,36}^{G}(\vecq_{1,2,3,4})&&\equiv H^{C}_{5,29}-H^{C}_{5,31}+H^{C}_{5,35}=2\alpha_{1234,5}(2\xi_{12,3}\alpha^s_{1,2} - \alpha_{3,12}\xi_{1,2})\xi_{123,4}, \\ 
	&h_{5,37}^{G}(\vecq_{1,2,3,4})&&\equiv H^{C}_{5,30}-H^{C}_{5,31}+H^{C}_{5,36}=4\alpha_{1234,5}\xi_{123,4}\xi_{12,3}\xi_{1,2}, \\ 
	&h_{5,38}^{G}(\vecq_{1,2,3,4})&&\equiv -H^{C}_{5,6}-H^{C}_{5,11}-H^{C}_{5,20}-H^{C}_{5,21}+H^{C}_{5,26}+H^{C}_{5,30}-H^{C}_{5,34}+H^{C}_{5,37}+H^{C}_{5,39}\\*
    &&&=(2\alpha^s_{1234,5}(-2(\xi_{123,4} + \alpha^s_{123,4})\alpha_{12,3} + \alpha_{123,4}\alpha_{3,12} + \alpha^s_{12,34}\xi_{3,4} + \alpha_{12,34}\alpha^s_{3,4}) \\*
    &&&\qquad+ \alpha_{1234,5}(4\xi_{123,4}\alpha^s_{12,3} - 2\alpha_{4,123}\xi_{12,3} + 2\xi_{12,34}\alpha^s_{3,4} - \alpha^s_{12,34}\xi_{3,4}))\xi_{1,2}, \\ 
	&h_{5,39}^{G}(\vecq_{1,2,3,4})&&\equiv (-H^{C}_{5,6}-H^{C}_{5,11}-H^{C}_{5,20}-H^{C}_{5,21}+H^{C}_{5,26}+H^{C}_{5,28}+2H^{C}_{5,29}+H^{C}_{5,32}+H^{C}_{5,37}+2H^{C}_{5,40})/2\\*
    &&&=\alpha^s_{1234,5}(-2(\xi_{123,4} + \alpha^s_{123,4})\alpha_{12,3} + \alpha_{123,4}\alpha_{3,12} + \alpha^s_{12,34}\xi_{3,4} + \alpha_{12,34}\alpha^s_{3,4})\xi_{1,2} \\*
    &&&\qquad+ \frac{1}{2}\alpha_{1234,5}(2\xi_{123,4}(2\alpha^s_{12,3}\alpha^s_{1,2} + \alpha_{12,3}\xi_{1,2}) + \alpha_{4,123}(-4\xi_{12,3}\alpha^s_{1,2} \\*
    &&&\qquad+ \alpha_{3,12}\xi_{1,2}) + 2\xi_{12,34}\alpha^s_{1,2}\alpha^s_{3,4} - (\alpha^s_{12,34}\xi_{3,4} + \alpha_{34,12}\alpha^s_{3,4})\xi_{1,2}), \\ 
	&h_{5,40}^{G}(\vecq_{1,2,3,4})&&\equiv H^{C}_{5,44}-H^{C}_{5,11}=(\alpha_{5,1234}\alpha^s_{12,34}\xi_{3,4} + \alpha_{123,45}\alpha_{12,3}\xi_{4,5})\xi_{1,2}, \\ 
	&h_{5,41}^{G}(\vecq_{1,2,3,4})&&\equiv H^{C}_{5,28}+H^{C}_{5,29}+H^{C}_{5,48}+H^{C}_{5,49}\\*
    &&&=\alpha_{1234,5}\alpha_{4,123}(-2\xi_{12,3}\alpha^s_{1,2} + \alpha_{3,12}\xi_{1,2}) + \alpha_{12,345}(2\xi_{45,3}\alpha^s_{4,5} - \alpha_{3,45}\xi_{4,5})\alpha^s_{1,2}, \\ 
	&h_{5,42}^{G}(\vecq_{1,2,3,4})&&\equiv H^{C}_{5,28}+H^{C}_{5,30}+H^{C}_{5,48}+H^{C}_{5,50}=-2\alpha_{1234,5}\alpha_{4,123}\xi_{12,3}\xi_{1,2} + 2\alpha_{12,345}\xi_{45,3}\alpha^s_{1,2}\xi_{4,5}, \\ 
	&h_{5,43}^{G}(\vecq_{1,2,3,4})&&\equiv-H^{C}_{5,28}-H^{C}_{5,29}+H^{C}_{5,51}+H^{C}_{5,52}\\*
    &&&=\alpha_{1234,5}\alpha_{4,123}(2\xi_{12,3}\alpha^s_{1,2} - \alpha_{3,12}\xi_{1,2}) + \alpha_{345,12}(2\xi_{45,3}\alpha^s_{4,5} - \alpha_{3,45}\xi_{4,5})\alpha^s_{1,2}, \\ 
	&h_{5,44}^{G}(\vecq_{1,2,3,4})&&\equiv -H^{C}_{5,28}-H^{C}_{5,30}+H^{C}_{5,51}+H^{C}_{5,53}=2(\alpha_{1234,5}\alpha_{4,123} + \alpha_{123,45}\alpha^s_{4,5})\xi_{12,3}\xi_{1,2}, \\ 
	&h_{5,45}^{G}(\vecq_{1,2,3,4})&&\equiv H^{C}_{5,6}+H^{C}_{5,11}+H^{C}_{5,20}+H^{C}_{5,21}-H^{C}_{5,26}+H^{C}_{5,28}-H^{C}_{5,37}-2H^{C}_{5,51}+H^{C}_{5,58}\\*
    &&&=(-2\alpha^s_{1234,5}(-2\alpha_{4,123}\alpha^s_{12,3} + \alpha_{123,4}(\alpha_{3,12} - \alpha_{12,3}) + (\alpha_{12,34} - \alpha_{34,12})\alpha^s_{3,4})  \\*
    &&&\qquad- \alpha_{5,1234}(-2\xi_{123,4}\alpha_{12,3} + \alpha_{4,123}\alpha_{3,12}+ \alpha^s_{12,34}\xi_{3,4} + \alpha_{34,12}\alpha^s_{3,4}) \\*
    &&&\qquad + 2(\alpha_{12,345}\alpha^s_{3,45} + \alpha_{123,45}\alpha_{3,12})\alpha^s_{4,5})\xi_{1,2}, \\ 
	&h_{5,46}^{G}(\vecq_{1,2,3,4})&&\equiv H^{C}_{5,11}-H^{C}_{5,45}+H^{C}_{5,57}+H^{C}_{5,59}\\*
    &&&=(-\alpha_{5,1234}\alpha^s_{12,34}\xi_{3,4} + 2\alpha^s_{123,45}\alpha_{3,12}\xi_{4,5} + \alpha_{45,123}\alpha_{12,3}\xi_{4,5})\xi_{1,2}, \\ 
	&h_{5,47}^{G}(\vecq_{1,2,3,4})&&\equiv (-H^{C}_{5,6}-H^{C}_{5,10}-H^{C}_{5,11}+H^{C}_{5,18}-H^{C}_{5,21}+2H^{C}_{5,26}-H^{C}_{5,28}+H^{C}_{5,43}-2H^{C}_{5,48}+2H^{C}_{5,62})/2\\*
    &&&=(2\alpha^s_{1234,5}(\alpha_{123,4}\alpha_{3,12} + \alpha_{12,34}\alpha^s_{3,4}) + \frac{1}{2}\alpha_{5,1234}(-2\xi_{123,4}\alpha_{12,3} + \alpha_{4,123}\alpha_{3,12} + \alpha^s_{12,34}\xi_{3,4}  \\*
    &&&\qquad+ \alpha_{34,12}\alpha^s_{3,4}))\xi_{1,2}+ 2\alpha_{12,345}\alpha^s_{3,45}\alpha^s_{1,2}\xi_{4,5} + \alpha_{345,12}(\alpha^s_{3,45}\xi_{1,2}\alpha^s_{4,5} + \alpha_{45,3}\alpha^s_{1,2}\xi_{4,5})
\end{alignat*}
The mapping from the naive to the reduced bases is the following:
\begin{alignat*}{8}
    & d^{F,G}_{5,1}\equiv &&{D}^{{\m F},{\m G}}_{5,22},  &\qquad& d^{F,G}_{5,2}\equiv &&{D}^{{\m F},{\m G}}_{5,23},  &\qquad& d^{F,G}_{5,3}\equiv &&{D}^{{\m F},{\m G}}_{5,37},  &\qquad& d^{F,G}_{5,4}\equiv &&{D}^{{\m F},{\m G}}_{5,38}-{D}^{{\m F},{\m G}}_{5,25},  \\
    & d^{F,G}_{5,5}\equiv &&{D}^{{\m F},{\m G}}_{5,41},  &\qquad& d^{F,G}_{5,6}\equiv &&{D}^{{\m F},{\m G}}_{5,59},  &\qquad& d^{F,G}_{5,7}\equiv &&{D}^{\m C}_{5,14} - {D}^{{\m F},{\m G}}_{5,14},  &\qquad& d^{F,G}_{5,8}\equiv &&{D}^{\m C}_{5,22},  \\
    & d^{F,G}_{5,9}\equiv &&{D}^{\m C}_{5,23},  &\qquad& d^{F,G}_{5,10}\equiv &&{D}^{\m C}_{5,37},  &\qquad& d^{F,G}_{5,11}\equiv && {D}^{\m C}_{5,38}-{D}^{\m C}_{5,25},  &\qquad& d^{F,G}_{5,12}\equiv &&{D}^{\m C}_{5,41}, \\  
    & d^{F,G}_{5,13}\equiv &&{D}^{\m C}_{5,46} - {D}^{{\m F},{\m G}}_{5,46},  &\qquad& d^{F,G}_{5,14}\equiv &&{D}^{\m C}_{5,47} - {D}^{{\m F},{\m G}}_{5,47},  &\qquad& d^{F,G}_{5,15}\equiv &&{D}^{\m C}_{5,55} - {D}^{{\m F},{\m G}}_{5,55},  &\qquad& d^{F,G}_{5,16}\equiv &&{D}^{\m C}_{5,56} - {D}^{{\m F},{\m G}}_{5,56}, \\
    & d^{F,G}_{5,17}\equiv &&{D}^{\m C}_{5,61},  &\qquad& d^{F,G}_{5,18}\equiv &&{D}^{{\m F},{\m G}}_{5,35},  &\qquad& d^{F,G}_{5,19}\equiv &&{D}^{{\m F},{\m G}}_{5,36},  &\qquad& d^{F,G}_{5,20}\equiv &&{D}^{{\m F},{\m G}}_{5,39}, \\
    & d^{F,G}_{5,21}\equiv &&{D}^{{\m F},{\m G}}_{5,40},  &\qquad& d^F_{5,22}\equiv &&{D}^{\m F}_{5,52},  &\qquad& d^F_{5,23}\equiv &&{D}^{\m F}_{5,53},  &\qquad& d^F_{5,24}\equiv &&{D}^{\m F}_{5,61}, \\
    & d^F_{5,25}\equiv &&{D}^{\m F}_{5,62},  &\qquad& d^F_{5,26}\equiv &&{D}^{\m F}_{5,63}-{D}^{\m F}_{5,57},  &\qquad& d^F_{5,27}\equiv &&{D}^{\m F}_{5,64},  &\qquad& d^F_{5,28}\equiv &&{D}^{\m F}_{5,65}, \\
    & d^F_{5,29}\equiv &&{D}^{\m C}_{5,35},  &\qquad& d^F_{5,30}\equiv &&{D}^{\m C}_{5,36},  &\qquad& d^F_{5,31}\equiv &&{D}^{\m C}_{5,39},  &\qquad& d^F_{5,32}\equiv &&{D}^{\m C}_{5,40}, \\
    & d^F_{5,33}\equiv &&{D}^{\m C}_{5,49},  &\qquad& d^F_{5,34}\equiv &&{D}^{\m C}_{5,50},  &\qquad& d^F_{5,35}\equiv &&{D}^{\m C}_{5,52},  &\qquad& d^F_{5,36}\equiv &&{D}^{\m C}_{5,53}, \\
    & d^F_{5,37}\equiv &&{D}^{\m C}_{5,58},  &\qquad& d^F_{5,38}\equiv &&{D}^{\m C}_{5,59},  &\qquad& d^F_{5,39}\equiv &&{D}^{\m C}_{5,62}, &\qquad& d^G_{5,22}\equiv &&{D}^{\m G}_{5,49}, \\
    & d^G_{5,23}\equiv &&{D}^{\m G}_{5,50},  &\qquad& d^G_{5,24}\equiv &&{D}^{\m G}_{5,52},  &\qquad& d^G_{5,25}\equiv &&{D}^{\m G}_{5,53},  &\qquad& d^G_{5,26}\equiv &&{D}^{\m G}_{5,58}, \\
    & d^G_{5,27}\equiv &&{D}^{\m G}_{5,61},  &\qquad& d^G_{5,28}\equiv &&{D}^{\m G}_{5,62},  &\qquad& d^G_{5,29}\equiv &&{D}^{\m G}_{5,63}-{D}^{\m G}_{5,57},  &\qquad& d^G_{5,30}\equiv &&{D}^{\m G}_{5,64}, \\
    & d^G_{5,31}\equiv &&{D}^{\m G}_{5,65},  &\qquad& d^G_{5,32}\equiv &&{D}^{\m C}_{5,8} - {D}^{\m G}_{5,8},  &\qquad& d^G_{5,33}\equiv &&{D}^{\m C}_{5,9} - {D}^{\m G}_{5,9},  &\qquad& d^G_{5,34}\equiv &&{D}^{\m C}_{5,12} - {D}^{\m G}_{5,12}, \\
    & d^G_{5,35}\equiv &&{D}^{\m C}_{5,13} - {D}^{\m G}_{5,13},  &\qquad& d^G_{5,36}\equiv &&{D}^{\m C}_{5,35},  &\qquad& d^G_{5,37}\equiv &&{D}^{\m C}_{5,36},  &\qquad& d^G_{5,38}\equiv &&{D}^{\m C}_{5,39}, \\
    & d^G_{5,39}\equiv &&{D}^{\m C}_{5,40},  &\qquad& d^G_{5,40}\equiv &&{D}^{\m C}_{5,44} - {D}^{\m G}_{5,44},  &\qquad& d^G_{5,41}\equiv &&{D}^{\m C}_{5,49},  &\qquad& d^G_{5,42}\equiv &&{D}^{\m C}_{5,50}, \\
    & d^G_{5,43}\equiv &&{D}^{\m C}_{5,52},  &\qquad& d^G_{5,44}\equiv &&{D}^{\m C}_{5,53},  &\qquad& d^G_{5,45}\equiv &&{D}^{\m C}_{5,58},  &\qquad& d^G_{5,46}\equiv &&{D}^{\m C}_{5,59}, \\
    & d^G_{5,47}\equiv &&{D}^{\m C}_{5,62}
\end{alignat*}

\section{Angle-averaged fourth order kernel}\label{F4AAappendix}

In this appendix, we provide an analytic result for the angular average
\be
  F_4^\Omega(k_1,k_2,k_3,q)\equiv \int\frac{d\Omega_q}{4\pi}F_4^{\text{EdS}}(\veck_1,\veck_2,\vecq,-\vecq)\,,
\ee
without taking any limit with respect to the relative size of $k_1=|\veck_1|, k_2=|\veck_2|, k_3=|\veck_1+\veck_2|$ and $q=|\vecq|$, respectively.
Results for the angle-average of the corresponding time-dependent kernel are not displayed here due to their length.
{\newpage
\begin{dgroup*}
\begin{dmath*}
    F_4^\Omega(k_1,k_2,k_3,q)=
\end{dmath*}
\begin{dmath*}
    \frac{1}{149022720 q^7 k_1^9 k_2^9 k_3^5}\bigg[
    4 q^3 k_1 k_2 k_3\Big[585 k_2^6 k_3^4 k_1^{14}-15 \left(-165 k_3^4 q^8+1949 k_2^2 k_3^4 q^6-3895 k_2^4 k_3^4 q^4+90 k_2^8 k_3^4+k_2^6 \left(189 q^6-210 k_3^2 q^4+1822 k_3^4 q^2+805 k_3^6\right)\right) k_1^{12}+\left(1530 k_3^4 k_2^{10}+15 \left(189 q^6-210 k_3^2 q^4+1318 k_3^4 q^2+1501 k_3^6\right) k_2^8+\left(-11340 q^8+54180 k_3^2 q^6-419226 k_3^4 q^4+222610 k_3^6 q^2+5655 k_3^8\right) k_2^6+30 \left(4347 q^6 k_3^4-3548 q^4 k_3^6\right) k_2^4+45 \left(1306 q^6 k_3^6-285 q^8 k_3^4\right) k_2^2-7425 q^8 k_3^6\right) k_1^{10}+\left(-1350 k_3^4 k_2^{12}+15 \left(189 q^6-210 k_3^2 q^4+1318 k_3^4 q^2+1501 k_3^6\right) k_2^{10}+2 \left(11340 q^8-50400 k_3^2 q^6+370461 k_3^4 q^4-227830 k_3^6 q^2+7050 k_3^8\right) k_2^8+k_3^2 \left(7560 q^8-148950 k_3^2 q^6+250234 k_3^4 q^4-120370 k_3^6 q^2+5415 k_3^8\right) k_2^6+15 \left(471 k_3^4 q^8-6566 k_3^6 q^6+2595 k_3^8 q^4\right) k_2^4+135 \left(146 q^8 k_3^6-221 q^6 k_3^8\right) k_2^2+7425 q^8 k_3^8\right) k_1^8+\left(585 k_3^4 k_2^{14}-15 \left(189 q^6-210 k_3^2 q^4+1822 k_3^4 q^2+805 k_3^6\right) k_2^{12}+\left(-11340 q^8+54180 k_3^2 q^6-419226 k_3^4 q^4+222610 k_3^6 q^2+5655 k_3^8\right) k_2^{10}+k_3^2 \left(7560 q^8-148950 k_3^2 q^6+250234 k_3^4 q^4-120370 k_3^6 q^2+5415 k_3^8\right) k_2^8+2 k_3^4 \left(5175 q^8-28605 k_3^2 q^6+4766 k_3^4 q^4-34455 k_3^6 q^2+210 k_3^8\right) k_2^6+45 q^4 k_3^6 \left(135 q^4+148 k_3^2 q^2+202 k_3^4\right) k_2^4+15 q^6 k_3^8 \left(20 k_3^2-459 q^2\right) k_2^2-2475 q^8 k_3^{10}\right) k_1^6+15 q^4 k_2^6 k_3^4 \left(3895 k_2^6+\left(8694 q^2-7096 k_3^2\right) k_2^4+\left(471 q^4-6566 k_3^2 q^2+2595 k_3^4\right) k_2^2+606 k_3^6+444 q^2 k_3^4+405 q^4 k_3^2\right) k_1^4-15 q^6 k_2^6 k_3^4 \left(k_2^2-k_3^2\right) \left(1949 k_2^4+\left(855 q^2-1969 k_3^2\right) k_2^2+20 k_3^4-459 q^2 k_3^2\right) k_1^2+2475 q^8 k_2^6 k_3^4 \left(k_2^2-k_3^2\right)^3\Big]\\
    -15 k_2^7 k_3^5 \left(q^3-q k_1^2\right)^2\Big[39 k_1^{12}-\left(1947 q^2+582 k_2^2+791 k_3^2\right) k_1^{10}+\left(1689 q^4+1787 k_3^2 q^2+51 k_2^4+601 k_3^4+k_2^2 \left(6891 q^2+2120 k_3^2\right)\right) k_1^8+\left(219 q^6-1113 k_3^2 q^4+179 k_3^4 q^2+492 k_2^6+151 k_3^6-3 k_2^4 \left(2155 q^2+211 k_3^2\right)+k_2^2 \left(-7284 q^4+758 k_3^2 q^2+390 k_3^4\right)\right) k_1^6+q^2 \left(1017 k_2^6+3 \left(2423 q^2-559 k_3^2\right) k_2^4+\left(471 q^4-4376 k_3^2 q^2+391 k_3^4\right) k_2^2+269 k_3^6-321 q^2 k_3^4+405 q^4 k_3^2\right) k_1^4-3 q^4 \left(k_2^2-k_3^2\right) \left(558 k_2^4+\left(285 q^2-473 k_3^2\right) k_2^2-17 \left(5 k_3^4+9 q^2 k_3^2\right)\right) k_1^2+165 q^6 \left(k_2^2-k_3^2\right)^3\Big] T_1(k_1)\\
    -15 k_1^7 k_3^5 \left(q^3-q k_2^2\right)^2 \Big[39 k_2^{12}-\left(1947 q^2+791 k_3^2\right) k_2^{10}+\left(1689 q^4+1787 k_3^2 q^2+601 k_3^4\right) k_2^8+\left(219 q^6-1113 k_3^2 q^4+179 k_3^4 q^2+151 k_3^6\right) k_2^6+\left(405 k_3^2 q^6-321 k_3^4 q^4+269 k_3^6 q^2\right) k_2^4-51 \left(9 k_3^4 q^6+5 k_3^6 q^4\right) k_2^2-165 q^6 k_3^6+3 k_1^6 \left(55 q^6-558 k_2^2 q^4+339 k_2^4 q^2+164 k_2^6\right)-3 k_1^4 \left(-17 k_2^8+\left(2155 q^2+211 k_3^2\right) k_2^6+\left(559 q^2 k_3^2-2423 q^4\right) k_2^4+\left(285 q^6-1031 q^4 k_3^2\right) k_2^2+165 q^6 k_3^2\right)+k_1^2 \left(-582 k_2^{10}+\left(6891 q^2+2120 k_3^2\right) k_2^8+\left(-7284 q^4+758 k_3^2 q^2+390 k_3^4\right) k_2^6+\left(471 q^6-4376 k_3^2 q^4+391 k_3^4 q^2\right) k_2^4+6 \left(219 q^6 k_3^2-194 q^4 k_3^4\right) k_2^2+495 q^6 k_3^4\right)\Big] T_1(k_2) \\
    -105 q^2 k_1^7 k_2^7 \left(q^2-k_3^2\right) \left(9 q^2+2 k_3^2\right) \Big[\left(-3 q^4+2 k_3^2 q^2+k_3^4\right) k_1^6+\left(-12 q^6+52 k_3^2 q^4-56 k_3^4 q^2+16 k_3^6+k_2^2 \left(3 q^4-2 k_3^2 q^2-9 k_3^4\right)\right) k_1^4+\left(-15 k_3^8+46 q^2 k_3^6-47 q^4 k_3^4+8 q^6 k_3^2+k_2^4 \left(3 q^4-2 k_3^2 q^2-9 k_3^4\right)+8 k_2^2 \left(3 q^6-12 k_3^2 q^4+17 k_3^4 q^2-4 k_3^6\right)\right) k_1^2+k_2^6 \left(-3 q^4+2 k_3^2 q^2+k_3^4\right)-2 k_3^4 \left(-2 q^6+k_3^2 q^4+k_3^6\right)+4 k_2^4 \left(-3 q^6+13 k_3^2 q^4-14 k_3^4 q^2+4 k_3^6\right)+k_2^2 \left(-15 k_3^8+46 q^2 k_3^6-47 q^4 k_3^4+8 q^6 k_3^2\right)\Big] T_1\left(k_3\right)\\
    +k_1^2 k_2^2 \left(9 q^2+2 k_3^2\right)\tilde{T}_2\left(k_1,k_2,k_3\right)+k_1^2 k_2^2 \left(9 q^2+2 k_3^2\right)\tilde{T}_2\left(k_2,k_1,k_3\right)+4 k_1^2 k_2^2 \left(q^2-k_3^2\right)\tilde{T}_2\left(k_3,k_1,k_2\right)+3840 q^2\tilde{T}_3 k_1^7 \left(q^2-k_1^2\right)^2 k_2^7 \left(q^2-k_2^2\right)^2 k_3^8\bigg],
\end{dmath*}
\end{dgroup*}}
with
\begin{align*}
    S(k_1,k_2,k_3) &\equiv \sqrt{q^2 \left(k_1^2-k_3^2+q^2\right)+k_2^2 \left(k_3^2-q^2\right)},\\
    \tilde{T}_2 (k_i,k_j,k_k)&\equiv \frac{840 q^2 k_1^5 k_2^5 k_3^5 k_i^3 \left(k_j^2-q^2\right)^2 \left(k_k^2-q^2\right)^2 T_2(k_i,k_j,k_k)}{\left(k_3^2-q^2\right) S(k_i,k_j,k_k)},\\
    \tilde{T}_3&\equiv T_3/S(k_3, k_1, k_2)
\end{align*}
and the inverse hyperbolic functions
\begin{align*}
    T_1(k)&\equiv2\text{ artanh}\left[\frac{2 k q}{k^2 + q^2}\right],\\
    T_2(k_1,k_2,k_3)&\equiv2\text{ artanh}\left[\frac{2 k_1 q S(k_1,k_2,k_3)}{-q^2 \left(-2 k_1^2+k_2^2+k_3^2\right)+k_2^2 k_3^2+q^4}\right],\\
    T_3&\equiv2\text{ artanh}\left[\frac{4 k_3 q \left(-q^2 \left(k_1^2+k_2^2-2 k_3^2\right)+k_1^2 k_2^2+q^4\right) S(k_3,k_1,k_2)}{8 k_3^2 q^2 \left(q^2-k_1^2\right) \left(q^2-k_2^2\right)+\left(k_1^2-q^2\right)^2 \left(q^2-k_2^2\right)^2+8 k_3^4 q^4}\right].
\end{align*}
These are the only trigonometric expressions appearing for the angular average of the corresponding individual $h$-functions as well.

\providecommand{\href}[2]{#2}\begingroup\raggedright\endgroup

\end{document}